\documentclass[%twocolumn,%showpacs,preprintnumbers,
amsmath,%amssymb,aps,
prd,nofootinbib,floatfix,11pt,%preprint,
]{revtex4}

\usepackage{hyperref}

\usepackage{placeins}
\usepackage{afterpage}

\usepackage{graphicx}% Include figure filesf
\usepackage{setspace}
\usepackage{bm}% bold math
\usepackage[dvips]{color}
\definecolor{Red}{cmyk}{0,1,1,0}
\definecolor{BrickRed}{cmyk}{0,0.89,0.94,0.28}
\definecolor{Blue}{cmyk}{1,1,0,0}
\definecolor{Green}{cmyk}{1,0,1,0}

\newcommand\beq{\begin{eqnarray}}
\newcommand\eeq{\end{eqnarray}}

\newcommand\signmudel{n}

\def\lsim{\mathrel{\rlap{\lower4pt\hbox{$\sim$}}
    \raise1pt\hbox{$<$}}}                % less than or approx. symbol
\def\gsim{\mathrel{\rlap{\lower4pt\hbox{$\sim$}}
    \raise1pt\hbox{$>$}}}            

\allowdisplaybreaks
\begin{document}

\renewcommand{\theequation}{\arabic{section}.\arabic{equation}}
\renewcommand{\thefigure}{\arabic{section}.\arabic{figure}}
\renewcommand{\thetable}{\arabic{section}.\arabic{table}}

\title{\Large \baselineskip=20pt 
The curtain lowers on directly detectable higgsino dark matter}
\author{Stephen P.~Martin}
\affiliation{\mbox{\it Department of Physics, Northern Illinois University, DeKalb IL 60115}}

\begin{abstract}\normalsize \baselineskip=16pt
A higgsino could be some or all of the dark matter, with a mass bounded from above by about 1.1 TeV assuming a thermal freezeout density,  and from below by collider searches. Direct detection experiments imply purity constraints on a dark matter higgsino, limiting the mixing with the electroweak gauginos. Using the new strong limits available as of the end of 2024 from  the LUX-ZEPLIN experiment, I quantify the resulting lower bounds on gaugino masses and upper bounds on higgsino mass splittings, assuming that the scalar superpartners and Higgs bosons of minimal supersymmetry are in the decoupling limit. Similar bounds are projected for the critical future scenario that direct detection experiments reach the neutrino fog that hampers discovery prospects. 
\end{abstract}

\maketitle

\tableofcontents

\baselineskip=15.7pt

\newpage

%%%%%%%%%%%%%%%%%%%%%%%%%%%%%%%%%%%%%%%%%%%%%%%%%%%%%%%%%%%%%%%
\section{Introduction\label{sec:intro}}
\setcounter{equation}{0}
\setcounter{figure}{0}
\setcounter{table}{0} 
\setcounter{footnote}{1}

Indirect experimental evidence, coming from astrophysical observations on distances 
ranging from spiral galaxies to the largest cosmological scales, strongly suggests
the existence of some form of dark matter, with
a total density, averaged over the largest distance scales, given  by \cite{Planck:2018vyg}
\beq
\Omega_{\rm DM} h^2 = 0.1200 \pm 0.0012,
\label{eq:observedOmega}
\eeq 
where $h = ({\rm Mpc} \cdot {\rm sec}/100 \>{\rm km}) H^0$
in terms of the present-time Hubble expansion rate $H_0$, and $\Omega_{\rm DM}$
is the ratio of the total dark matter density to the critical density.
(For a recent comprehensive review of dark matter, see \cite{Cirelli:2024ssz}.) 
One of the historical motivations for considering softly broken supersymmetry
as an extension of the Standard Model is the fact that if $R$-parity is conserved,
the lightest neutral superpartner will be a dark matter particle, and the density predicted by thermal freezeout could, at least very roughly, accommodate the value in eq.~(\ref{eq:observedOmega}). However, direct searches for superpartners at colliders, including the CERN LEP $e^+ e^-$ collider, the Fermilab Tevatron $p \overline p$ collider, and the $pp$ Large Hadron Collider (LHC), have reduced the parameter space within the Minimal Supersymmetric Standard Model (MSSM)
in which the dark matter could be explained as a neutralino. At the same time, direct
and indirect dark matter detection experiments have also reduced the available parameter
space.

One supersymmetric dark matter possibility that remains viable and theoretically motivated is that the lightest supersymmetric particle
(LSP) could be a mostly higgsino neutralino, symbolized below as $\tilde N_1$. The other higgsino-like mass eigenstates consist of a slightly heavier chargino $\tilde C_1$ and another neutralino $N_2$.
To the extent that the mixing with the heavier gaugino-like neutralinos and charginos is small,
these states should be approximately degenerate in mass. Besides the direct searches at the LHC, there are several other independent indications
suggesting that the other superpartners should be much heavier. First, flavor-mixing and CP-violating bounds are most easily accommodated by squark and slepton masses well above the 1 TeV scale \cite{Wells:2003tf,Arkani-Hamed:2004zhs,Wells:2004di}. 
Second, the fact that the lightest Higgs boson mass of 125.1 GeV is in the upper part of the range allowed by the MSSM is easiest to explain \cite{Haber:1990aw,Okada:1990vk,Ellis:1990nz,Haber:1996fp} by radiative corrections involving top squarks  in the range of at least several TeV. Third, the precise unification of gauge couplings favors superpartner masses in the several TeV range or higher \cite{Bhattiprolu:2023lfh}. This will be the baseline scenario for the following.
The case of light higgsinos and much heavier gauginos, squarks, and sleptons is a notoriously difficult one to probe at colliders, with current bounds restricted to below about 200 GeV for the LSP. 

Assuming that the LSP is mostly higgsino, and that its density today is determined by thermal freezeout in the standard cosmology (i.e., without significant impacts from
nonthermal sources such as out-of-equilibrium decays of heavier particles), the prediction $\Omega h^2 = 0.12$ is realized
for $m_{\tilde N_1}$ near 1.1 TeV, with some weak dependence on other MSSM parameters. If the mass is less than this critical value, then the predicted higgsino contribution to the density today is correspondingly lower, $\Omega_{\mbox{\scriptsize LSP}} h^2 < 0.12$, but with a more significant dependence on the MSSM parameters. This is certainly an allowed possibility, since the rest of the dark matter density could be some other species such as axions. This has been argued for and studied in detail in refs.~\cite{Bae:2013bva,Bae:2014yta,Bae:2017hlp,Baer:2019uom}, for example.
Below, I will refer to this as the ``thermal" $\Omega_{\mbox{\scriptsize LSP}} h^2$ case, and the predicted direct detection rates are accordingly reduced by a factor of the ratio
\beq
\xi = \Omega_{\mbox{\scriptsize LSP}} h^2/0.12,
\eeq 
while most indirect detection rates (excepting annihilations to neutrinos in the Sun) are suppressed by $\xi^2$. If $m_{\tilde N_1}$ is of order 100 GeV (near the low end of its allowed range because of collider limits on charginos), then $\xi$ could easily be at the per cent level.

Another possibility for $m_{\tilde N_1} < 1.1$ TeV is that there is some nonthermal source for the higgsino-like dark matter, such as a late-decaying particle, which can 
\cite{Moroi:1999zb,Fujii:2001xp,Gelmini:2006pw,Gelmini:2006pq,Han:2019vxi,Fukuda:2024ddb}
bring the density up to the cosmologically favored value $\Omega_{\mbox{\scriptsize LSP}} h^2 = 0.12$ without need for another dark matter particle. I will refer to this possibility as the ``nonthermal" scenario; it is much more strongly constrained than the thermal case if $m_{\tilde N_1}$ is less than 1.1 TeV, because it assumes $\xi=1$ so that there is no rate suppression proportional to $\xi$  or $\xi^2$ for direct or indirect detection processes.
Other important works studying various aspects of higgsino dark matter, including their collider signatures, are found in 
\cite{Drees:1996pk,Thomas:1998wy,Feng:2000gh,Giudice:2004tc,Profumo:2004at,Hisano:2004ds,Baer:2011ec,Baer:2012cf,Baer:2013yha,Schwaller:2013baa,Baer:2014cua,Han:2014kaa,Low:2014cba,Nagata:2014wma,Evans:2014pxa,Bae:2015jea,Mahbubani:2017gjh,Fukuda:2017jmk,Kowalska:2018toh,Baer:2018rhs,Han:2018wus,Fukuda:2019kbp,Baer:2020sgm,Rinchiuso:2020skh,Delgado:2020url,Co:2021ion,Baer:2021srt,Carpenter:2021jbd,Evans:2022gom,Baer:2022qrw,Dessert:2022evk,Carpenter:2023agq,Bisal:2023fgb,Bisal:2024ezn,Ibe:2023dcu,Rodd:2024qsi}.

Indirect dark matter detection could occur in a variety of ways, including searches by the IceCube experiment for neutrinos from the Sun \cite{IceCube:2016dgk}, gamma rays from the galactic center using the ground-based Cherenkov telescope array H.E.S.S.
\cite{HESS:2016mib,HESS:2018cbt,HESS:2022ygk}, and from annihilation in dwarf galaxies using the Fermi-LAT space-based gamma-ray telescope \cite{Fermi-LAT:2015ycq,Fermi-LAT:2015att,MAGIC:2016xys,Fermi-LAT:2016uux}. In the future, the 
Cherenkov Telescope Array \cite{CTAConsortium:2017dvg}
and the Southern Wide-field Gamma-ray Observatory \cite{Albert:2019afb} may be able 
\cite{Rinchiuso:2020skh,Rodd:2024qsi}
to discover a thermal relic higgsino with the critical mass near 1.1 TeV, if the dark matter density profile is favorable.

However, the strongest present direct detection constraints on a dark matter higgsino,
relying on its density in the local region and its couplings to nucleons, have been obtained from liquid xenon detectors, most recently \cite{PandaX-4T:2021bab,XENON:2023cxc,LZ:2022lsv,PandaX:2024qfu,LZCollaboration:2024lux}. 
If the gauginos masses are extremely large, more than a few times $10^7$ GeV, then the light higgsino masses are split by less than about 200 keV, and higgsino dark matter would be ruled out by $Z$-mediated inelastic scattering of the LSP to $m_{\tilde N_2}$. 
For the more natural MSSM case of lighter gauginos, as assumed here, inelastic scattering is not an issue, and the bounds become weaker as the gaugino mass parameters are taken larger, corresponding to purer higgsino states with smaller mass splittings 
\beq
\Delta M_+ &\equiv& m_{\tilde C_1} - m_{\tilde N_1},
\\
\Delta M_0 &\equiv& m_{\tilde N_2} - m_{\tilde N_1}.
\eeq 
These mass differences are important for collider searches, since they affect the kinematics of decay products
and the possibility of observing displaced decays or disappearing tracks from the 
macroscopic decay length of the higgsino-like chargino. In ref.~\cite{Martin:2024pxx}, I studied
the constraints on higgsino purity and mass splittings implied by the 2022 LUX-ZEPLIN 
direct detection limits \cite{LZ:2022lsv}, quantifying the impact on collider searches. Since then, the cross-section limits released by the LUX-ZEPLIN collaboration in 2024 (LZ2024) \cite{LZCollaboration:2024lux} have pushed the limits approximately an order of magnitude stronger. 
Looking to the future, the prospects for discovery in direct detection experiments will become background-limited 
due to scattering of astrophysical neutrinos \cite{Billard:2013qya}, the neutrino fog, for which the criteria in ref.~\cite{OHare:2021utq} is used in the following. 
The purpose of the present paper is to update and extend the results of \cite{Martin:2024pxx}, using the LZ2024 limits, and to project what these results will be in the future if the neutrino fog is reached.

%%%%%%%%%%%%%%%%%%%%%%%%%%%%%%%%%%%%%%%%%%%%%%%%%%%%%%%%%%%%%%%
\section{Parameterization of mixed higgsinos\label{sec:mixing}}
\setcounter{equation}{0}
\setcounter{figure}{0}
\setcounter{table}{0} 
\setcounter{footnote}{1}

In the MSSM, the parameters governing the neutralino and chargino masses and couplings
at tree level consist of the $U(1)_Y$ and $SU(2)_L$ gauge couplings $g'$ and $g$ and gaugino masses $M_1$ and $M_2$, the Higgsino mass parameter $\mu$, the mixing angle $\alpha$
in the neutral $CP$-even Higgs sector, and the angle $\beta$ related to the ratio of Higgs vacuum expectation values by $\tan\beta = \langle H_u^0 \rangle/\langle H_d^0 \rangle$. In the following, I take the lightest Higgs boson mass to be fixed at 125.1 GeV, and the other Higgs bosons, all squarks and sleptons, and the gluino to be decoupled with masses at 10 TeV, as motivated (but certainly not proved) by the considerations
reviewed in the Introduction. I also impose the decoupling relation $\alpha = \beta - \pi/2$. 

Following the conventions in ref.~\cite{Martin:1997ns},
the tree-level mass matrix for the neutralinos are given in the basis $(\tilde B, \tilde W, \tilde H_d^0, \tilde H_u^0)$ by
\beq
{\bf M}_{\tilde N} &=& 
\begin{pmatrix} 
M_1 & 0 & -s_W c_\beta m_Z & s_W s_\beta m_Z \\
0 & M_2 & c_W c_\beta m_Z & -c_W s_\beta m_Z \\
-s_W c_\beta m_Z & c_W c_\beta m_Z & 0 & -\mu \\
s_W s_\beta m_Z & -c_W s_\beta m_Z & -\mu & 0
\end{pmatrix}
,
\eeq
and the mass matrix for the charginos in the basis $(\tilde W^+, \tilde H_u^+, \tilde W^-, \tilde H_d^-)$ is
\beq
{\bf M}_{\tilde C} = \begin{pmatrix}
0 & {\bf X}^T \\
{\bf X} & 0\end{pmatrix},
\qquad\quad
{\bf X} = \begin{pmatrix}
M_2 & \sqrt{2} s_\beta c_W m_Z \\
\sqrt{2} c_\beta c_W m_Z & \mu
\end{pmatrix}.
\eeq
Here $s_W = g'/\sqrt{g^2 + g^{\prime 2}}$ and $c_W = g/\sqrt{g^2 + g^{\prime 2}}$ are the sine and cosine of the weak mixing angle, and $s_\beta$ and $c_\beta$ are the sine and cosine of $\beta$, and $s_{2\beta} = \sin(2 \beta)$ in the following. Now one can treat electroweak symmetry breaking as a perturbation, and expand the masses and couplings in small $m_Z$. (In many works on light higgsinos the less accurate special case hierarchical assumption $|M_1|, |M_2| \gg |\mu| \gg m_Z$ is adopted instead.) 

In general, $M_1$, $M_2$, and $\mu$ could have arbitrary phases, but there are strong
constraints on this coming from CP violating constraints, particularly from the electric
dipole moment of the electron \cite{Roussy:2022cmp}. These constraints are quite stringent  even if the gauginos and other superpartners
are in the multi-TeV range; for discussions in the context of the MSSM, see \cite{Cesarotti:2018huy,Co:2021ion}. Therefore for simplicity I will assume that these mass parameters are all (approximately) real in the convention where the Higgs VEVs are real,
even though it is possible to evade CP-violating constraints for general phases by an
extreme decoupling of superpartners through large masses. However, I now generalize the treatment in ref.~\cite{Martin:2024pxx} by allowing their signs
to be arbitrary, instead of assuming that $M_1$ and $M_2$ have the same sign.
Since the focus is on the possibility of a higgsino-like LSP,
I assume $|\mu| < |M_1|, |M_2|$.  
Now, define the real (but possibly negative) quantity
\beq
\delta \,\equiv\, 
 \frac{c_W^2 (M_2 + s_{2\beta} \mu)}{M_2^2 - \mu^2} +
\frac{s_W^2 (M_1 + s_{2\beta} \mu)}{M_1^2 - \mu^2} 
,
\eeq
in terms of which the neutral higgsino mass splitting, at leading order in small $m_Z$, is 
\beq
\Delta M_0 \,=\, m_Z^2 |\delta|,
\label{eq:DeltaM0delta}
\eeq
for any signs of $\mu$, $M_1$, and $M_2$. Note that $\delta$ can be negative only if
at least one of $M_1$ and $M_2$ is negative, regardless of the sign of $\mu$, due to the assumption that the LSP is higgsino-like.
Let us now define
\beq
\signmudel  &\equiv& {\rm sign}(\mu \delta) \>=\> \pm 1.
\label{eq:definesigma}
\eeq
[Note that if $M_1$ and $M_2$ are both positive, then $\signmudel  = {\rm sign}(\mu)$.]
The tree-level mass difference between the charged higgsino-liked particle and the LSP,
at leading order in small $m_Z$, can be written as
\beq
\Delta M_+ 
%\equiv m_{\tilde C_1} - m_{\tilde N_1} 
\,=\, 
{\rm sign}(\delta) \, \frac{m_Z^2}{2}
\left [ 
\frac{c_W^2 (1 - \signmudel  s_{2 \beta})}{M_2 + \signmudel  \mu}
+ 
\frac{s_W^2 (1 + \signmudel  s_{2 \beta})}{M_1 - \signmudel  \mu}
\right ]
.
\eeq
Note that it is sometimes possible for $\Delta M_+$ to be negative, implying that the chargino is the LSP, which of course would be fatal for a dark matter interpretation.

The Lagrangian governing the couplings of the dark matter LSP to the
125.1 GeV Higgs boson (denoted $h$) and the $Z$ boson is
\beq
{\cal L} &=& 
-\sqrt{g^2 + g^{\prime 2}} \left ( \frac{1}{2} y_{h}^{\phantom{Z}} h \tilde N_1 \tilde N_1 + {\rm c.c.}  + g_{Z}^{\phantom{Z}} 
Z_\mu \tilde N_1^\dagger \overline\sigma^\mu \tilde N_1 \right )
,
\eeq 
where, in general, the tree-level LSP-boson couplings in the MSSM are
\beq
y_h &=& 
\left (
s_W N_{11}^* - c_W N_{12}^*
\right )\left (
s_\alpha N_{13}^* + c_\alpha N_{14}^* 
\right),
\\
g_Z &=& \left (|N_{14}|^2 - |N_{13}|^2 \right )/2,
\eeq
where the unitary neutralino mixing matrix elements $N_{ij}$ are chosen such that $N^* {\bf M}_{\tilde N} N^{-1}$ is diagonal with positive real eigenvalues.
Defining a common factor
\beq
\chi \,=\, \frac{c_W^2}{M_2 - \signmudel  \mu} +  \frac{s_W^2}{M_1 - \signmudel  \mu},
\label{eq:defchi}
\eeq
one finds
in the decoupling limit $\alpha = \beta - \pi/2$ and the small $m_Z$ limit that the squares
of the couplings are 
\beq
y_h^2 &=& \frac{m_Z^2}{4} \left ( 1 + \signmudel  s_{2\beta} \right )^2 \chi^2
,
\label{eq:yh2}
\\
g_Z^2 &=& \frac{m_Z^4}{16 \mu^2} \left (1 - 2 s_\beta^2 \right )^2 \chi^2
.
\label{eq:yZ2}
\eeq
These are in turn proportional to the most important contributions to, respectively, the spin-independent (SI) and spin-dependent (SD) cross-sections for scattering of the LSP off of nucleons.

It is clear from eqs.~(\ref{eq:defchi})-(\ref{eq:yZ2}) that both the SI and SD cross-sections are suppressed
in the limit of purer higgsinos, corresponding to $m_Z \ll |M_1 - \signmudel  \mu|$ and
$|M_2 - \signmudel  \mu|$, which is also correlated with small $\Delta M_0$ and $\Delta M_+$.

\begin{figure}[!t]
\begin{minipage}[]{0.57\linewidth}
\begin{flushleft}
\includegraphics[width=0.95\linewidth]{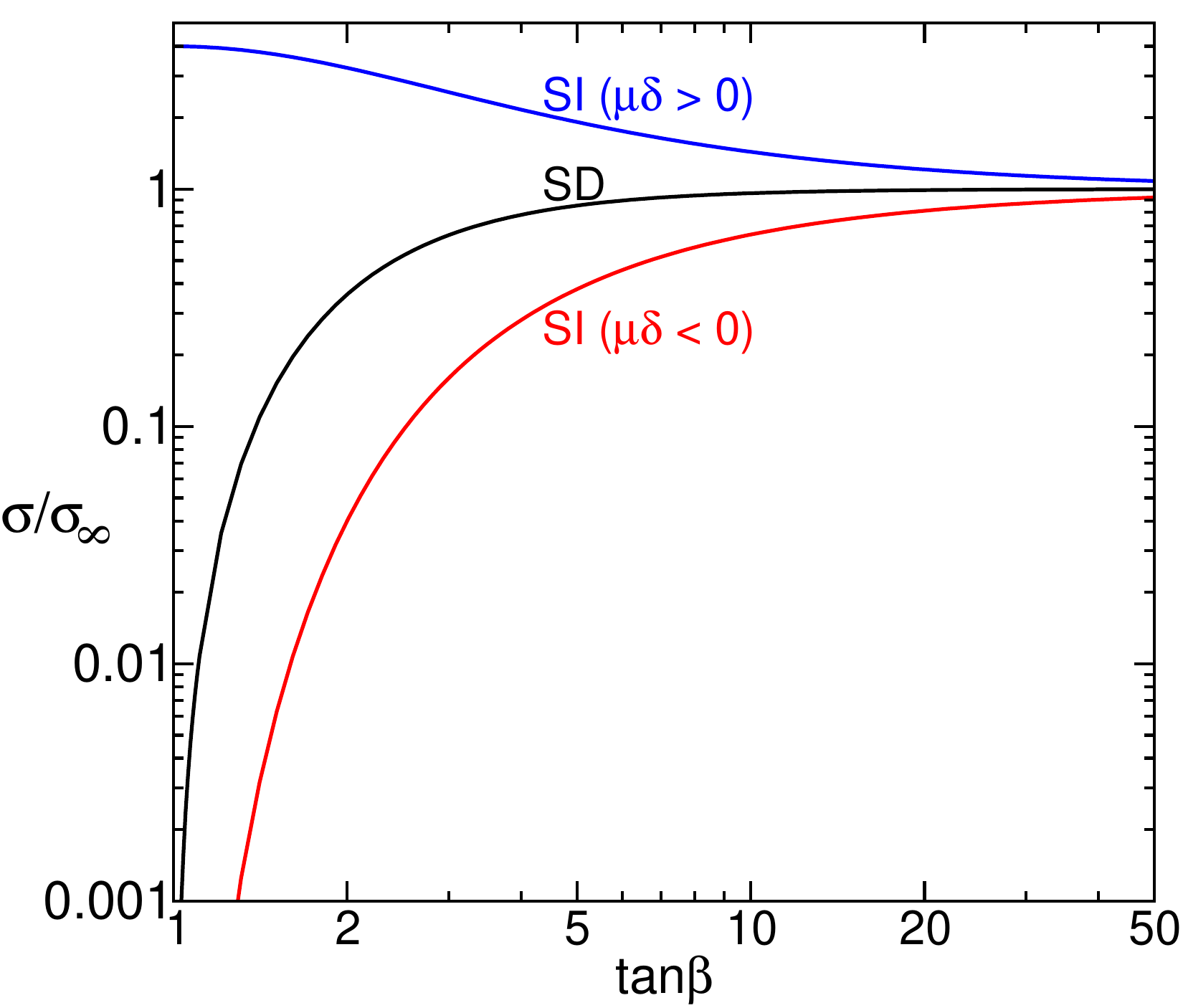}
\end{flushleft}
\end{minipage}
\begin{minipage}[]{0.42\linewidth}
\caption{\label{fig:sigmaratio}The dependences of spin-independent (SI) and spin-dependent (SD) cross-sections on $\tan\beta$,
compared to their respective large $\tan\beta$ limits, for all other parameters held fixed, in the approximation that only $h$ exchange is important for the SI cross-section and only $Z$
exchange for the SD cross-section.
The SI cross-section has two branches, depending on the sign of $\mu\delta$. It has a blind spot for negative $\signmudel  = {\rm sign}(\mu \delta)$ and small $\tan\beta \rightarrow 1$, while the SD cross-section has a weaker suppression for small $\tan\beta$, independent of the sign of $\signmudel $.}
\end{minipage}
\end{figure}

There are also blind spots \cite{Cheung:2012qy} 
for small\footnote{However, approaching too closely to the exact blind spot $\tan\beta=1$ is well-known to be problematic in the MSSM, for two reasons. First, it is difficult to achieve $M_h = 125.1$ GeV, since the tree-level contribution to the lightest Higgs boson squared mass is $m_Z^2 \cos^2(2\beta)$. A possible way to evade this is to add vectorlike quark supermultiplets \cite{Moroi:1992zk,Moroi:1991mg,Babu:2004xg,Babu:2008ge,Martin:2009bg,Graham:2009gy}
with dominantly singlet mass terms but large Yukawa couplings, which supplement the top-stop radiative corrections to $M_h$. The second problem is that the top-quark Yukawa coupling is proportional to $m_t/\sin\beta$, and so will become nonperturbatively large when renormalization group evolved into the UV if $\tan\beta$ is too small. Therefore, in the following I will restrict attention to $\tan\beta \geq 1.6$, although it may be possible to evade this as well by modifying the theory at high energy scales.}
$\tan\beta$ in the SI cross-section if $\signmudel  = -1$, and
in the SD cross-section regardless of $n$. In fact,
for fixed $\mu$, $M_1$, and $M_2$, the SI cross-section monotonically decreases for smaller $\signmudel /\tan\beta$. This is illustrated in Figure \ref{fig:sigmaratio}, which shows the ratios
of the SI and SD cross-sections as a function of $\tan\beta$, compared to their limits
as $\tan\beta \rightarrow \infty$. 
%%%%%%%%%%%%%%%%%%%%%%%%%%%%%%%%%%%%%%%%%%%%%%%%%%%%%%%%%%%%%%%%%%%%%%%%%%%%%%%%%
%%%%%%%%%%%%%%%%%%%%%%%%%%%%%%%%%%%%%%%%%%%%%%%%%%%%%%%%%%%%%%%%%%%%%%%%%%%%%%%%%
This shows that 
direct detection bounds always have a lesser impact on the $M_1, M_2$ parameter space for small $\tan\beta$ and negative $\signmudel $, which in particular corresponds to negative $\mu$ if $M_1$ and $M_2$ are both positive. 
Note that for small $\tan\beta$ the suppression of the SI cross-section is stronger than that of the SD cross-section, so that the latter can have a greater impact provided
that $\signmudel $ is negative. Conversely, the SI cross-section is enhanced by as much as a factor of 4 for small $\tan\beta$ if $\signmudel $ is positive, compared to the large $\tan\beta$ limit. 

\begin{figure}[!t]
\begin{minipage}[]{0.55\linewidth}
\begin{flushleft}
\includegraphics[width=0.95\linewidth]{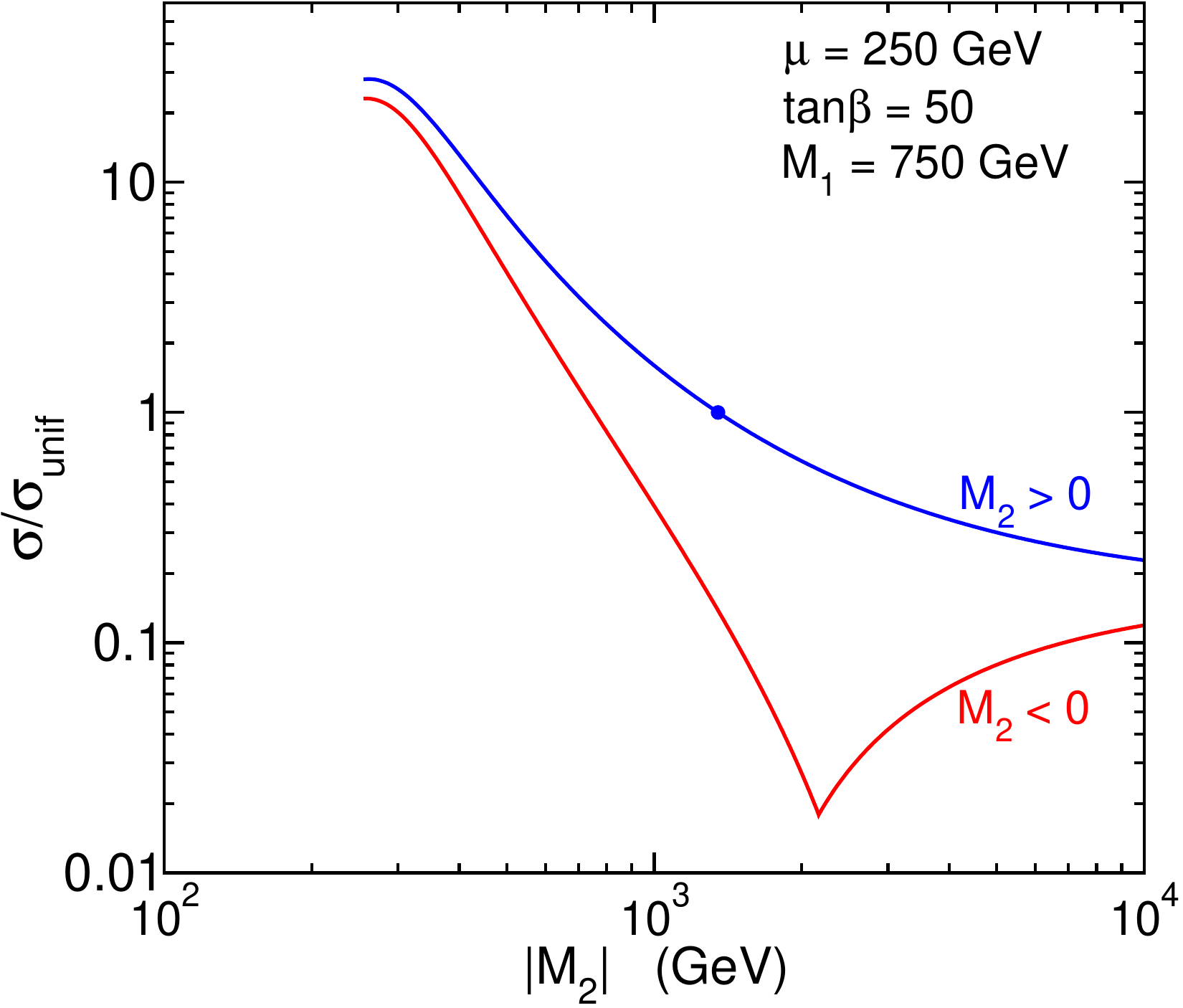}
\end{flushleft}
\end{minipage}
\begin{minipage}[]{0.44\linewidth}
\caption{\label{fig:sigmaratioM2}
The SI cross-section as a function of $|M_2|$, relative to its value for the gaugino
mass unification case $M_2 = 1.8 M_1$, for an illustrative choice with fixed $\mu = 250$ GeV, $\tan\beta=50$, and $M_1 = 750$ GeV. The upper curve has $M_2 > 0$, with the reference point $M_2 = 1.8 M_1$ indicated by a dot. The lower curve has $M_2 < 0$, showing the quasiblind spot for 
$M_2 \approx -2165$ GeV, near the point where $\delta = 0$. At this quasiblind-spot minimum, the cross-section is highly suppressed but does not vanish. The SD cross-section behaves very similarly.}
\end{minipage}
\end{figure}

Equations (\ref{eq:defchi})-(\ref{eq:yZ2}) suggest that there can also be an approximate blind spot in both the SI and SD cross-sections if $M_1$ and $M_2$ have opposite signs in such a way that $\chi$ is small. While this can indeed lead to highly suppressed cross-sections, it is not possible to achieve a perfect blind spot by tuning $\chi=0$ exactly. To prove it, note that we can use $\signmudel^2 = 1$ to write 
\beq
 \chi = \delta
+ (\signmudel  - s_{2\beta}) \mu 
\left (\frac{c_W^2}{M_2^2 - \mu^2} + \frac{s_W^2}{M_1^2 - \mu^2} \right )
.
\label{eq:deltarewritten}
\eeq
Now if one attempts to tune $\chi=0$ exactly, then from eq.~(\ref{eq:deltarewritten}) we see that $\delta$ would necessarily have the opposite sign of
$\signmudel  \mu$, which contradicts the definition of $\signmudel $ in eq.~(\ref{eq:definesigma}). Instead, the highest degree of suppression in the SI and SD
cross-sections is achieved for $\delta \approx 0$, where $\chi$ (and therefore $y_h$) is also small, but non-zero.
Another way to say the same thing is that although one of the higgsino-like neutralinos can be tuned to have vanishing tree-level couplings to $h$ and $Z$, that neutralino can only be $\tilde N_2$, never the LSP. As an illustration, Figure \ref{fig:sigmaratioM2} shows how the SI cross-section $\sigma$ (approximately proportional to $\chi^2$)
depends on $M_2$, for fixed values $\mu = 250$ GeV, $M_1 = 750$ GeV, and $\tan\beta = 50$. The plotted quantity is the ratio of the SI cross-section $\sigma$ to the value $\sigma_{\rm unif}$ it takes when $M_2 = 1.8 M_1 = 1350$ GeV, the approximate condition implied by gaugino mass unification models. 
%%%%%%%%%%%%%%%%%%%%%%%%%%%%%%%%%%%%%%%%%%%%%%%%%%%%%%%%%%%%%%%%%%%%%%%%%%%%%%%%%
%%%%%%%%%%%%%%%%%%%%%%%%%%%%%%%%%%%%%%%%%%%%%%%%%%%%%%%%%%%%%%%%%%%%%%%%%%%%%%%%%
Note that the cross-section is highly suppressed when $M_2 \approx -2165$ GeV (corresponding very nearly to $\delta = 0$),
but it never vanishes, as $\chi=0$ is not possible. 

It is also important that, unlike the $\tan\beta=1$ blind spot, this SI and SD  
cross-section suppression is closely associated with a similarly suppressed mass splitting among the neutral higgsinos, as it corresponds to the crossing of the two smallest neutralino mass matrix eigenvalues. This can be seen by noting that from eq.~(\ref{eq:DeltaM0delta}), $\delta=0$ corresponds to $\Delta M_0 = 0$ exactly at tree-level and lowest order in the expansion in small $m_Z$. This shows that if one wants to find the maximum allowed higgsino mass splitting consistent with a given direct detection limit, it will be realized by approaching the ($\tan\beta=1$, $n=-1$) blind spot, not the $\delta \approx 0$ quasiblind spot.\footnote{Note that these blind spots are quite distinct from other supersymmetric dark matter SI blind spots 
\cite{Ellis:2000ds,Baer:2003jb,Baer:2006te,Huang:2014xua,Baum:2023inl,Arganda:2024tqo} that rely
on cancellation between Higgs boson-mediated amplitudes. The latter play no role in the present paper, as I am assuming that the heavy Higgs bosons are decoupled.}
For this reason, in the following sections, I will concentrate on the
cases in which $M_1$ and $M_2$ are both positive.

The above results are qualitatively correct, but only tree-level. One particularly important correction is the 1-loop contribution \cite{Thomas:1998wy} to $\Delta M_+$, given by $F(|\mu|/m_Z)\> \mbox{355 MeV}$, where
\beq
F(x) = \frac{x}{\pi} \int_0^1 dt\>\, (2 - t) \ln [1 + t/x^2 (1-t)^2],
\eeq
which varies between $F(1.10) =  0.724$ for $|\mu| =  100$ GeV to the asymptotic value $F(\infty) = 1$ for very heavy higgsinos. In realistic supersymmetric models, this
contribution to $\Delta M_+$ is subdominant unless the gaugino mass parameters are at least 10 TeV,
but it is nevertheless an important contribution. In the following, I use the public code
{\tt SOFTSUSY v4.1.12} \cite{Allanach:2001kg} to generate the couplings and masses with one-loop corrections. This is interfaced with 
{\tt micrOMEGAs v6.0} 
\cite{Belanger:2001fz,Belanger:2004yn,Belanger:2020gnr,Alguero:2023zol}
to obtain the predicted thermal freezeout higgsino contribution to the density and the SI and SD cross-sections. 
These can be compared to the LZ2024 bounds on SI and SD scattering from xenon nuclei.
The default {\tt micrOMEGAs v6.0} settings for nuclear matrix elements are used, which are a bit on the conservative side (i.e., predicting slightly smaller cross-sections, and therefore implying slightly weaker limits) compared to other public codes \cite{Gondolo:2004sc,Harz:2023llw}.

In the numerical results to follow, I always take 
the masses of all squarks and sleptons, the gluino, and all Higgs bosons except for
the 125.1 GeV scalar to be set to 10 TeV, so that they are essentially decoupled. The properties of the higgsino-like particles, including the LSP, are then determined by
their mixings with the gauginos. In order to have well-defined model lines, in the rest of this paper I 
concentrate on two cases with different
ratios of gaugino masses. First, the gaugino mass unification scenario is inspired by the apparent unification
of gauge couplings in supersymmetry near $2 \times 10^{16}$ GeV to impose
\beq
M_2 = 1.8 M_1\qquad\mbox{(gaugino mass unification)}
\label{eq:M2gauginomassunif}
\eeq 
at the 10 TeV scale. Second, the anomaly-mediated supersymmetry breaking (AMSB) scenario is inspired by the boundary conditions \cite{Randall:1998uk,Giudice:1998xp}
to instead take
\beq
M_1 = 3.2 M_2\qquad\mbox{(AMSB)}
,
\label{eq:M1AMSB}
\eeq
again at the 10 TeV scale. These can be viewed
as representative proxies for the cases of a relatively light bino and a relatively light wino, respectively, although in each case the impacts of the heavier gaugino are not completely negligible, especially in the gaugino mass unification case.

%%%%%%%%%%%%%%%%%%%%%%%%%%%%%%%%%%%%%%%%%%%%%%%%%%%%%%%%%%%%%%%
\section{Constraints on higgsino purity and gaugino masses from LZ2024\label{sec:gauginomassconstraints}}
\setcounter{equation}{0}
\setcounter{figure}{0}
\setcounter{table}{0} 
\setcounter{footnote}{1}

\baselineskip=16.5pt

\begin{figure}[!t]
\centering
\vspace{-0.7cm}
\mbox{
\includegraphics[width=0.51\linewidth]{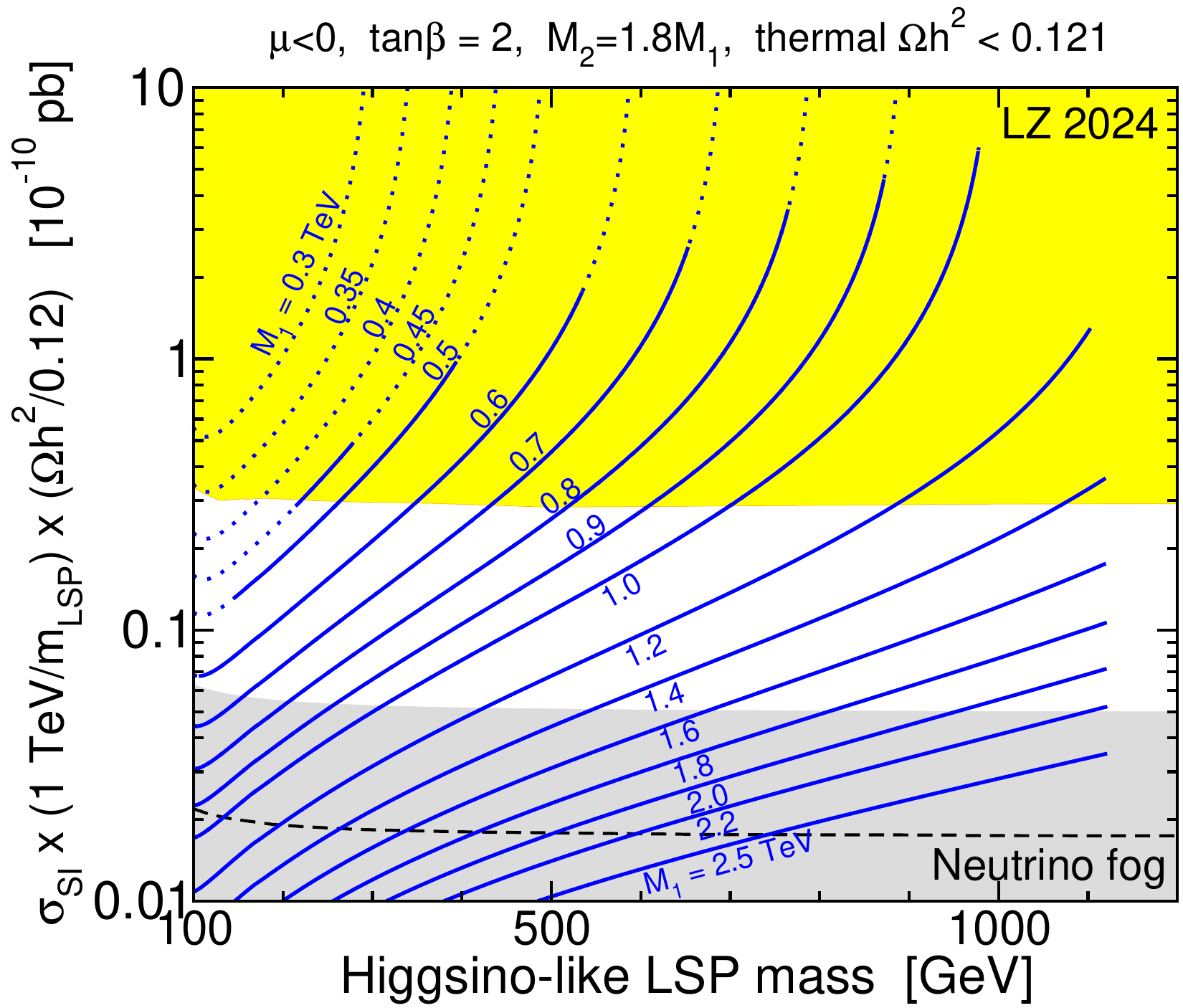}
\includegraphics[width=0.51\linewidth]{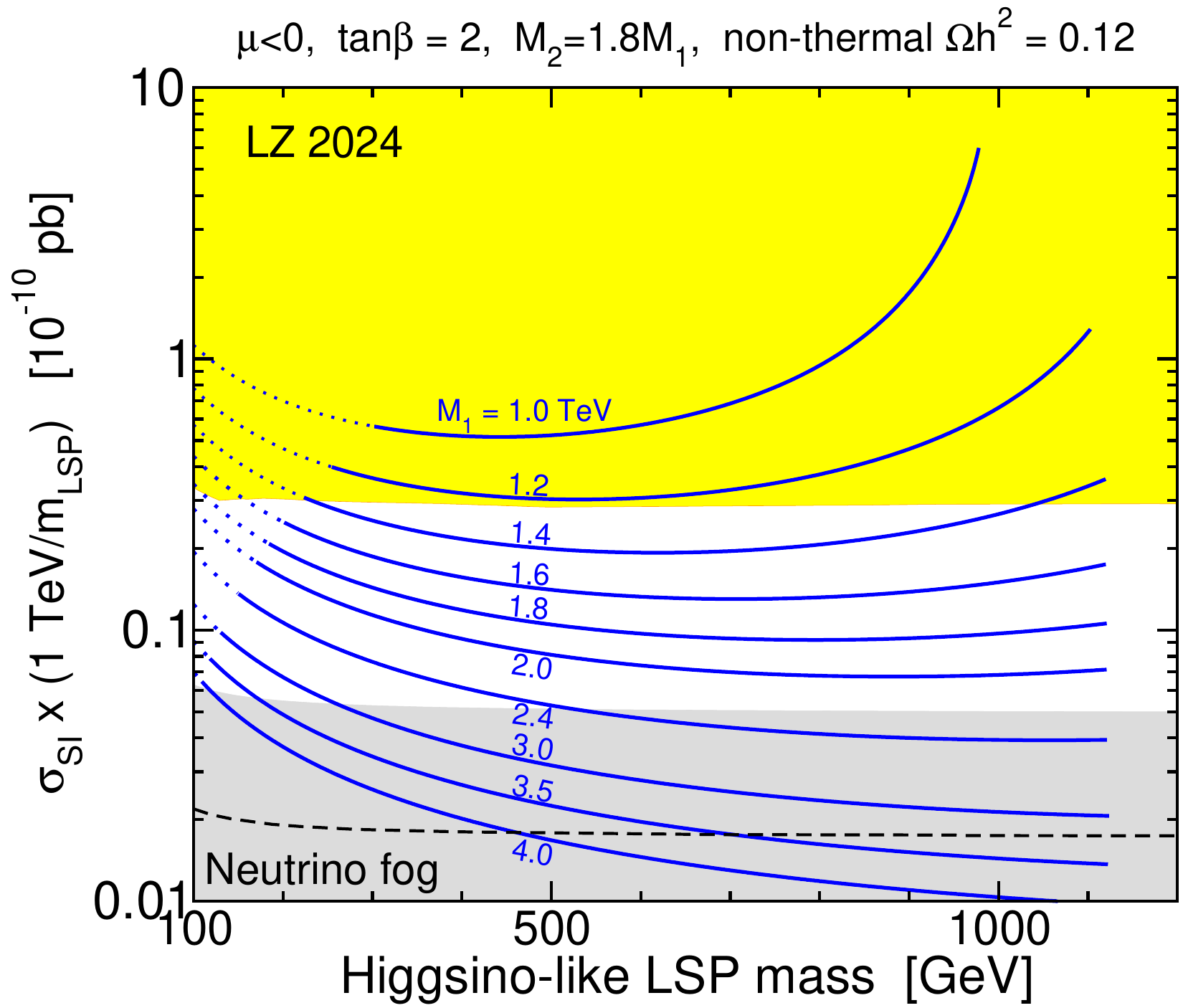}
}
\mbox{
\includegraphics[width=0.51\linewidth]{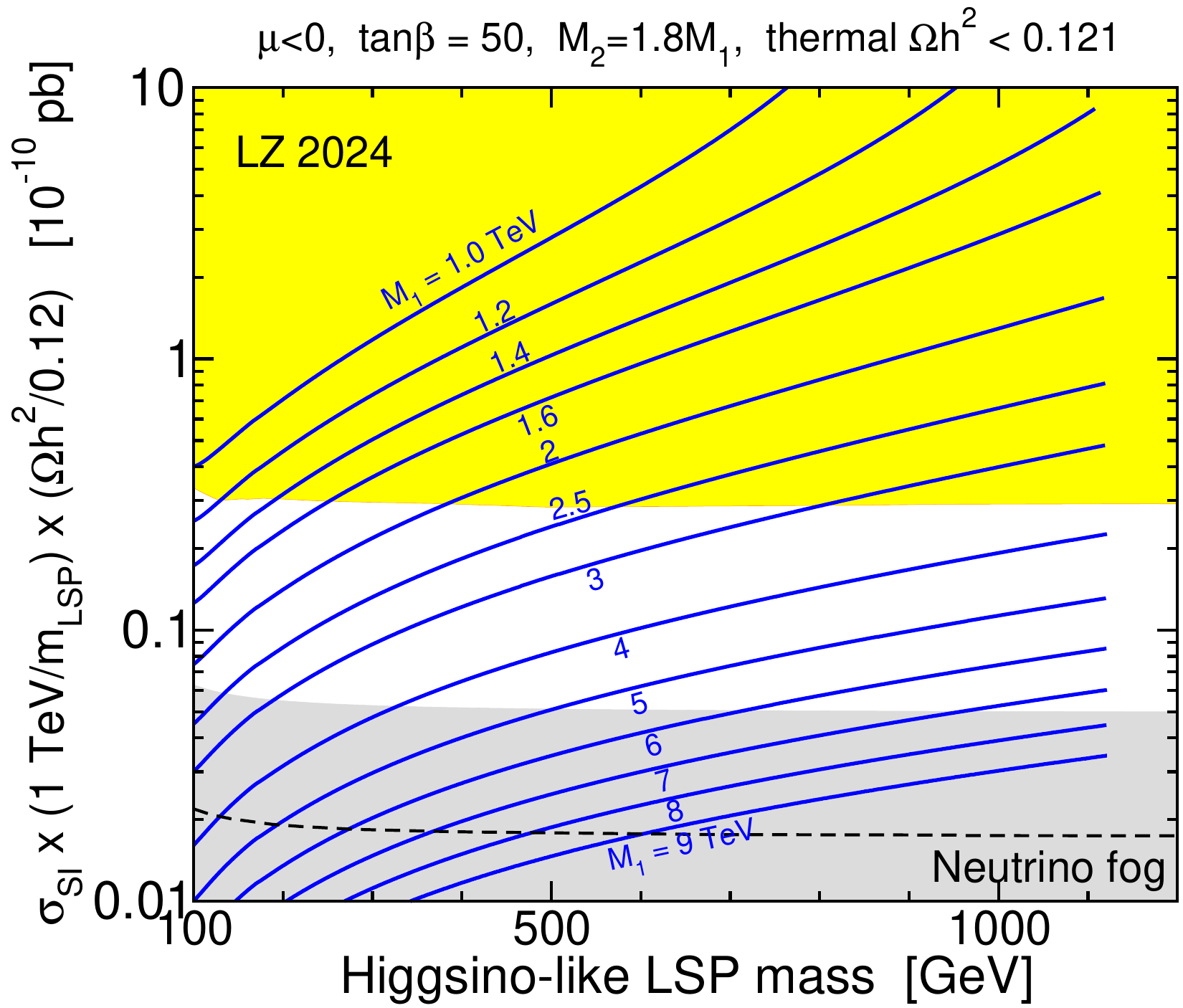}
\includegraphics[width=0.51\linewidth]{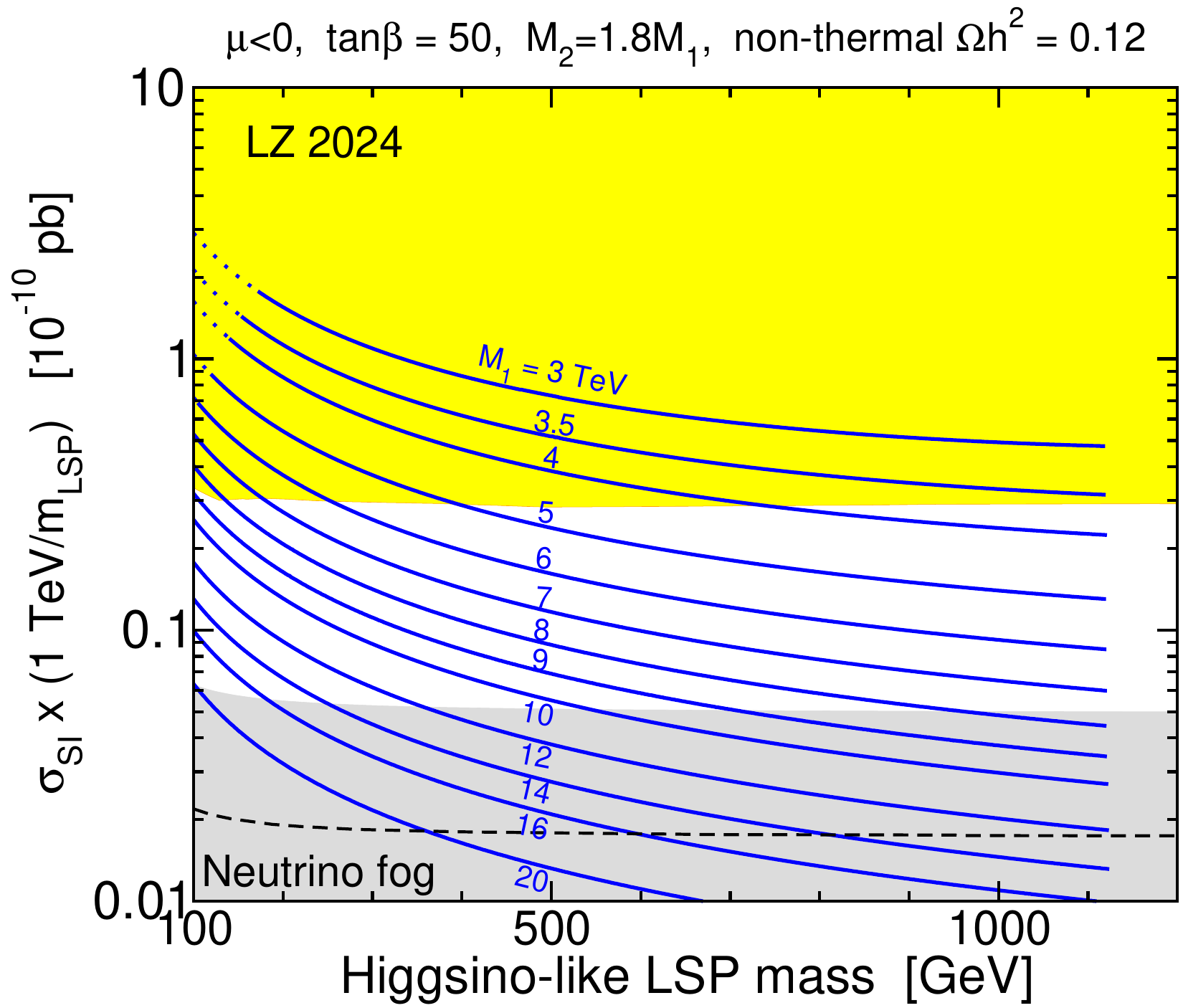}
}
\mbox{
\includegraphics[width=0.51\linewidth]{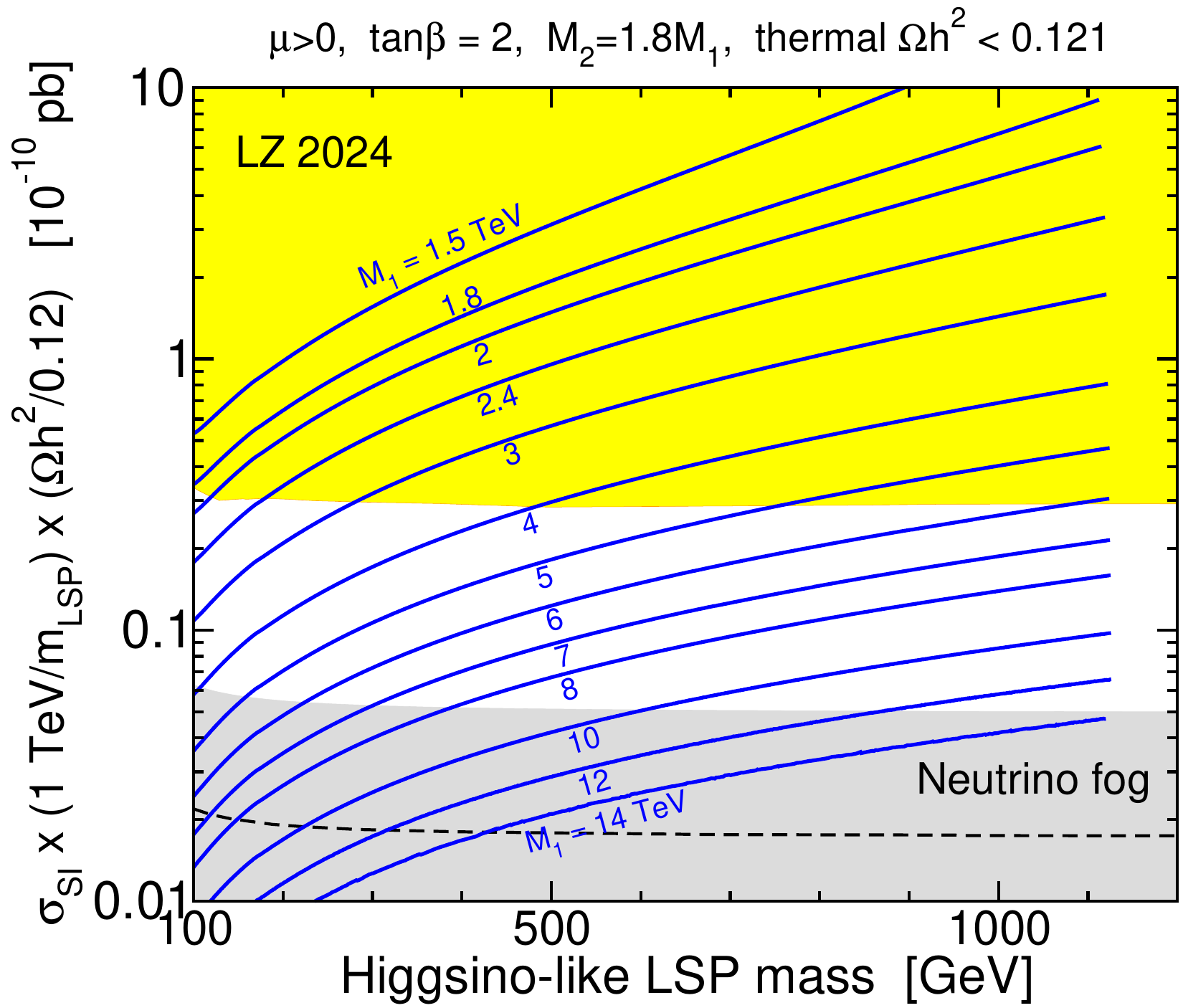}
\includegraphics[width=0.51\linewidth]{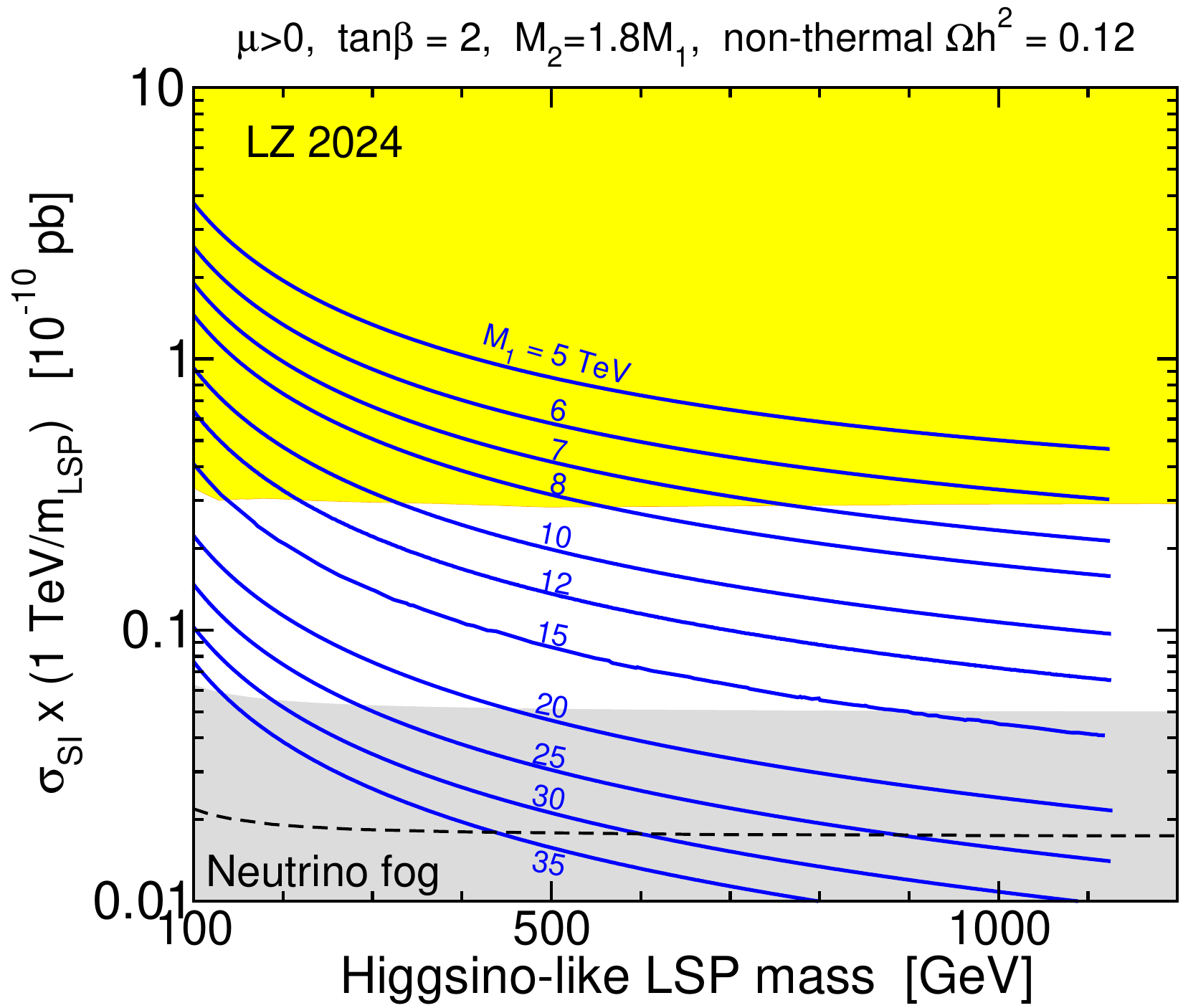}
}
\caption{\label{fig:unif}
SI LSP-nucleon cross-section, scaled by 1 TeV$/M_{\rm LSP}$ and $\Omega_{\mbox{\scriptsize LSP}} h^2/0.12$, as a function of the higgsino-like LSP mass $m_{\tilde N_1}$, with separate curves for various $M_1$ values as labeled in TeV. The wino mass is $M_2 = 1.8 M_1$,
as motivated by gaugino mass unification, and $\tan\beta$, sign($\mu$) = $2,-$ (top row),
$50,-$ (middle row) and $2,+$ (bottom row). The left panels assume the thermal prediction for $\Omega_{\mbox{\scriptsize LSP}} h^2 < 0.121$,
while the right panels assume that some nonthermal source increases the LSP density to
$\Omega_{\mbox{\scriptsize LSP}} h^2 = 0.12$.
The shaded (yellow) band at top is excluded by the LZ2024 SI limit \cite{LZCollaboration:2024lux}. The (gray) shaded band at bottom is the discovery neutrino fog, with the exclusion neutrino fog indicated by the dashed line. The dotted portions of curves are excluded by the LZ2024 SD limit \cite{LZCollaboration:2024lux}.}
\end{figure}

\begin{figure}[!p]
\centering
\vspace{-0.7cm}
\mbox{
\includegraphics[width=0.51\linewidth]{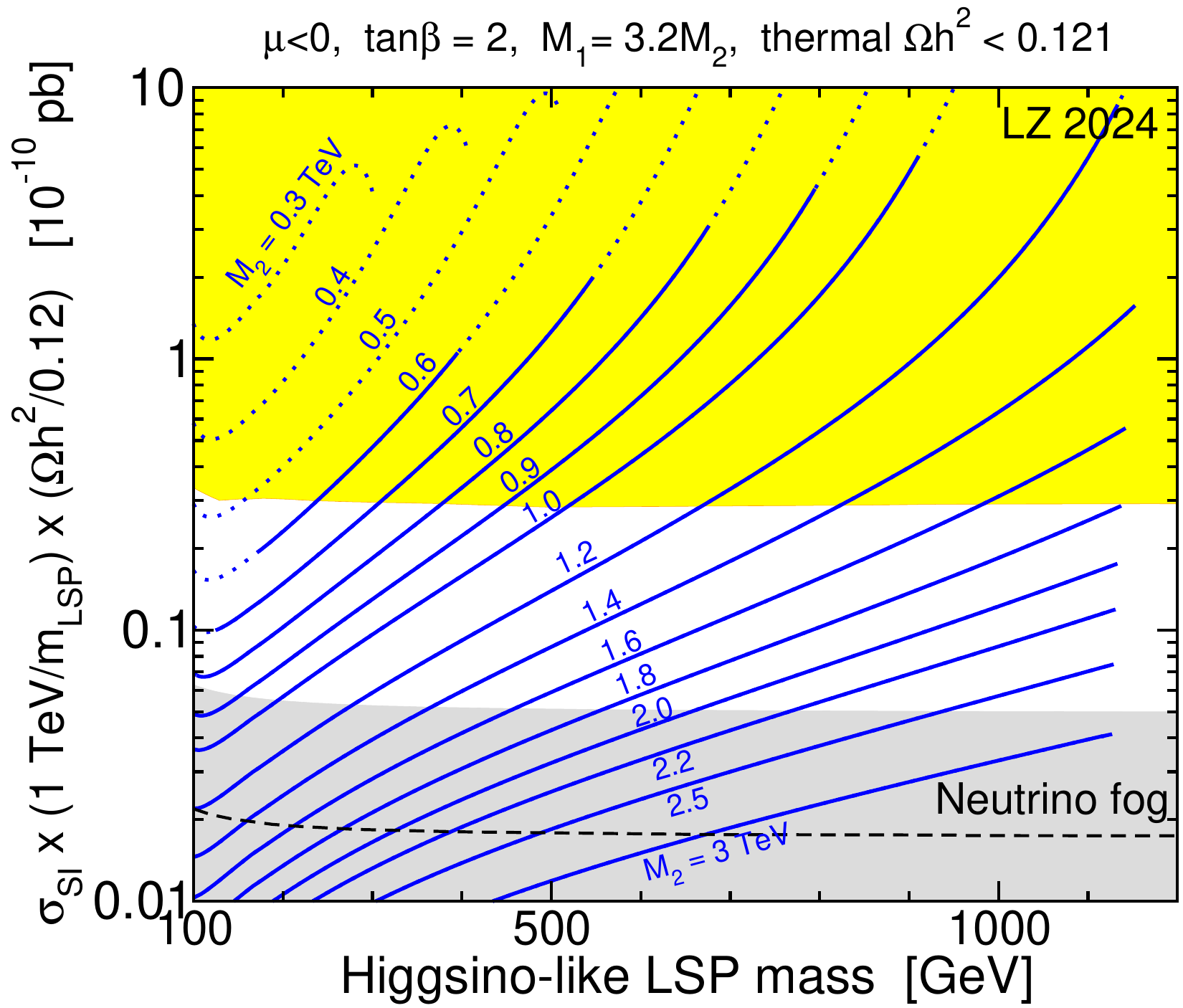}
\includegraphics[width=0.51\linewidth]{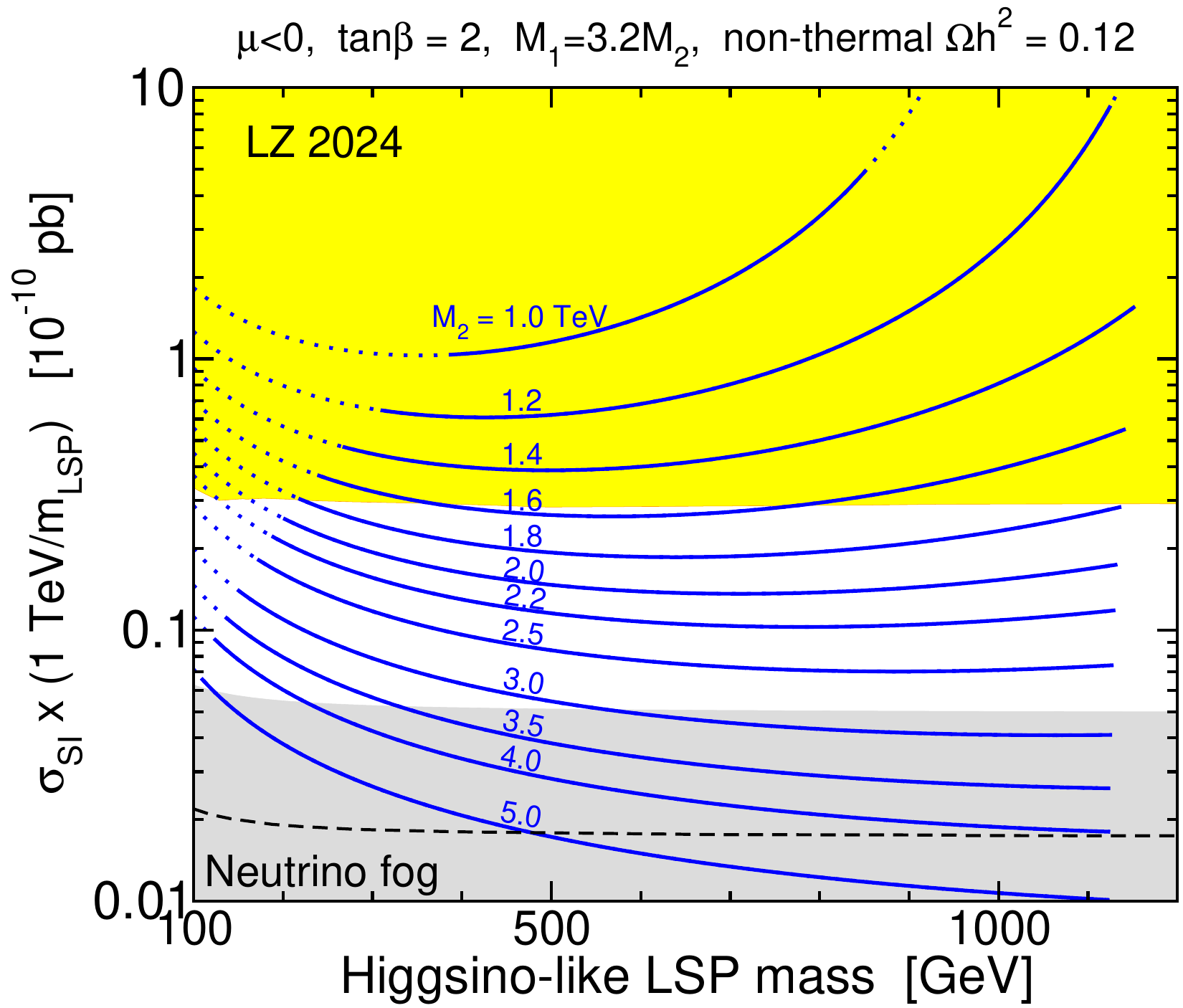}
}
\mbox{
\includegraphics[width=0.51\linewidth]{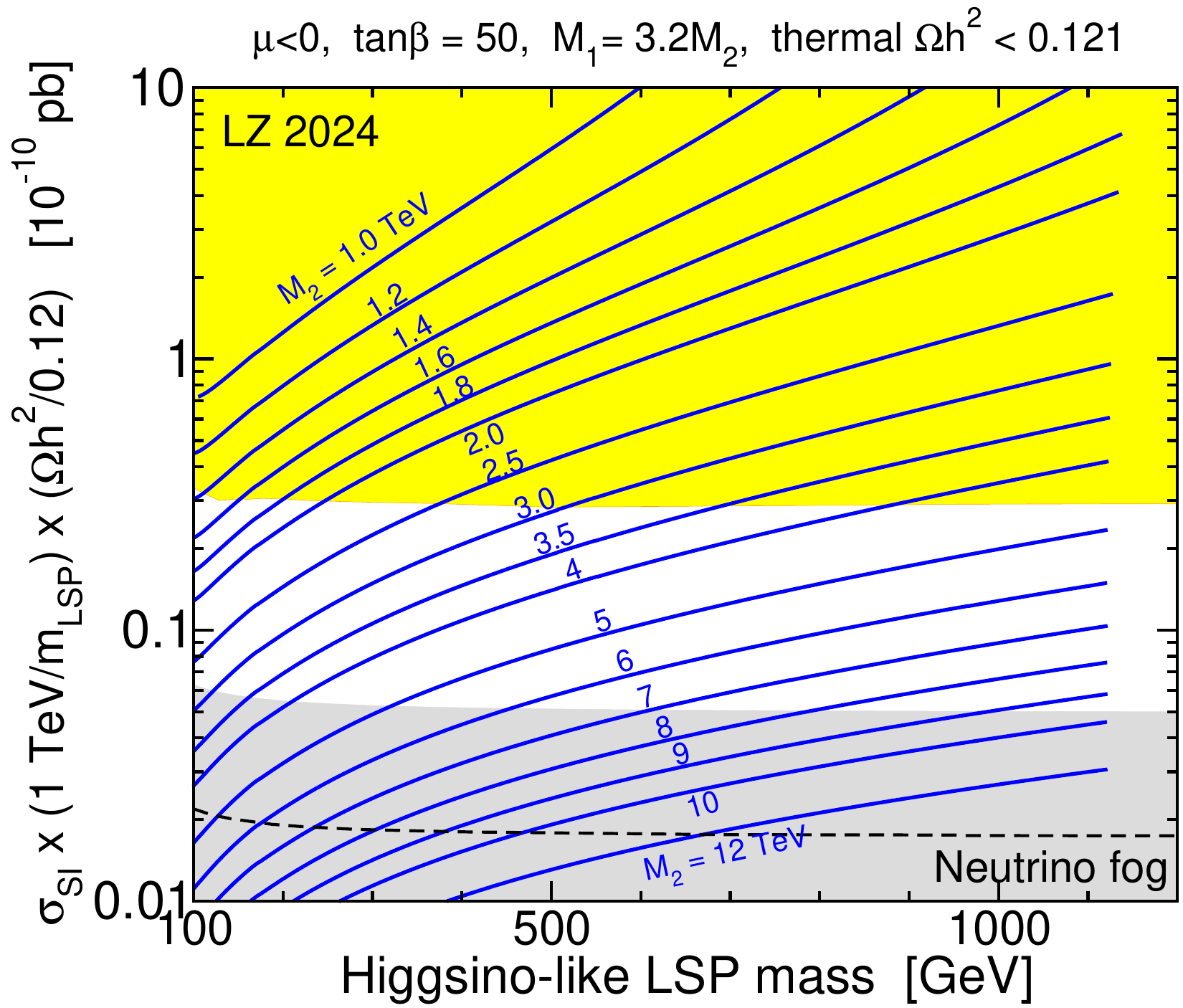}
\includegraphics[width=0.51\linewidth]{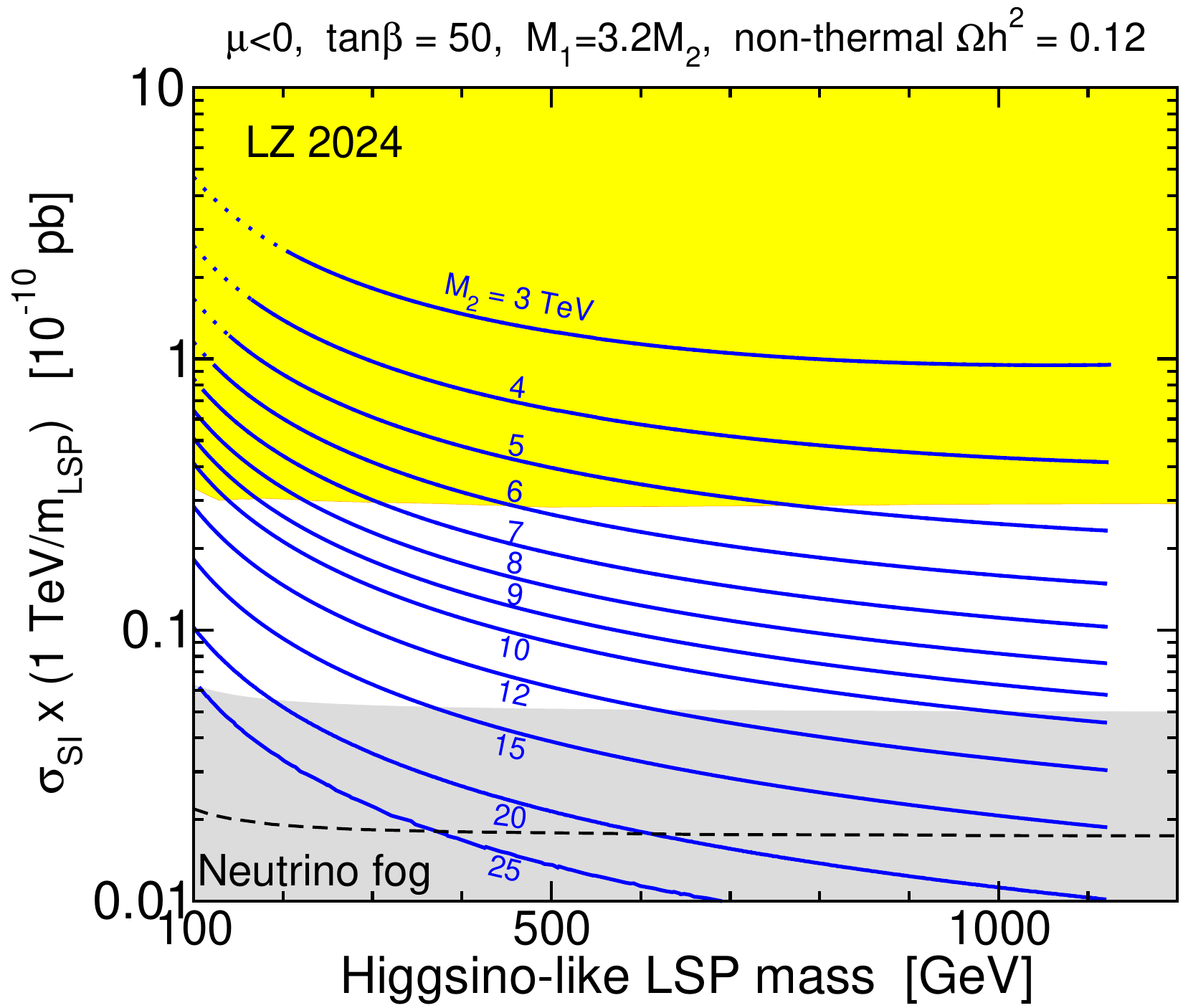}
}
\mbox{
\includegraphics[width=0.51\linewidth]{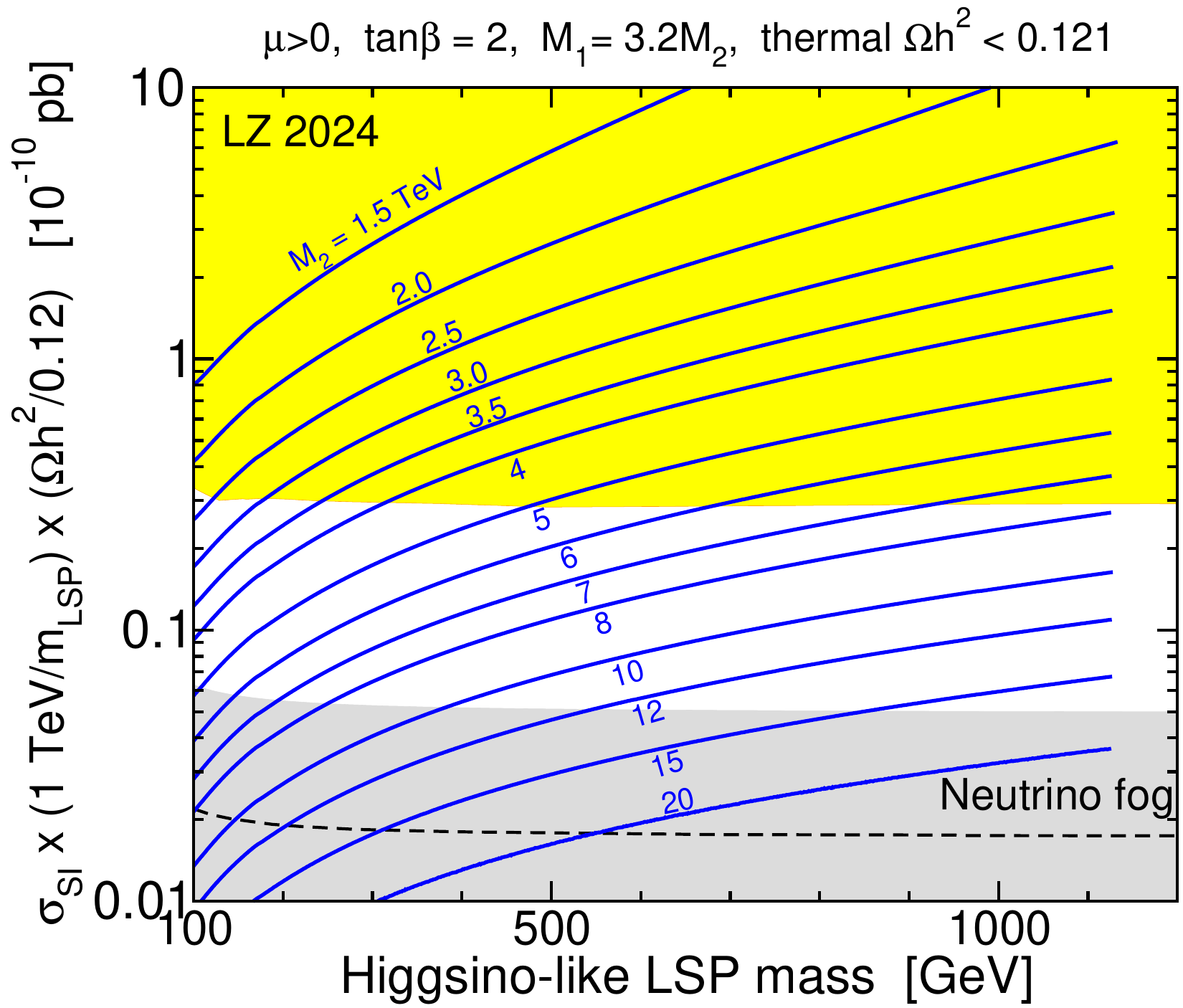}
\includegraphics[width=0.51\linewidth]{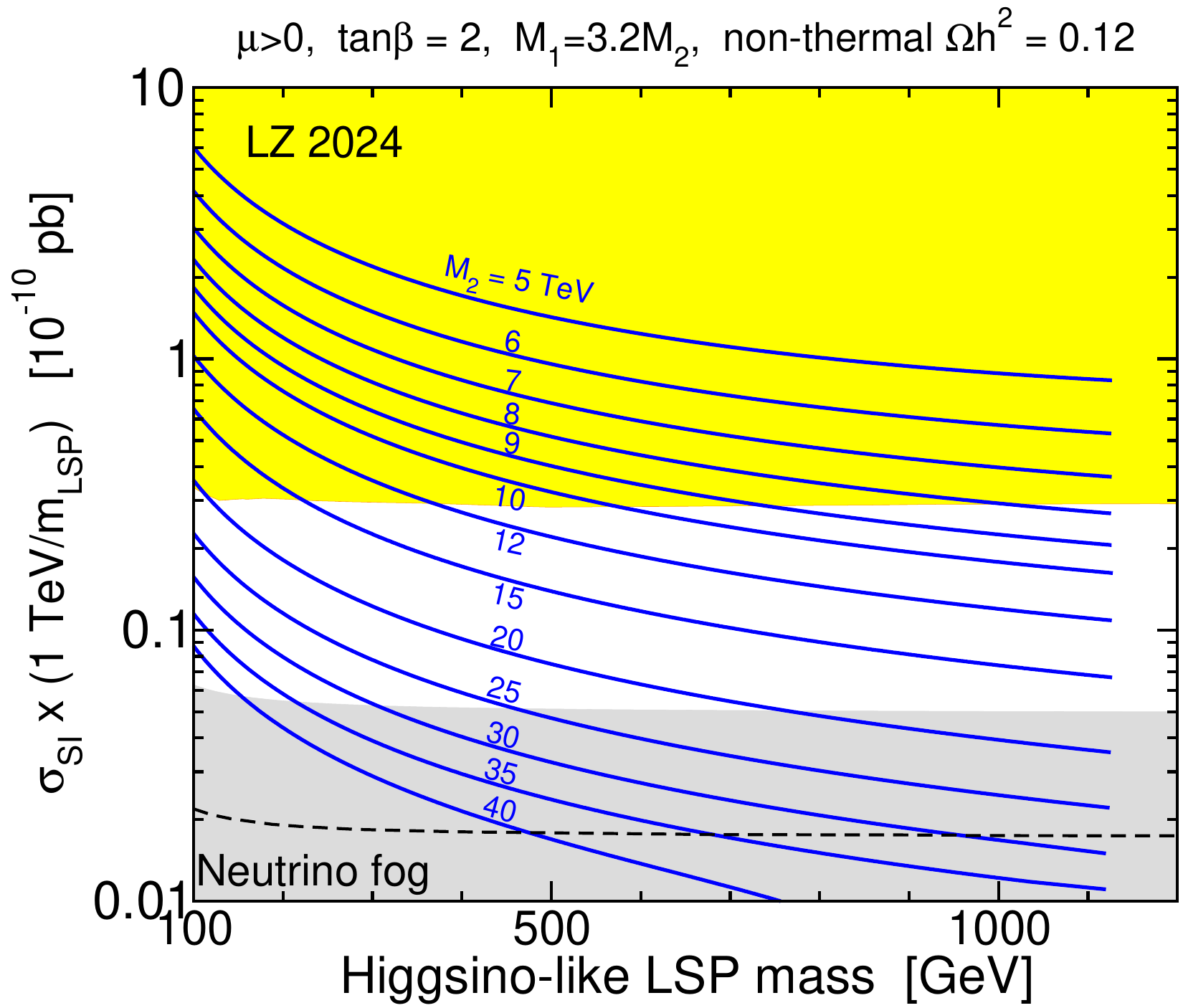}
}
\caption{\label{fig:AMSB}
As in Figure \ref{fig:unif}, but for $M_1 = 3.2 M_2$ as motivated by AMSB models,
with separate curves for various values of the wino mass parameter $M_2$ as labeled (in TeV).}
\end{figure}

I first present results for the cross-sections for the gaugino mass unification case
of eq.~(\ref{eq:M2gauginomassunif}). Figure \ref{fig:unif} shows the predicted SI cross-sections for three cases:
$\tan\beta = 2$ with negative $\mu$ (top row), 
$\tan\beta = 50$ with negative $\mu$ (middle row), 
and
$\tan\beta = 2$ with positive $\mu$ (bottom row). 
%%%%%%%%%%%%%%%%%%%%%%%%%%%%%%%%%%%%%%%%%%%%%%%%%%%%%%%%%%%%%%%%%%%%%%%%%%%%%%%%%
%%%%%%%%%%%%%%%%%%%%%%%%%%%%%%%%%%%%%%%%%%%%%%%%%%%%%%%%%%%%%%%%%%%%%%%%%%%%%%%%%
From the discussion in the previous
section, the (top, middle, bottom) rows therefore have a (suppressed, intermediate, enhanced) strength of higgsino coupling to nucleons through $h$ exchange. Each model curve
assumes a different value of $M_1$, as labeled, with larger $M_1$ corresponding
to a more pure higgsino LSP. The cross-sections are scaled by a factor of
(1 TeV$/m_{\tilde N_1}$), so that past, present, and future bounds from direct
detection experiments are approximately horizontal lines. The upper shaded (yellow)
region\footnote{This may be regarded as the metaphorical curtain referred to in the title of this paper.} in each figure is the nominal SI exclusion from LZ2024, while the lower
shaded (gray) region is the neutrino fog that will hamper a 
discovery\footnote{The neutrino fog given in ref.~\cite{OHare:2021utq} is defined in terms of the logarithmic derivative of the $3\sigma$ discovery limit cross-section with respect to the exposure. Using instead a $90$\% exclusion confidence level criterion gives a neutrino fog cross-section that is about a factor of 3 lower, as given in Figure 2a in ref.~\cite{XLZD:2024nsu}. To distinguish the two, I will refer to the former as the ``discovery neutrino fog", and the latter as the ``exclusion neutrino fog". They are shown, respectively, as the (gray) shaded region and the dashed line in Figures \ref{fig:unif} and \ref{fig:AMSB} of the present paper.} 
as defined in ref.~\cite{OHare:2021utq}. 
The difference between the two columns of panels in Figure \ref{fig:unif} is that in
the left column it is assumed that the density of higgsino dark matter comes entirely
from thermal freezeout, with the remainder made up of some other undetected particle, so that there is an additional suppression factor proportional to
$\xi = \Omega_{\mbox{\scriptsize LSP}} h^2/0.12$. The right column is the nonthermal case, which assumes that
some unspecified nonthermal source enhances the higgsino density to  $\Omega_{\mbox{\scriptsize LSP}} h^2 = 0.12$.
As noted in the previous section, for $\tan\beta=2$ and $\mu < 0$, the SD bounds are
sometimes stronger than the SI bounds. To indicate this, the portion of each model line
that is excluded by the LZ2024 SD limit is dotted rather than solid. The endpoint of each model line, near $m_{\tilde N_1} = 1100$ GeV, is where one finds $\Omega_{\mbox{\scriptsize LSP}} h^2 = 0.121$.

From Figure \ref{fig:unif}, we see that in the thermal higgsino case with 
$\tan\beta=2$ and $\mu < 0$, one can have $M_1$
as low as about 450 GeV if the higgsino LSP mass is around 200 GeV. The region to be explored before encountering the discovery neutrino fog goes up to about $M_1 = 2.2$ TeV if
$\mu = -1.1$ TeV. The stronger constraints in the nonthermal higgsino case imply that
$M_1$ must exceed 1.2 TeV for any LSP mass, with a significant impact from the SD limit for higgsino mass below 250 GeV.  

In the case of $\tan\beta = 2$ and $\mu > 0$, the purity constraint on higgsinos implied by LZ2024 requires $M_1 > 2$ TeV in the thermal case and $M_1 > 6$ TeV in the nonthermal case. For the intermediate case of large $\tan\beta$, as illustrated in the middle row, we see that the region remaining to be probed before encountering the discovery neutrino fog includes a range from slightly over 1.2 TeV to about 20 TeV, depending on whether one assumes the thermal or nonthermal cases.

In the same way, Figure \ref{fig:AMSB} shows the results for AMSB-inspired models
obeying eq.~(\ref{eq:M1AMSB}). 
%%%%%%%%%%%%%%%%%%%%%%%%%%%%%%%%%%%%%%%%%%%%%%%%%%%%%%%%%%%%%%%%%%%%%%%%%%%%%%%%%
%%%%%%%%%%%%%%%%%%%%%%%%%%%%%%%%%%%%%%%%%%%%%%%%%%%%%%%%%%%%%%%%%%%%%%%%%%%%%%%%%
The results are seen to be qualitatively similar to the gaugino mass unification case, but it is notable that mixing of higgsinos predominantly with the winos is more constrained than mixing with a bino of the same mass. For example, if $\tan\beta=2$ and $\mu<0$, the AMSB thermal higgsino requires $M_2 > 500$ GeV, as seen in the upper left panel of 
Figure \ref{fig:AMSB}.

In terms of the SI cross-section, the current limits from LZ2024 are approximately a factor of 6 above the neutrino fog hampering a possible discovery (and about a factor of 18 above the exclusion neutrino fog). Another way to describe this reach
is in terms of the gaugino masses that are indirectly probed by their mixing with the 
higgsino LSP. The unshaded ranges in Figures \ref{fig:unif} and \ref{fig:AMSB} evidently correspond to a factor of only roughly 1.25 to 2.4 in gaugino masses, for fixed $\tan\beta$ and $\mu$ and $M_2/M_1$. However, these ranges are particularly important,
as they mostly correspond to regions of parameter space that are difficult or impossible to directly probe at the Large Hadron Collider, and even challenging for proposed future colliders. In the thermal higgsino case with low $\Omega_{\mbox{\scriptsize LSP}} h^2$ (implying $m_{\tilde N_1}$ well below 1 TeV), it would be interesting to make a detailed comparison with the prospects for
indirect detection using gamma-ray astronomy, which is beyond the scope of the present work.

\begin{figure}[!p]
\centering
\mbox{
\includegraphics[width=0.51\linewidth]{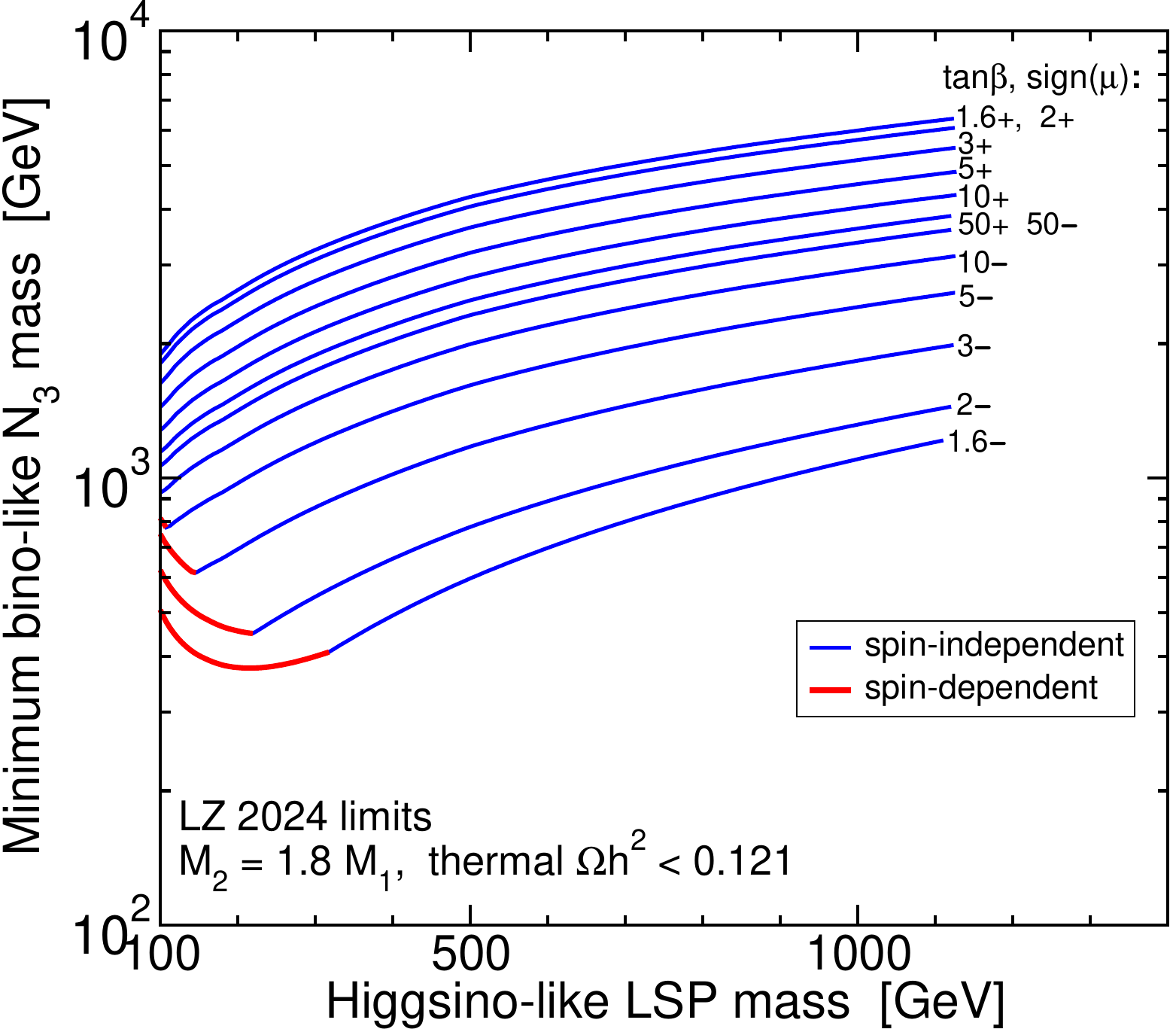}
\includegraphics[width=0.51\linewidth]{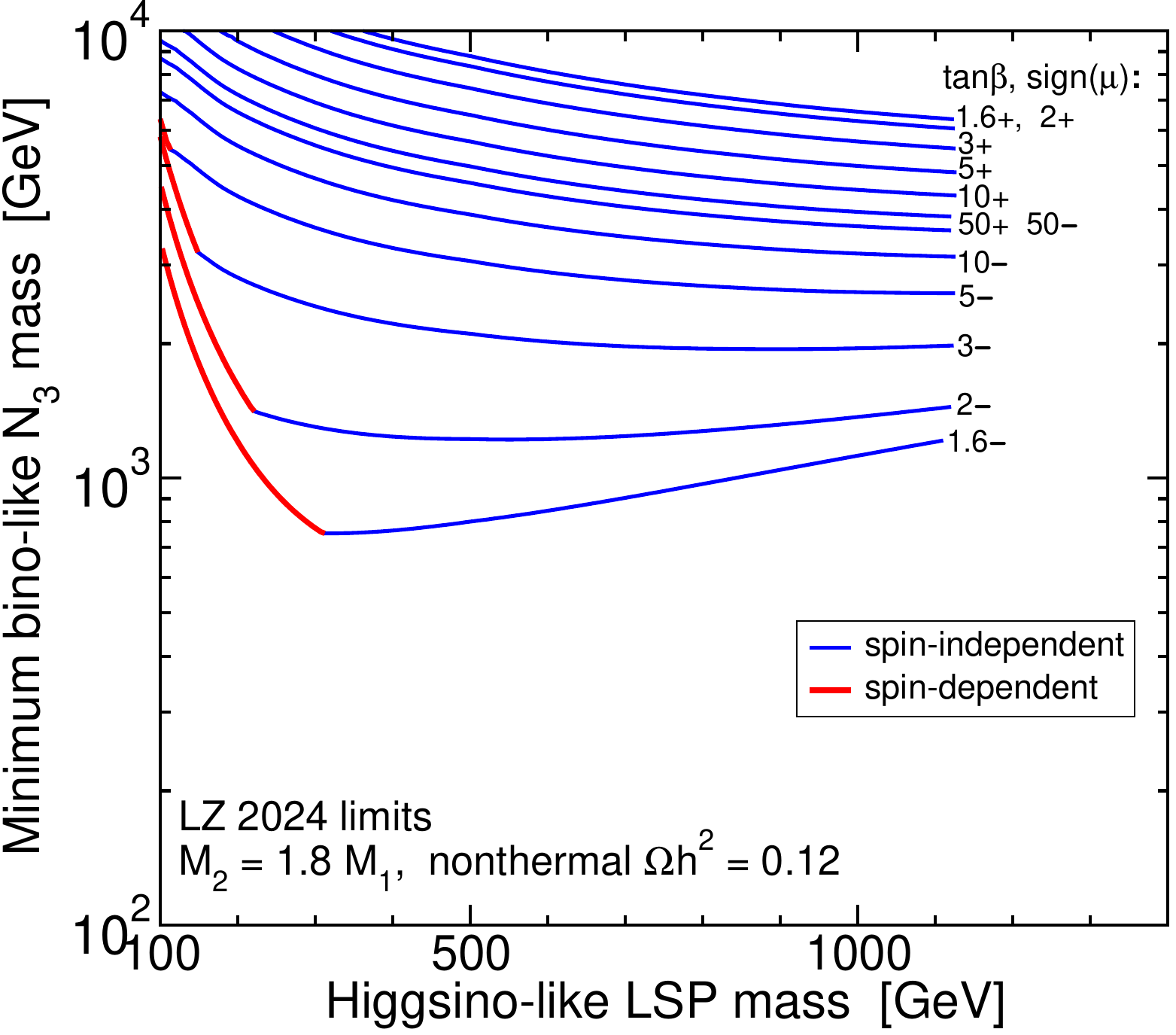}
}
\caption{\label{fig:contoursM1unif}
The minimum bino-like neutralino mass $m_{\tilde N_3}$ (related to $M_1$) allowed by the LZ2024 dark matter limits \cite{LZCollaboration:2024lux}, as a function of the higgsino-like LSP mass $m_{\tilde N_1}$
(related to $|\mu|$).
The wino mass parameter is taken to be $M_2 = 1.8 M_1$, as motivated by gaugino mass unification models. The different curves have various values of $\tan\beta$ and sign$(\mu)$,
as labeled. In the left panel, the LSP density is set by thermal freezeout with $\Omega_{\mbox{\scriptsize LSP}} h^2 < 0.121$. In the right panel, it is assumed that some nonthermal source increases
the LSP density to $\Omega_{\mbox{\scriptsize LSP}} h^2 = 0.12$.
The thinner (blue) portions of the curves are set by the spin-independent LSP-xenon cross-section limit, while the thicker (red) portions are set by the spin-dependent cross-section limit. The lightest Higgs boson mass is fixed to $M_h = 125.1$ GeV, and all other
scalar masses are set to 10 TeV.
}
\end{figure}

As another way to present the purity bounds on higgsinos from LZ2024, I show in Figure
\ref{fig:contoursM1unif} the minimum allowed bino mass $M_1$ in the unified gaugino
mass case, for various combinations of $\tan\beta$ and sign$(\mu)$, as labeled.
The bounds are strongest for small $\tan\beta$ and positive $\mu$, and weakest
for small $\tan\beta$ and negative $\mu$, for reasons explained in terms of tree-level
kinematics and couplings in the previous section.
The left panel shows the results for the thermal higgsino density $\Omega_{\mbox{\scriptsize LSP}} h^2$ less than $0.12$, while the right panel has the results for the nonthermal higgsino case. Of course, these coincide at the endpoints of the curves near $m_{\tilde N_1} = 1.1$ TeV, where the predicted density is $\Omega_{\mbox{\scriptsize LSP}} h^2 = 0.121$ in either case. For the cases of $\tan\beta = 1.6$, 2, 3, and 5, with $\mu < 0$, for sufficiently low LSP mass, the minimum $M_1$ is set by the SD cross-section limit,
as indicated by the thicker (red) portions of the lines. This occurs only for LSP masses
less than about 300 GeV. The other lines (thinner, blue) are set by the SI cross-section bound. Note that for positive $\mu$ and small $\tan\beta$, in the nonthermal case the bino mass parameter is already required to be above 10 TeV when the LSP is light, corresponding to an ultra-pure higgsino.

\begin{figure}[!p]
\centering

\mbox{
\includegraphics[width=0.51\linewidth]{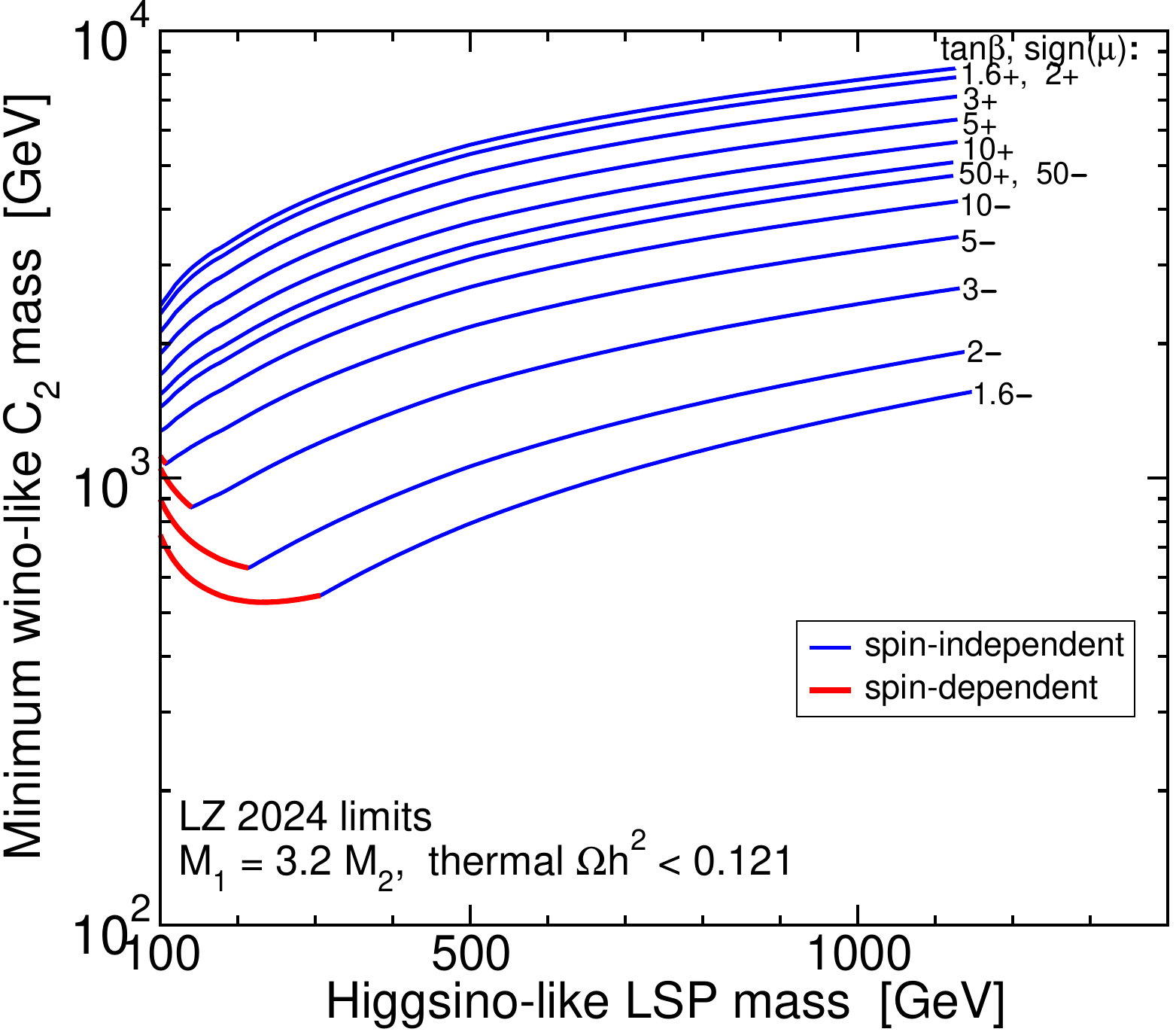}
\includegraphics[width=0.51\linewidth]{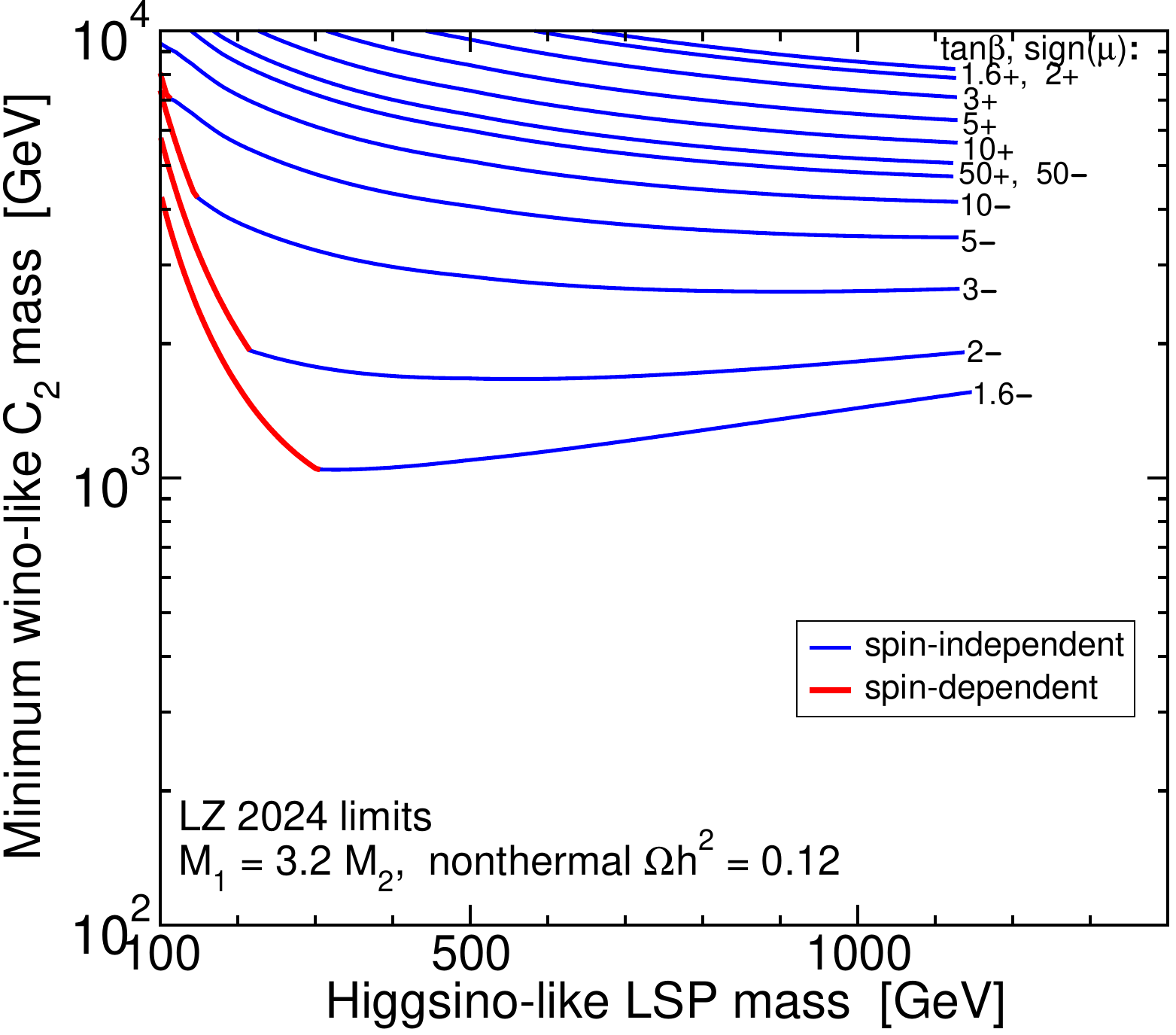}
}
\caption{\label{fig:contoursM2amsb}
The minimum wino-like chargino mass $m_{\tilde C_2}$ (related to $M_2$) allowed by the LZ2024 dark matter limits \cite{LZCollaboration:2024lux}, as a function of the higgsino-like LSP mass $m_{\tilde N_1}$ (related to $|\mu|$).
The bino mass parameter is taken to be $M_1 = 3.2 M_2$, as motivated by AMSB models. The different curves have various values of $\tan\beta$ and sign$(\mu)$,
as labeled. In the left panel, the LSP density is set by thermal freezeout with $\Omega_{\mbox{\scriptsize LSP}} h^2 < 0.121$. In the right panel, it is assumed that some nonthermal source increases
the LSP density to $\Omega_{\mbox{\scriptsize LSP}} h^2 = 0.12$.
The thinner (blue) portions of the curves are set by the spin-independent LSP-xenon cross-section limit, while the thicker (red) portions are set by the spin-dependent cross-section limit.
The lightest Higgs boson mass is fixed to $M_h = 125.1$ GeV, and all other
scalar masses are set to 10 TeV.
}
\end{figure}

The corresponding results for the minimum allowed wino mass $M_2$ in the AMSB-inspired scenario are shown in Figure \ref{fig:contoursM2amsb}. As already noted above, the results
are qualitatively similar, but stronger, as a slight wino mixture provides a larger coupling to the $h$ and $Z$ bosons than a corresponding bino mixture with the same mass. 
%%%%%%%%%%%%%%%%%%%%%%%%%%%%%%%%%%%%%%%%%%%%%%%%%%%%%%%%%%%%%%%%%%%%%%%%%%%%%%%%%
%%%%%%%%%%%%%%%%%%%%%%%%%%%%%%%%%%%%%%%%%%%%%%%%%%%%%%%%%%%%%%%%%%%%%%%%%%%%%%%%%
%%%%%%%%%%%%%%%%%%%%%%%%%%%%%%%%%%%%%%%%%%%%%%%%%%%%%%%%%%%%%%%%%%%%%%%%%%%%%%%%%
%%%%%%%%%%%%%%%%%%%%%%%%%%%%%%%%%%%%%%%%%%%%%%%%%%%%%%%%%%%%%%%%%%%%%%%%%%%%%%%%%
In the non-thermal case, it is evidently now not possible for the wino-like states to have masses less than 1 TeV, assuming $\tan\beta > 1.6$. Even in the thermal case, a wino with mass less than 1 TeV would require $\tan\beta$ less than about 4, and negative $\mu$.

\FloatBarrier 

\baselineskip=15.6pt

%%%%%%%%%%%%%%%%%%%%%%%%%%%%%%%%%%%%%%%%%%%%%%%%%%%%%%%%%%%%%%%
\section{Constraints on higgsino mass splittings from LZ2024\label{sec:higgsinomassconstraints}}
\setcounter{equation}{0}
\setcounter{figure}{0}
\setcounter{table}{0} 
\setcounter{footnote}{1}

\begin{figure}[!b]
\centering
\mbox{
\includegraphics[width=0.51\linewidth]{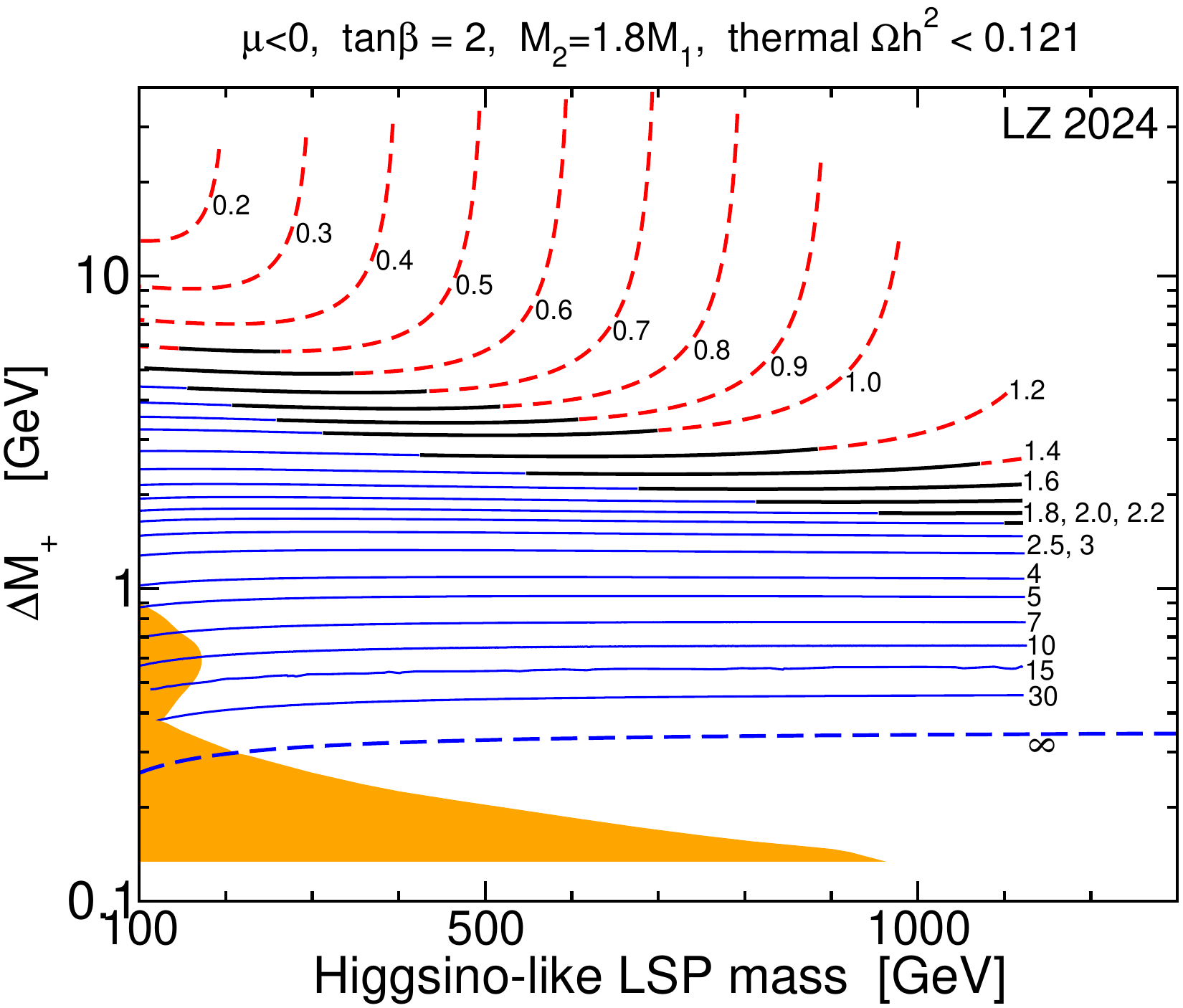}
\includegraphics[width=0.51\linewidth]{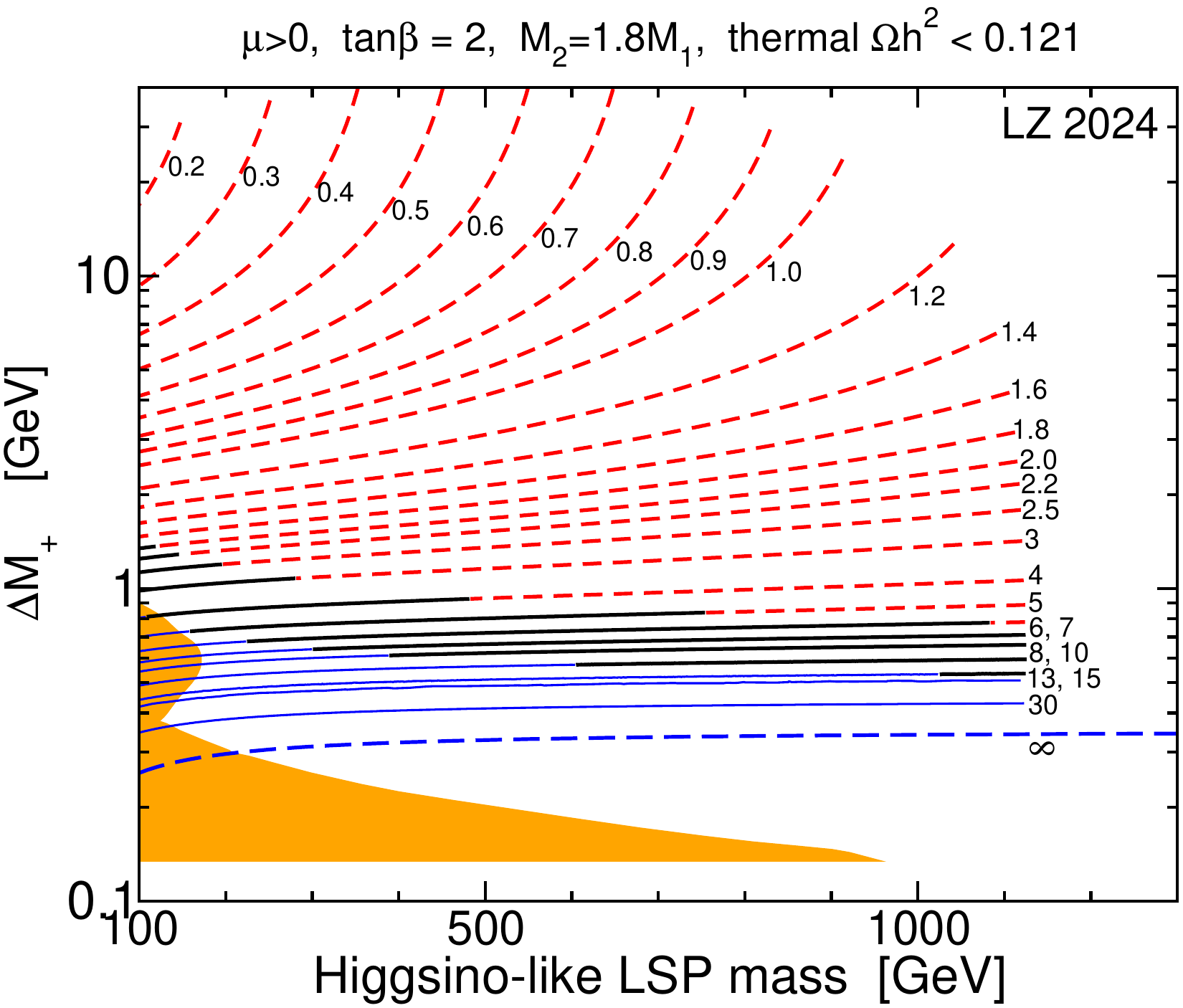}
}
\caption{\label{fig:contoursunif}
The higgsino-like chargino-LSP mass splitting $\Delta M_+ = m_{\tilde C_1} - m_{\tilde N_1}$ as a function of the LSP mass $m_{\tilde N_1}$, for various choices of the bino mass parameter $M_1$ from 0.2 TeV to 30 TeV as labeled,
assuming that the gaugino masses satisfy the unification relation $M_2 = 1.8 M_1$.
The left and right panels have negative and positive $\mu$, respectively, with $\tan\beta=2$, $M_h$ fixed at 125.1 GeV, and 
all other
scalar masses set to 10 TeV. The dashed (red) portions of each curve are ruled out by the LZ2024 limits \cite{LZCollaboration:2024lux} assuming that the dark matter has a relic abundance set by thermal freezeout. The lighter solid (blue) portions lie within the discovery neutrino fog region.
The shaded (orange) regions are excluded by  
ATLAS and CMS disappearing track searches (lower part) and
the ATLAS low-momentum mildly displaced track search (upper part).}
\end{figure}

As a corollary of the lower bounds on gaugino masses, their limited mixings
with the higgsino-like states implies stringent upper bounds on the mass differences
$M_+$ and $M_0$. For example, in Figure \ref{fig:contoursunif}, I show the chargino-LSP mass splitting $M_+$ for models with gaugino mass unification, as a function of the LSP mass, for various values of the bino mass parameter $M_1$ as labeled.  Here I have chosen
$\tan\beta=2$, $\mu<0$ for the left panel and $\tan\beta=2$, $\mu>0$ for the right panel, so that
results for larger $\tan\beta$ will always be between these two choices.
%%%%%%%%%%%%%%%%%%%%%%%%%%%%%%%%%%%%%%%%%%%%%%%%%%%%%%%%%%%%%%%%%%%%%%%%%%%%%%%%%
%%%%%%%%%%%%%%%%%%%%%%%%%%%%%%%%%%%%%%%%%%%%%%%%%%%%%%%%%%%%%%%%%%%%%%%%%%%%%%%%%
For each model curve with constant gaugino mass parameters, the dashed (red) portion
is excluded by LZ2024, and the lighter (blue) portion is below the discovery neutrino fog,
with the thermal freezeout prediction for $\Omega_{\mbox{\scriptsize LSP}} h^2$ assumed. The thicker (black) portion of each curve corresponds to parameters that are not yet excluded
by LZ2024, but still above the discovery neutrino fog and therefore likely to be accessible to direct detection experiments in the near future.
Also shown as the shaded (orange) regions are the exclusions by ATLAS \cite{ATLAS:2024umc} for low-momentum mildly displaced tracks \cite{Fukuda:2019kbp} (the upper part), and by
the union of the ATLAS \cite{ATLAS:2022rme} and CMS \cite{CMS:2023mny} exclusion regions for disappearing track searches (the lower part). It is notable that 
the disappearing track searches do not impact the supersymmetric parameter space at all
except for ultrapure higgsinos with $M_1 > 25$ TeV. The search based on mildly displaced tracks is limited, so far, to higgsinos that mix with gaugino masses of at least several TeV, and LSP masses below 200 GeV. 

\begin{figure}[!b]
\centering
\mbox{
\includegraphics[width=0.51\linewidth]{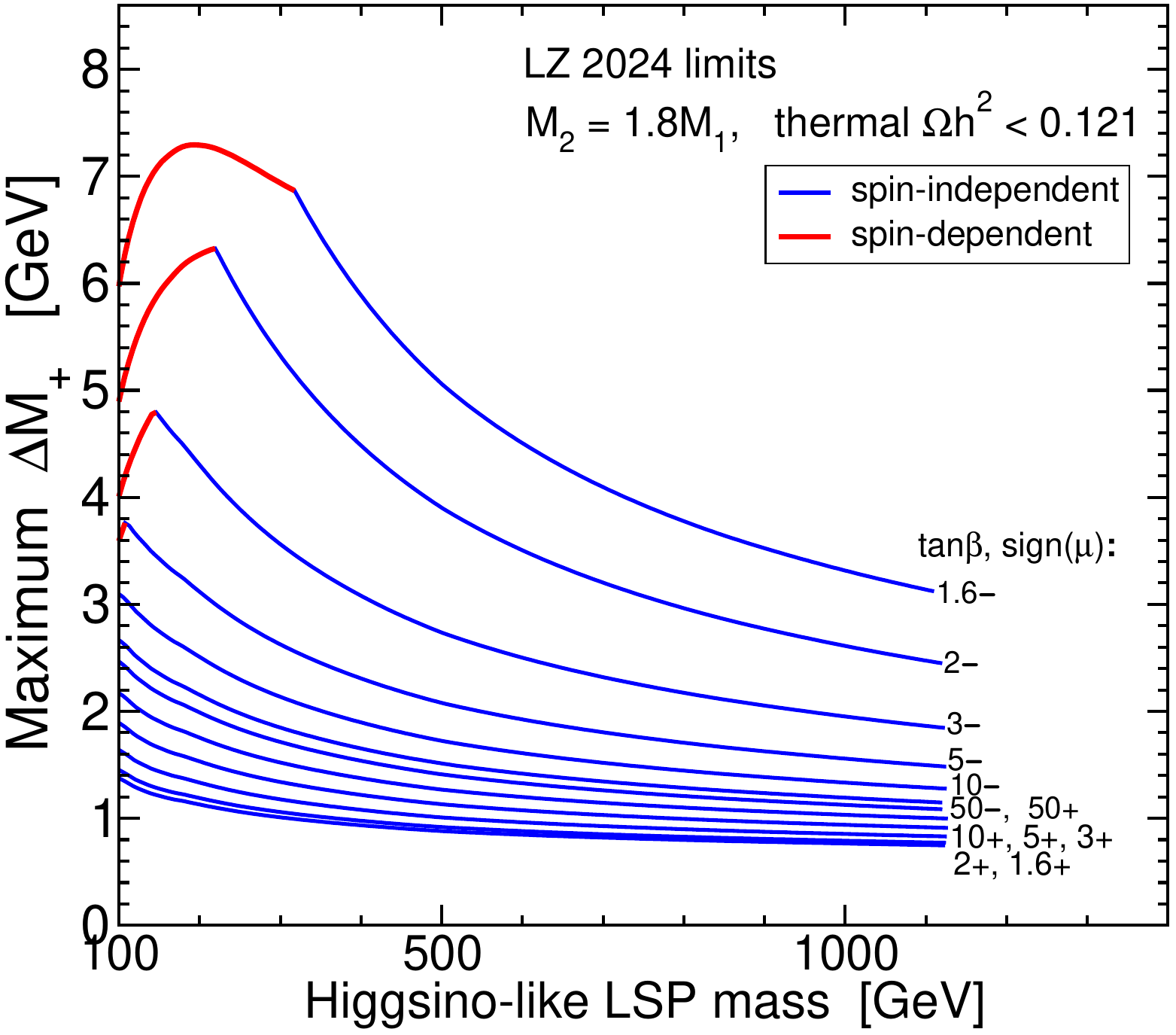}
\includegraphics[width=0.51\linewidth]{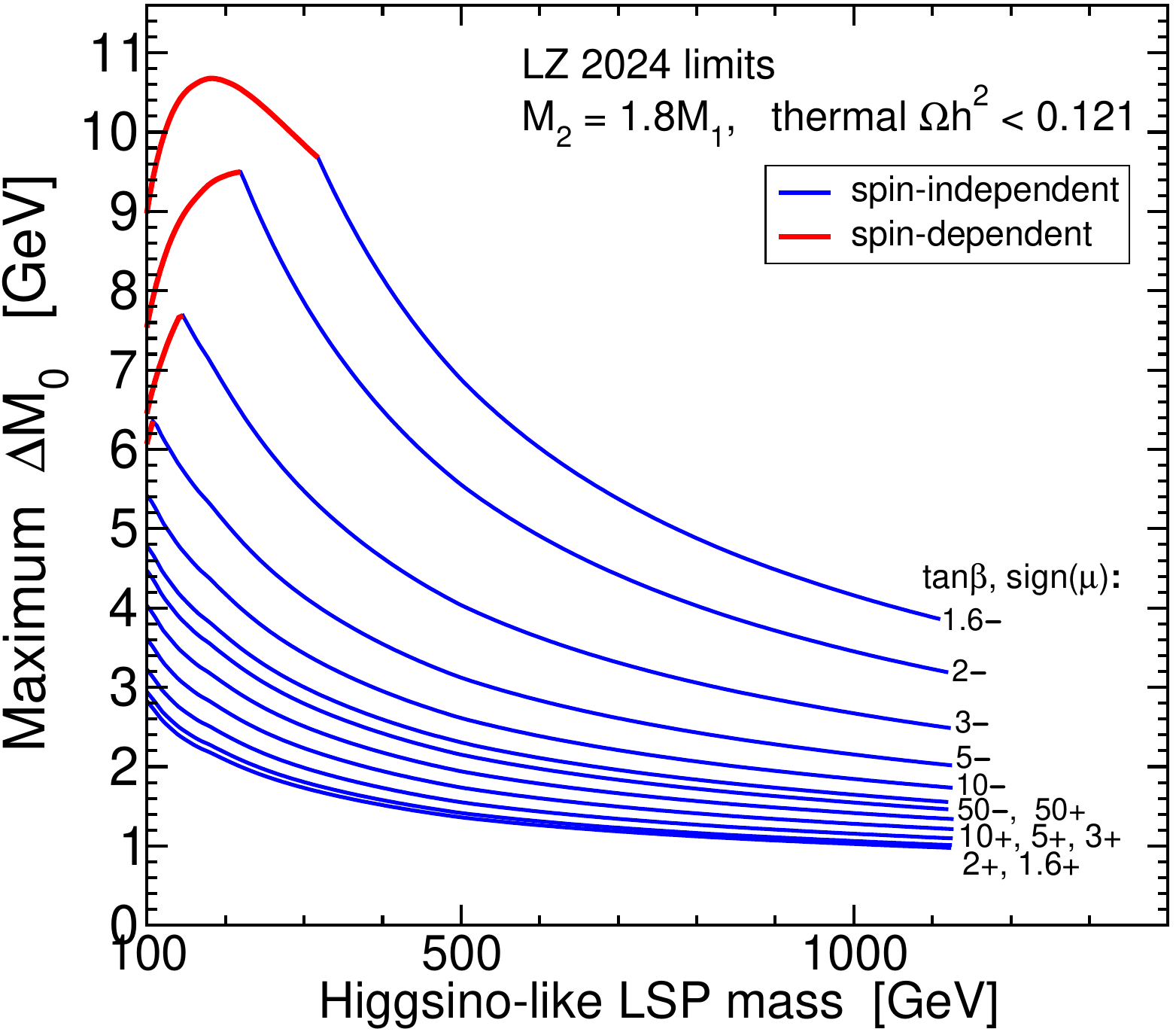}
}
\mbox{
\includegraphics[width=0.51\linewidth]{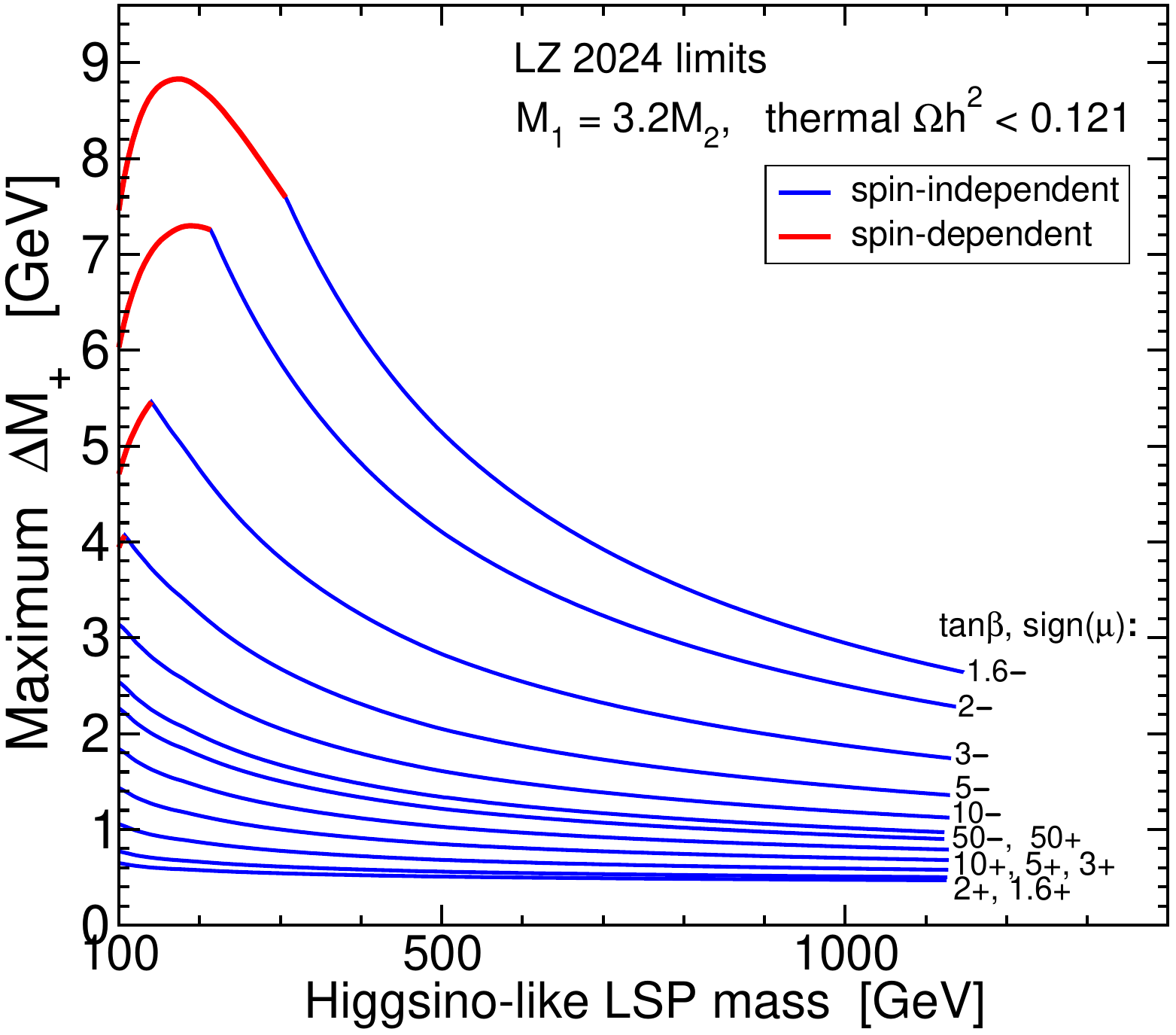}
\includegraphics[width=0.51\linewidth]{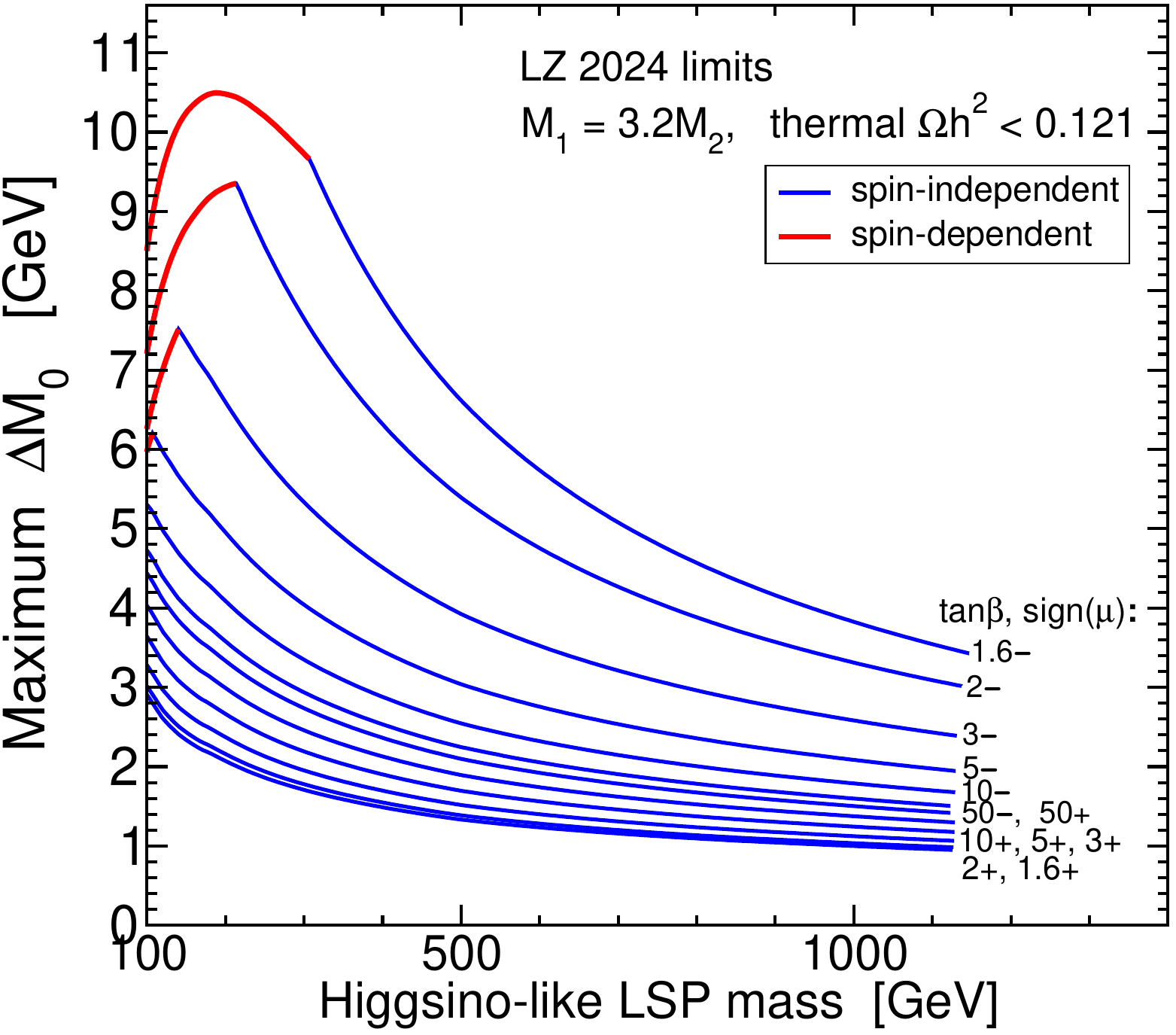}
}
\caption{\label{fig:deltaMunif}
The maximum mass splittings $\Delta M_+ = m_{\tilde C_1} - m_{\tilde N_1}$ (left panels) and $\Delta M_0 = m_{\tilde N_2} - m_{\tilde N_1}$ (right panels) allowed by the LZ2024 limits \cite{LZCollaboration:2024lux} for the higgsino-like states, for various choices of $\tan\beta$ and sign($\mu$) as labeled.
In the top two panels, the gaugino mass parameters are assumed to be related by the gaugino mass unification condition $M_2 = 1.8 M_1$, and in the bottom two panels
they are related by the AMSB-inspired condition $M_1 = 3.2 M_1$.
The thinner (blue) portions of the curves are set by the SI LSP-xenon cross-section limit, while the thicker (red) portions are set by the 
SD cross-section limit. The LSP density is set by thermal freezeout, with $\Omega_{\mbox{\scriptsize LSP}} h^2 = 0.121$ for the points on the far right of each curve.
The lightest Higgs boson mass is fixed to $M_h = 125.1$ GeV, and all other
scalar masses are set to 10 TeV.}
\end{figure}

More generally, I find upper bounds on the mass splittings $\Delta M_+$ and $\Delta M_0$
as shown in Figure \ref{fig:deltaMunif}, for the gaugino
mass unification and AMSB scenarios. The model line curves have different values
of $\tan\beta$ and sign$(\mu)$, as labeled. 
The thinner (blue) portions of the curves are set by the LZ2024 SI LSP-xenon cross-section limit, while the thicker (red) portions are set by the SD cross-section limit. We see that for $\tan\beta > 1.6$, one has $\Delta M_0$ less than 11 GeV, in both cases, and $\Delta M_+$ is less than 8 GeV in the gaugino mass unification case, and less than 9 GeV in the AMSB-inspired case. 
%%%%%%%%%%%%%%%%%%%%%%%%%%%%%%%%%%%%%%%%%%%%%%%%%%%%%%%%%%%%%%%%%%%%%%%%%%%%%%%%%
%%%%%%%%%%%%%%%%%%%%%%%%%%%%%%%%%%%%%%%%%%%%%%%%%%%%%%%%%%%%%%%%%%%%%%%%%%%%%%%%%
For larger $\tan\beta$, or for $\mu > 0$, or for larger $|\mu|$, the constraints become much more stringent, as shown. For higgsinos with mass near 1.1 TeV, such that the thermal freezeout density realizes the observed value $\Omega h^2 = 0.12$, we see that both $\Delta M_+$ and $\Delta M_0$ cannot exceed a few GeV. 
These bounds can be relaxed by going to $\mu<0$ with $\tan\beta < 1.6$, at the cost of 
difficulties achieving $M_h = 125.1$ GeV within the MSSM, and
a top-quark Yukawa coupling that may run non-perturbatively large in the ultraviolet.

At this writing, the latest ATLAS and CMS searches for neutralinos and charginos with compressed mass spectra leading to soft trilepton and dilepton events have observed 
\cite{ATLAS:2021moa,CMS:2023qhl,CMS:2021edw} 
mild excesses, corresponding to weaker-than-expected limits. 
Possible explanations of this in terms of neutralinos and charginos in compressed supersymmetry have appeared in refs.~\cite{Agin:2023yoq,Chakraborti:2024pdn,Agin:2024yfs}.
In ref.~\cite{Martin:2024pxx}, it was claimed that the excess regions could be made consistent with higgsino dark matter satisfying the LUX-ZEPLIN bounds from 2022 \cite{LZ:2022lsv}, which were the strongest ones at that time. However, with the newer LZ2024 bounds, the regions identified in ref.~\cite{Martin:2024pxx} are now essentially eliminated, since the higgsino mass splittings are now constrained to be too small. Of course, if the higgsino LSP is unstable, then the dark matter bounds simply do not apply, and it is easy to accommodate the LHC soft lepton excess region. Other possibilities to maintain the higgsino dark matter interpretation include $\tan\beta$ very close to 1 with $n=-1$, or interference coming from the contributions of other Higgs bosons. In any case, new results from the LHC should be dispositive.
 
\FloatBarrier

%%%%%%%%%%%%%%%%%%%%%%%%%%%%%%%%%%%%%%%%%%%%%%%%%%%%%%%%%%%%%%%
\section{Projections for higgsino dark matter after reaching the neutrino fog\label{sec:foggy}}
\setcounter{equation}{0}
\setcounter{figure}{0}
\setcounter{table}{0} 
\setcounter{footnote}{1}

Suppose that there is indeed no dark matter species with mass $m$ greater than 100 GeV,
density characterized by $\Omega h^2$, and SI
cross-section in the range
\beq
5 \times 10^{-12} \>{\rm pb} 
\,<\,
 \left (\frac{\mbox{1 TeV}}{m} \right ) 
\left (\frac{\Omega h^2}{0.12} \right ) \sigma_{\rm SI}
\,<\,
3 \times 10^{-11} \>{\rm pb},
\eeq
corresponding approximately to the unshaded region in any of the panels in Figures \ref{fig:unif} and \ref{fig:AMSB}.
Given the rapid pace of improved limits, it seems probable that the direct detection curtain will continue to lower, eliminating the unshaded region. After the discovery neutrino fog is reached, experimental progress is likely to be slowed considerably, so it is useful to project the constraints that will then apply for gaugino masses and higgsino mass splittings.

\begin{figure}[!t]
\centering
\mbox{
\includegraphics[width=0.51\linewidth]{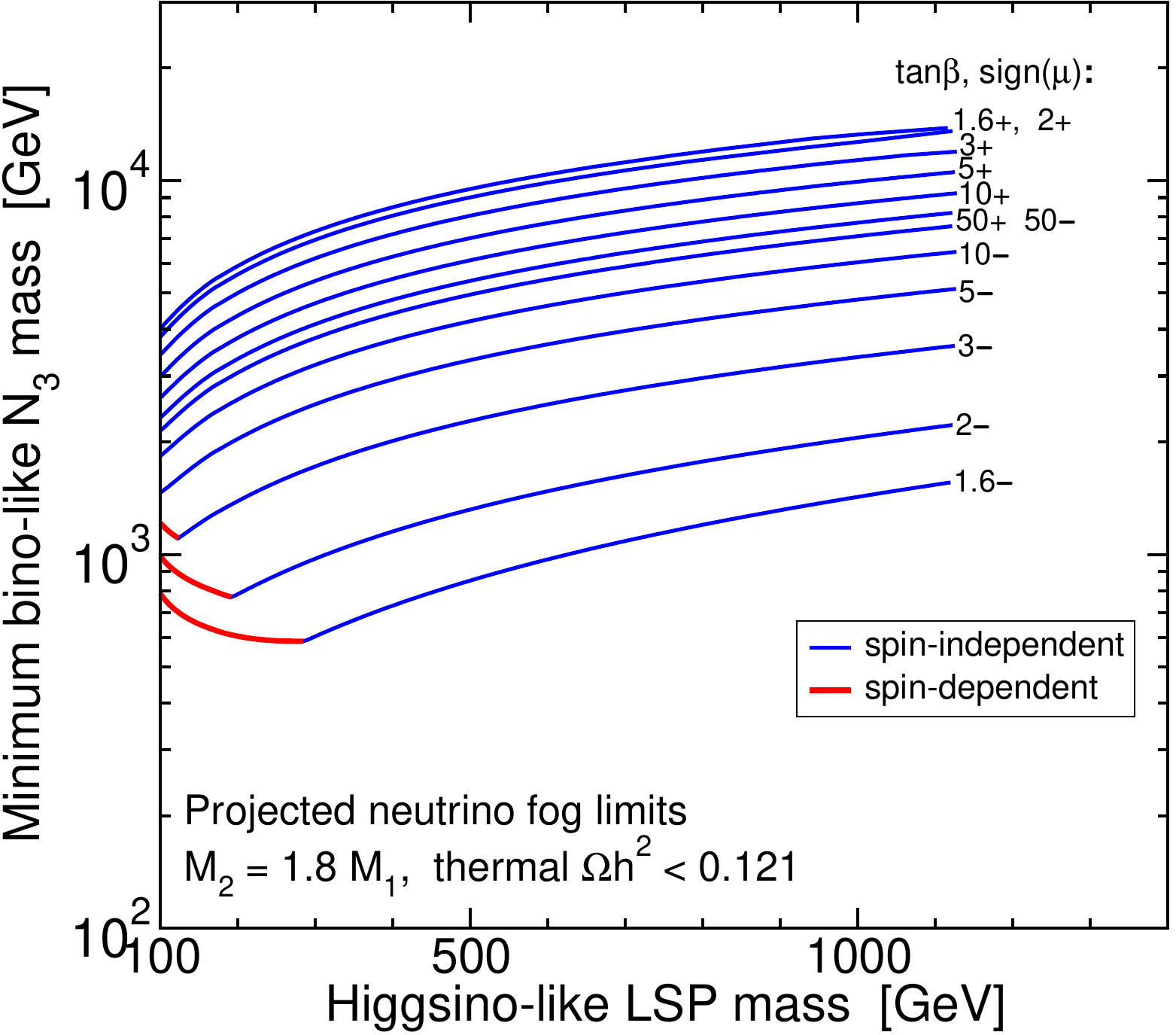}
\includegraphics[width=0.51\linewidth]{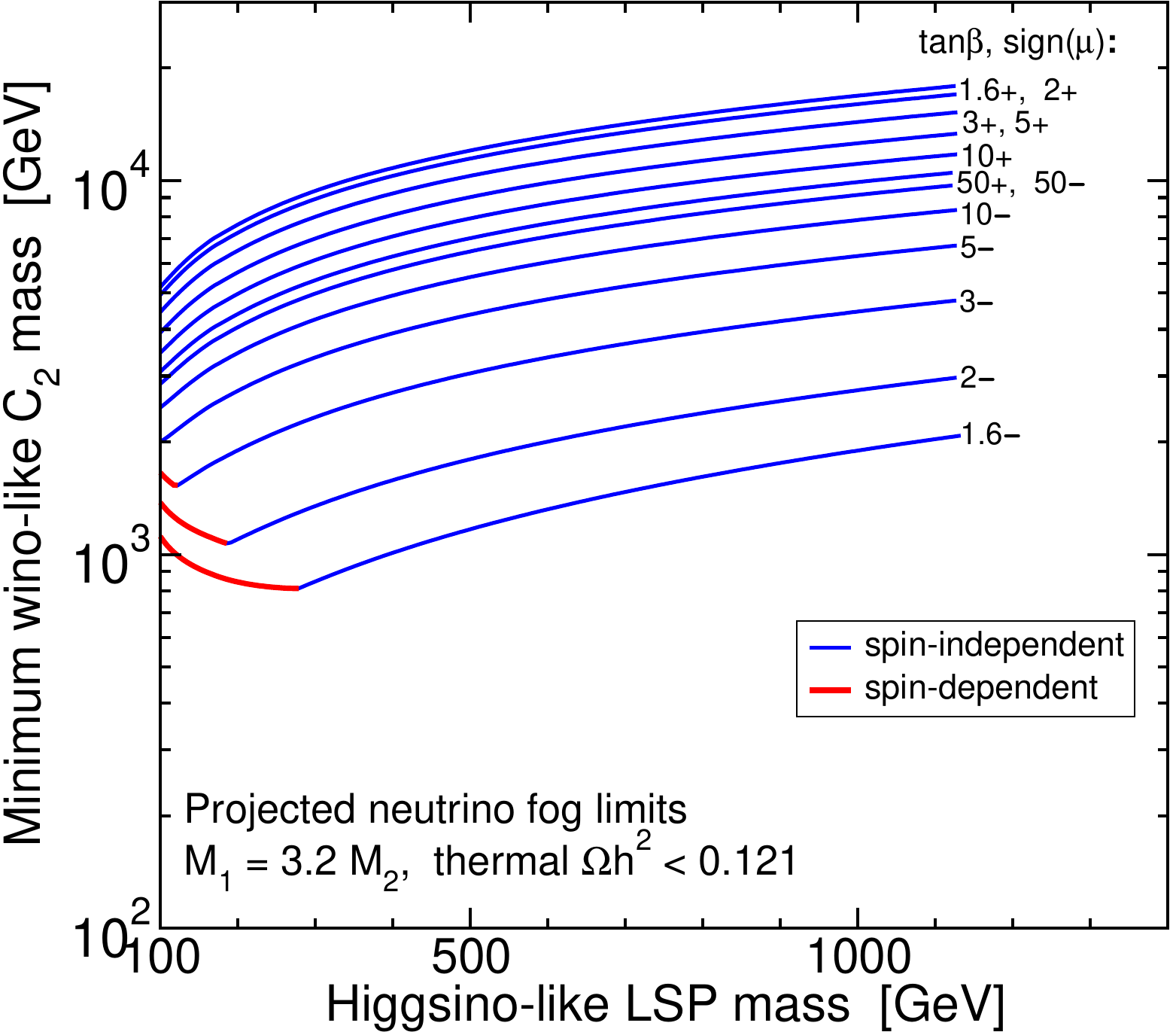}
}
\caption{\label{fig:deltaMunif_fog}Projected limits, if the discovery neutrino fog is reached, for the minimum bino-like neutralino $\tilde N_3$ mass in models with $M_2 = 1.8 M_1$ (left panel) and wino-like chargino $\tilde C_2$ mass in models with $M_1 = 3.2 M_1$ (right panel). These two choices are inspired by, respectively, gaugino mass unification and AMSB models. The curves have fixed choices of $\tan\beta$ and sign($\mu$) as labeled. The thinner (blue) portions of the curves are set by the SI cross-section limit, and the thicker (red) portions by the SD cross-section limit, in each case taken to equal the discovery neutrino fog level as given in ref.~\cite{OHare:2021utq}. The LSP density is determined by thermal freezeout, with $\Omega_{\mbox{\scriptsize LSP}} h^2 = 0.121$ for the points on the far right of each curve. The lightest Higgs boson mass is fixed to $M_h = 125.1$ GeV, and all other scalar masses are set to 10 TeV.}
\end{figure}

Figure \ref{fig:deltaMunif_fog} shows the projected limits, if the discovery neutrino fog is reached, for the minimum bino-like neutralino $\tilde N_3$ mass in models with $M_2 = 1.8 M_1$ (left panel) and for the wino-like chargino $\tilde C_2$ mass in models with $M_1 = 3.2 M_1$ (right panel). 
%%%%%%%%%%%%%%%%%%%%%%%%%%%%%%%%%%%%%%%%%%%%%%%%%%%%%%%%%%%%%%%%%%%%%%%%%%%%%%%%%
%%%%%%%%%%%%%%%%%%%%%%%%%%%%%%%%%%%%%%%%%%%%%%%%%%%%%%%%%%%%%%%%%%%%%%%%%%%%%%%%%
In this figure, it is assumed that the density $\Omega_{\mbox{\scriptsize LSP}} h^2$ is the thermal one, so that the higgsino LSP is only one component of the dark matter. All other model choices are the same as in the preceding, including decoupling of the superpartners other than neutralinos and charginos and the heavy Higgs scalars. 
The curves have fixed values of $\tan\beta$ and sign$(\mu)$, as labeled. 
As usual, the bounds become monotonically weaker for lower sign$(\mu)/\tan\beta$, and
for the lowest values the SD cross-section sets the limit when $m_{\tilde N_1}$ is sufficiently small. As mentioned earlier, the resulting lower bounds on the gaugino masses when the discovery neutrino fog is reached will be strengthened compared to the current ones (in Figures \ref{fig:contoursM1unif} and \ref{fig:contoursM2amsb}) by only a factor ranging from about 1.25 to 2.4,
depending on the choice of $\tan\beta$ and sign$(\mu)$ and the mass. 

\begin{figure}[!t]
\centering
\mbox{
\includegraphics[width=0.51\linewidth]{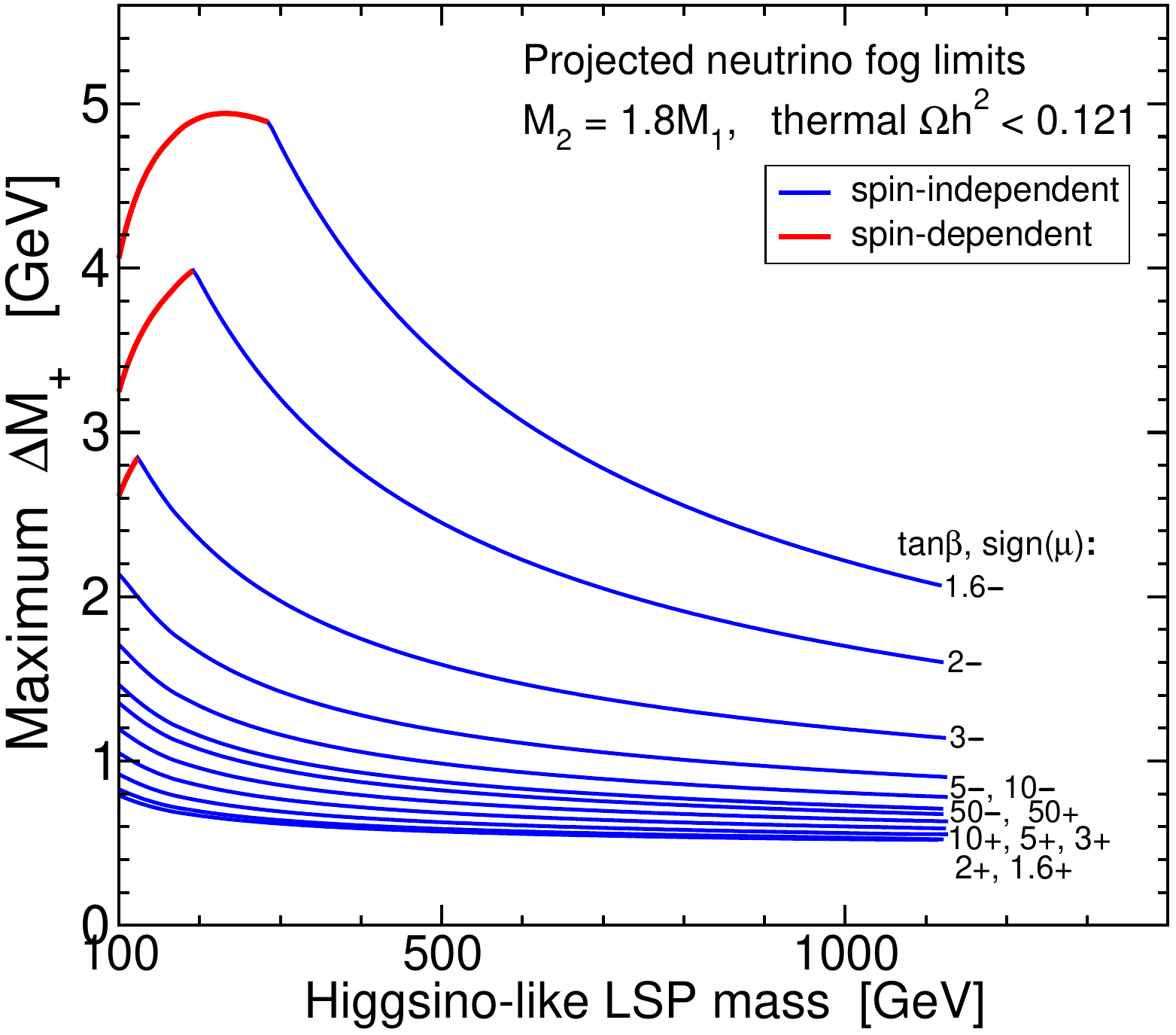}
\includegraphics[width=0.51\linewidth]{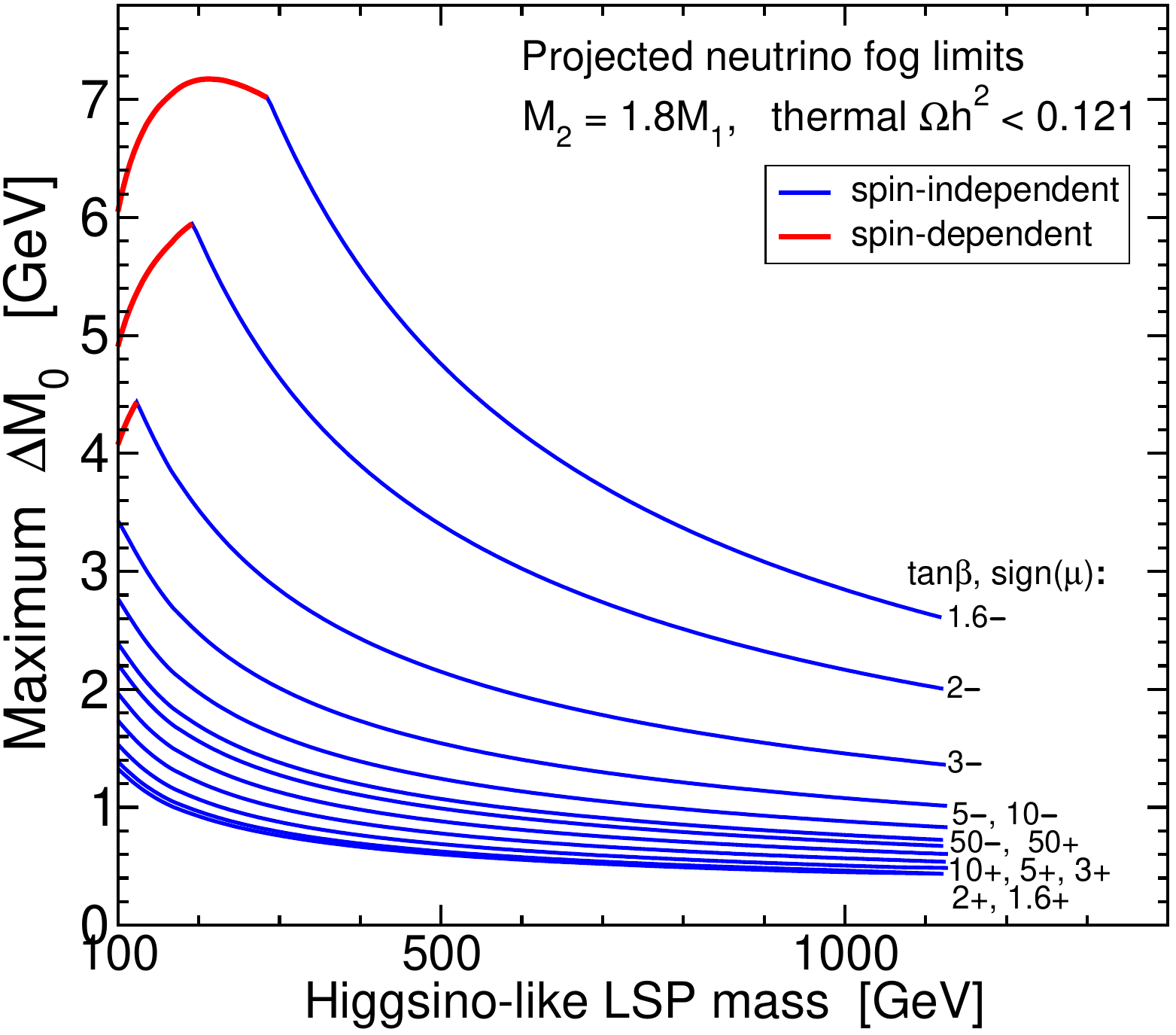}
}
\mbox{
\includegraphics[width=0.51\linewidth]{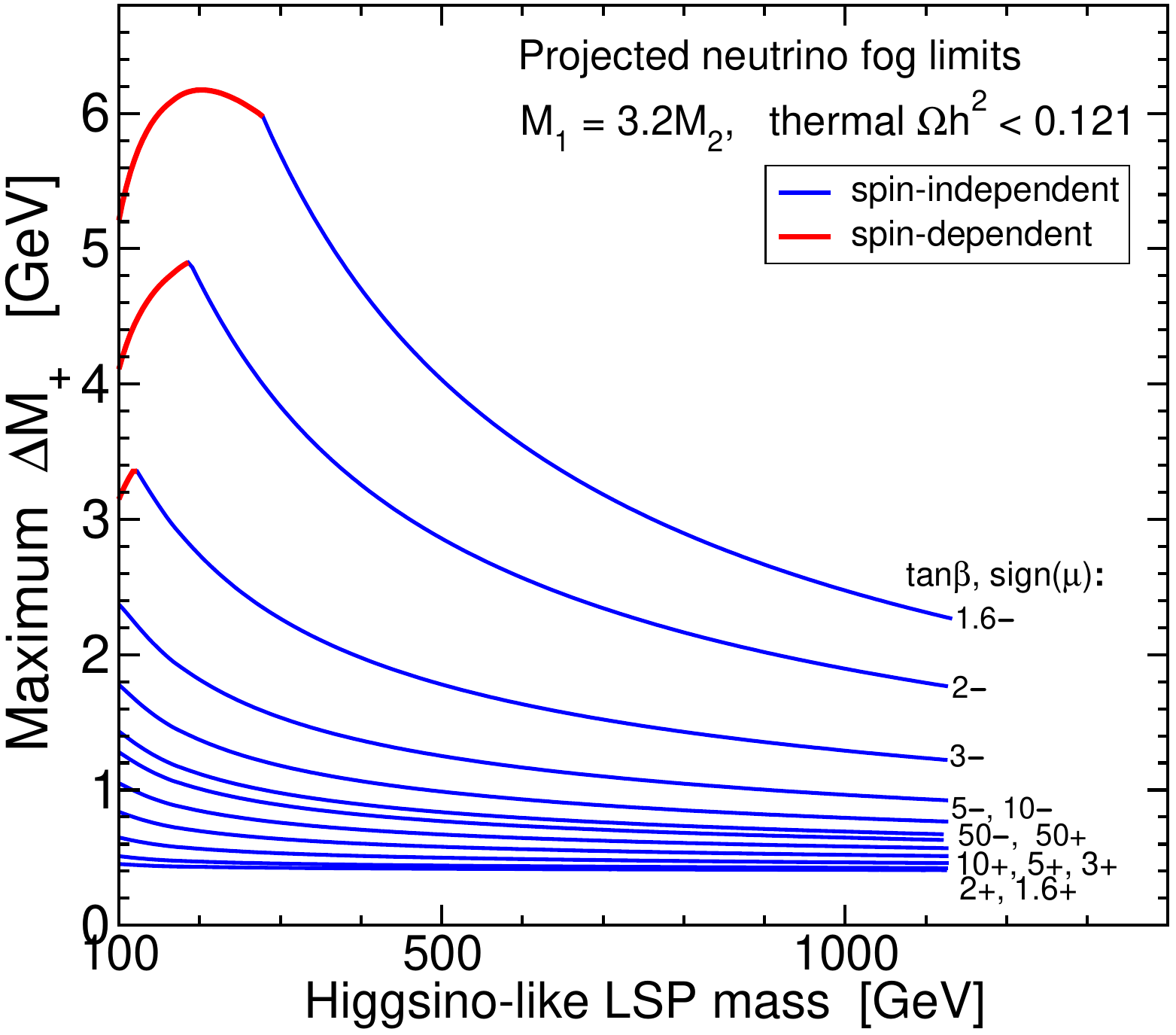}
\includegraphics[width=0.51\linewidth]{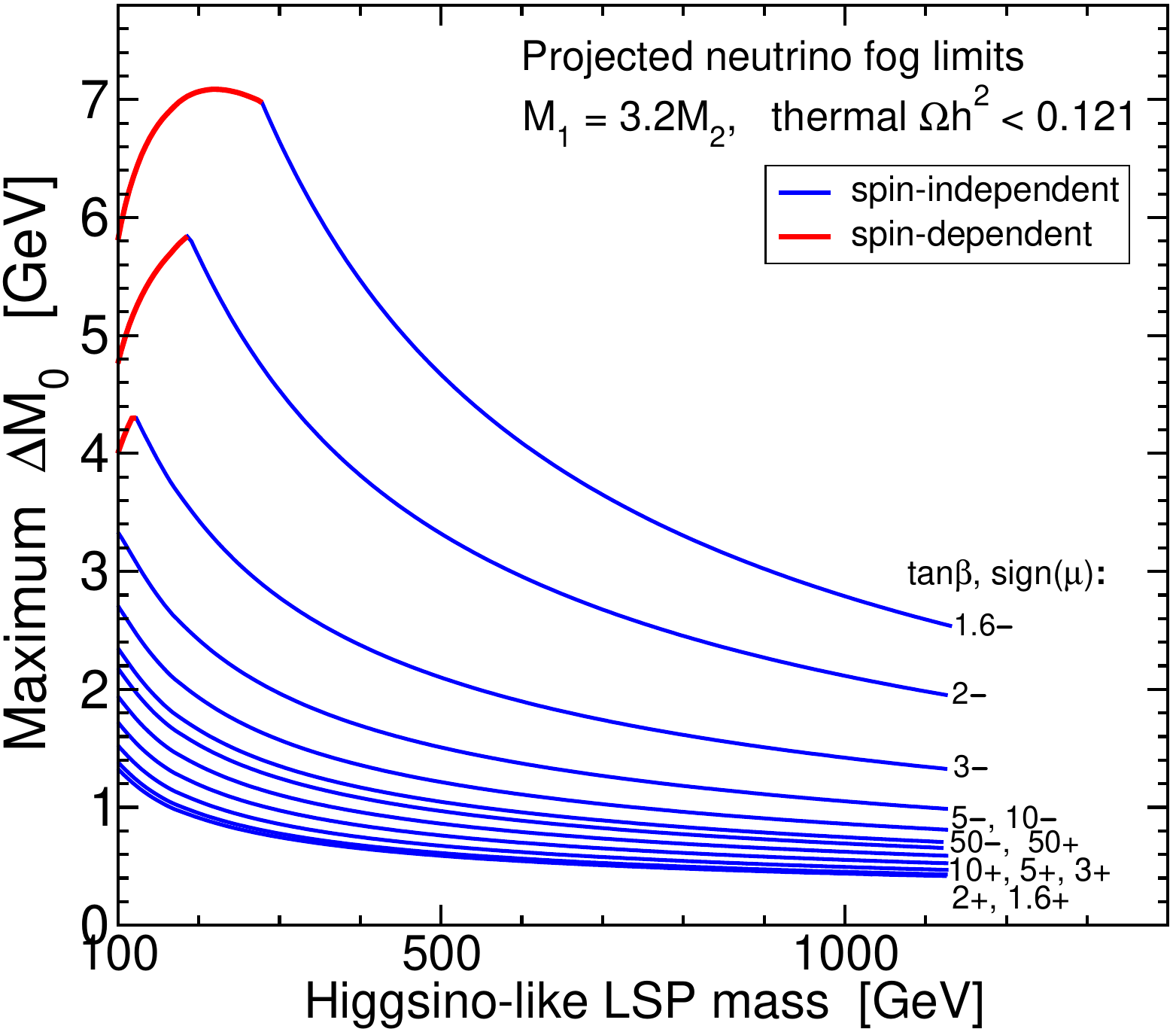}
}
\caption{\label{fig:deltaMfog}Projected limits, if the discovery neutrino fog is reached, for the maximum mass splittings $\Delta M_+ = m_{\tilde C_1} - m_{\tilde N_1}$ (left panels) and $\Delta M_0 = m_{\tilde N_2} - m_{\tilde N_1}$ (right panels) for the higgsino-like states, for various choices of $\tan\beta$ and sign($\mu$) as labeled. In the top two panels, the gaugino mass unification condition $M_2 = 1.8 M_1$ is assumed, and in the bottom two panels the gaugino masses are related by the AMSB-inspired condition $M_1 = 3.2 M_1$. The thinner (blue) portions of the curves are set by the SI LSP-xenon cross-section, and the thicker (red) portions by the SD cross-section, in each case taken to equal the discovery neutrino fog level as given in ref.~\cite{OHare:2021utq}. The LSP density is set by thermal freezeout, with $\Omega_{\mbox{\scriptsize LSP}} h^2 = 0.121$ at the far right of each curve. The lightest Higgs boson mass is fixed to $M_h = 125.1$ GeV, and all other
scalar masses are set to 10 TeV.}
\end{figure}

The corresponding projected limits for the higgsino mass splittings $\Delta M_+$ and
$\Delta M_0$ are shown in Figure \ref{fig:deltaMfog}. Taking into account the important SD
cross-section limits, we see that even if $m_{\tilde N_1}$ is near its LEP lower bound of approximately 100 GeV,  $\Delta M_0$ will be limited above by about 7 GeV if and when the discovery neutrino fog is reached, provided $\tan\beta > 1.6$, while $\Delta M_+$ is limited
to be somewhat less. The bounds for $\Delta M_0$ and $\Delta M_+$ decrease to about
2.6 GeV if the dark matter density is saturated with $m_{\tilde N_1} \approx 1.1$ TeV.
For large $\tan\beta$, or if $\mu > 0$, the mass splittings
will have to satisfy much more stringent limits, as shown. Present LHC limits
\cite{ATLAS:2021moa,CMS:2023qhl,CMS:2021edw}
on compressed neutralinos and charginos are not shown here because they are quite sensitive to assumptions made about the mass splittings, which in the experimental papers differ from the models considered here,
but in any case exclusions are only in the 100-200 GeV range for $m_{\tilde N_1}$.
%%%%%%%%%%%%%%%%%%%%%%%%%%%%%%%%%%%%%%%%%%%%%%%%%%%%%%%%%%%%%%%%%%%%%%%%%%%%%%%%%
%%%%%%%%%%%%%%%%%%%%%%%%%%%%%%%%%%%%%%%%%%%%%%%%%%%%%%%%%%%%%%%%%%%%%%%%%%%%%%%%%

The range of mass differences in Figure \ref{fig:deltaMfog} present a severe challenge for future collider searches. 
A future $e^+e^-$ collider should be able to probe higgsinos up to the kinematic limit at best, so that nearer-term Higgs factories will not have much (or perhaps any) reach beyond the LHC, and on longer time scales, at least about $\sqrt{s} = 2.3$ TeV would be required to cover the thermal freezeout range. 
Considering the disappearing track searches from a long-lived $m_{\tilde C_1}$, 
for a high-luminosity LHC with 3000 fb$^{-1}$ at $\sqrt{s} = 14$ TeV, it was estimated in ref.~\cite{ATLAS:2018jjf}
that the exclusion reach will extend up to about
$m_{\tilde N_1} = 270$ GeV. 
For future $pp$ colliders, disappearing track searches for higgsinos have been studied in refs.~\cite{Mahbubani:2017gjh,Fukuda:2017jmk,Saito:2019rtg}.
For a $\sqrt{s}=100$ TeV $pp$ collider with 30 ab$^{-1}$ of integrated luminosity, ref.~\cite{Saito:2019rtg} found that the discovery reach should extend beyond the critical masses near 1.1 TeV set by thermal freezeout. Reference \cite{Capdevilla:2021fmj} found that the same is true for a future $\mu^+\mu^-$ collider with 10 ab$^{-1}$ at $\sqrt{s} = 10$ TeV. 
However, all of these disappearing track signals reaches are valid only in the case of an extremely pure higgsino
with gaugino masses above 25 TeV or so. For the realistic case of a less pure higgsino, the lifetime of $m_{\tilde C_1}$ is too short for disappearing track searches to be efficient. As we have seen, there is a significant range of gaugino masses for which the dark matter direct detection will be below the projected discovery neutrino fog level but the higgsinos are not pure enough for disappearing track searches to be effective.

Considering larger mass splittings, projections for a soft-lepton signal at a high-luminosity LHC with 3000 fb$^{-1}$ at $\sqrt{s} = 14$ TeV have been presented in \cite{ATLAS:2018jjf}. It was found that the 5$\sigma$ discovery reach is restricted to $m_{\tilde N_1} < 210$ GeV and in a narrow range of $\Delta M_0$ near
4-6 GeV. Exclusion is possible up to at most $m_{\tilde N_1} = 350$ GeV, but with a significantly diminished reach for $\Delta M_0$ outside of the range 4-6 GeV. 
Monojet signals have been investigated at the LHC in refs.~\cite{Schwaller:2013baa,Baer:2014cua,Han:2014kaa,Low:2014cba}
and for future $pp$ colliders in refs.~\cite{Low:2014cba,Han:2018wus}, with varying results.
Future muon collider signatures for higgsinos without perfect purity have been investigated in refs.~\cite{Han:2020uak,Han:2022ubw,Capdevilla:2024bwt}.
A mono-muon signature unique to muon colliders was used 
in ref.~\cite{Han:2020uak,Han:2022ubw} 
to show that, essentially independent of
the small mass differences $\Delta M_0$ and $\Delta M_+$, a 95\% exclusion for the whole range up to $m_{\tilde N_1} \approx 1.1$ TeV can be accomplished with a $\sqrt{s} = 10$ TeV muon collider with 10 ab$^{-1}$, and a 5$\sigma$ discovery would require $\sqrt{s} = 14$ TeV with 20 ab$^{-1}$. In ref.~\cite{Capdevilla:2024bwt}, it was argued that a search for soft charged tracks at a $\sqrt{s} = 3$ TeV muon collider with 1 ab$^{-1}$ may be able to explore the region with $\Delta M_+ \lsim 1.5$ GeV. More detailed studies and proposals for innovative search strategies for future colliders would be useful in order to confirm and improve these results.
 
\FloatBarrier

%%%%%%%%%%%%%%%%%%%%%%%%%%%%%%%%%%%%%%%%%%%%%%%%%%%%%%%%%%%%%%%
\section{Outlook\label{sec:outlook}}
\setcounter{equation}{0}
\setcounter{figure}{0}
\setcounter{table}{0} 
\setcounter{footnote}{1}

Searches for higgsino-like dark matter have significantly reduced
the parameter space available where it could be directly detected easily above the neutrino fog, leaving 
regions with a high degree of higgsino purity and small mass splittings. 
In this paper, I have quantified this both for the current LZ2024 bounds and for the
projected future case that the discovery neutrino fog is reached. Assuming that the latter scenario is indeed realized without evidence for direct detection, further progress in direct detection exploration will continue, but will be slowed. 
The weakest upper bounds on higgsino mass splittings occur for small $\tan\beta\rightarrow 1$ and sign$(\mu\delta) = -1$, and the strongest for small $\tan\beta$ and sign$(\mu\delta) = +1$. The remaining parameter space includes gaugino masses in the interesting multi-TeV range.
It is unfortunate that this will be difficult to probe with collider experiments due to higgsino mass splittings that are small, but quite possibly not small enough to enable efficient disappearing track searches.

The results of the present paper were based on a simple, and well-motivated, assumption that all superpartners and heavy Higgs bosons are assumed to be decoupled due to large masses. It is important to keep in mind that the bounds displayed above are not mathematical theorems, as there are some possible ways to evade them by generalizing the assumptions made here. First, one might approach the $\tan\beta \rightarrow 1$ blind spot even more closely, despite the large top-quark Yukawa coupling and small Higgs mass problems. Second, there is another quasi-blind spot when $M_1$ and $M_2$ have opposite signs, identified above as $\delta \approx 0$; although this cannot lead to larger mass splittings among the higgsinos, it would at least allow the gauginos to be lighter. This suppression mechanism can be combined with the suppression from $\tan\beta\rightarrow 1$ and sign$(\mu\delta) = -1$, as they arise from independent multiplicative factors in the couplings. Third, one could invoke the more traditional blind spots in which there is interference between different Higgs scalar-mediated amplitudes for LSP-nucleon scattering. Fourth, it may be that the assumed density of dark matter in our neighborhood is smaller than the standard assumption. Finally, it should always be noted that the higgsino LSP associated with
a small $\mu$ parameter may not be stable at all, or has a density much smaller than the prediction of thermal freezeout, either of which would leave alive a much wider spectrum of possibilities for collider searches to look forward to.

Acknowledgments: This work is supported in part by the National Science Foundation grant with award number 2310533.

%%%%%%%%%%%%%%%%%%%%%%%%%%%%%%%%%%%%%%%%%%%%%%%%%%%%%%%%%%%%%%%%%%%%


\begin{thebibliography}{90}
\baselineskip=12.35pt

\bibitem{Planck:2018vyg}
N.~Aghanim \textit{et al.} [Planck],
``Planck 2018 results. VI. Cosmological parameters,''
Astron. Astrophys. \textbf{641}, A6 (2020)
[erratum: Astron. Astrophys. \textbf{652}, C4 (2021)]
%doi:10.1051/0004-6361/201833910
[arXiv:1807.06209 [astro-ph.CO]].

\bibitem{Cirelli:2024ssz}
M.~Cirelli, A.~Strumia and J.~Zupan,
``Dark Matter,''
[arXiv:2406.01705 [hep-ph]].

\bibitem{Wells:2003tf}
J.~D.~Wells,
``Implications of supersymmetry breaking with a little hierarchy between gauginos and scalars,''
[arXiv:hep-ph/0306127].

\bibitem{Arkani-Hamed:2004zhs}
N.~Arkani-Hamed, S.~Dimopoulos, G.~F.~Giudice and A.~Romanino,
``Aspects of split supersymmetry,''
Nucl. Phys. B \textbf{709}, 3-46 (2005)
%doi:10.1016/j.nuclphysb.2004.12.026
[arXiv:hep-ph/0409232].

\bibitem{Wells:2004di}
J.~D.~Wells,
``PeV-scale supersymmetry,''
Phys. Rev. D \textbf{71}, 015013 (2005)
%doi:10.1103/PhysRevD.71.015013
[arXiv:hep-ph/0411041].

\bibitem{Haber:1990aw}
H.~E.~Haber and R.~Hempfling,
``Can the mass of the lightest Higgs boson of the minimal supersymmetric model be larger than m(Z)?,''
Phys. Rev. Lett. \textbf{66}, 1815-1818 (1991)
doi:10.1103/PhysRevLett.66.1815

\bibitem{Okada:1990vk}
Y.~Okada, M.~Yamaguchi and T.~Yanagida,
``Upper bound of the lightest Higgs boson mass in the minimal supersymmetric standard model,''
Prog. Theor. Phys. \textbf{85}, 1-6 (1991)
doi:10.1143/ptp/85.1.1

\bibitem{Ellis:1990nz}
J.~R.~Ellis, G.~Ridolfi and F.~Zwirner,
``Radiative corrections to the masses of supersymmetric Higgs bosons,''
Phys. Lett. B \textbf{257}, 83-91 (1991)
doi:10.1016/0370-2693(91)90863-L

\bibitem{Haber:1996fp}
H.~E.~Haber, R.~Hempfling and A.~H.~Hoang,
``Approximating the radiatively corrected Higgs mass in the minimal supersymmetric model,''
Z. Phys. C \textbf{75}, 539-554 (1997)
%doi:10.1007/s002880050498
[arXiv:hep-ph/9609331 [hep-ph]].

\bibitem{Bhattiprolu:2023lfh}
P.~N.~Bhattiprolu and J.~D.~Wells,
``Precision unification and the scale of supersymmetry,''
Phys. Rev. D \textbf{109}, no.1, L011704 (2024)
%doi:10.1103/PhysRevD.109.L011704
[arXiv:2309.12954 [hep-ph]].

\bibitem{Bae:2013bva}
K.~J.~Bae, H.~Baer and E.~J.~Chun,
``Mainly axion cold dark matter from natural supersymmetry,''
Phys. Rev. D \textbf{89}, no.3, 031701 (2014)
%doi:10.1103/PhysRevD.89.031701
[arXiv:1309.0519 [hep-ph]].

\bibitem{Bae:2014yta}
K.~J.~Bae, H.~Baer and H.~Serce,
``Natural little hierarchy for SUSY from radiative breaking of the Peccei-Quinn symmetry,''
Phys. Rev. D \textbf{91}, no.1, 015003 (2015)
%doi:10.1103/PhysRevD.91.015003
[arXiv:1410.7500 [hep-ph]].

\bibitem{Bae:2017hlp}
K.~J.~Bae, H.~Baer and H.~Serce,
``Prospects for axion detection in natural SUSY with mixed axion-higgsino dark matter: back to invisible?,''
JCAP \textbf{06}, 024 (2017)
%doi:10.1088/1475-7516/2017/06/024
[arXiv:1705.01134 [hep-ph]].

\bibitem{Baer:2019uom}
H.~Baer, V.~Barger, D.~Sengupta, H.~Serce, K.~Sinha and R.~W.~Deal,
``Is the magnitude of the Peccei\textendash{}Quinn scale set by the landscape?,''
Eur. Phys. J. C \textbf{79}, no.11, 897 (2019)
%doi:10.1140/epjc/s10052-019-7408-x
[arXiv:1905.00443 [hep-ph]].

%%%%%%%%%%%%%%%%%%%%%%%%%%%%%%%%%%%%%%%%%

\bibitem{Moroi:1999zb}
T.~Moroi and L.~Randall,
``Wino cold dark matter from anomaly mediated SUSY breaking,''
Nucl. Phys. B \textbf{570}, 455-472 (2000)
%doi:10.1016/S0550-3213(99)00748-8
[arXiv:hep-ph/9906527].

\bibitem{Fujii:2001xp}
M.~Fujii and K.~Hamaguchi,
``Higgsino and wino dark matter from Q ball decay,''
Phys. Lett. B \textbf{525}, 143-149 (2002)
%doi:10.1016/S0370-2693(01)01412-5
[arXiv:hep-ph/0110072 [hep-ph]].

\bibitem{Gelmini:2006pw}
G.~B.~Gelmini and P.~Gondolo,
``Neutralino with the right cold dark matter abundance in (almost) any supersymmetric model,''
Phys. Rev. D \textbf{74}, 023510 (2006)
%doi:10.1103/PhysRevD.74.023510
[arXiv:hep-ph/0602230].

\bibitem{Gelmini:2006pq}
G.~Gelmini, P.~Gondolo, A.~Soldatenko and C.~E.~Yaguna,
``The Effect of a late decaying scalar on the neutralino relic density,''
Phys. Rev. D \textbf{74}, 083514 (2006)
%doi:10.1103/PhysRevD.74.083514
[arXiv:hep-ph/0605016 [hep-ph]].

\bibitem{Han:2019vxi}
C.~Han,
``Higgsino Dark Matter in a Non-Standard History of the Universe,''
Phys. Lett. B \textbf{798}, 134997 (2019)
%doi:10.1016/j.physletb.2019.134997
[arXiv:1907.09235 [hep-ph]].

\bibitem{Fukuda:2024ddb}
H.~Fukuda, Q.~Li, T.~Moroi and A.~Niki,
``Non-thermal production of Higgsino dark matter by late-decaying scalar fields,''
[arXiv:2410.15733 [hep-ph]].



%%%%%%%%%%%%%%%%%%%%%%%%%%%%%%%%%%%%%

\bibitem{Drees:1996pk}
M.~Drees, M.~M.~Nojiri, D.~P.~Roy and Y.~Yamada,
``Light Higgsino dark matter,''
Phys. Rev. D \textbf{56}, 276-290 (1997)
[erratum: Phys. Rev. D \textbf{64}, 039901 (2001)]
%doi:10.1103/PhysRevD.64.039901
[arXiv:hep-ph/9701219].

\bibitem{Thomas:1998wy}
S.~D.~Thomas and J.~D.~Wells,
``Phenomenology of Massive Vectorlike Doublet Leptons,''
Phys. Rev. Lett. \textbf{81}, 34-37 (1998)
%doi:10.1103/PhysRevLett.81.34
[arXiv:hep-ph/9804359].

\bibitem{Feng:2000gh}
J.~L.~Feng, K.~T.~Matchev and F.~Wilczek,
``Neutralino dark matter in focus point supersymmetry,''
Phys. Lett. B \textbf{482}, 388-399 (2000)
%doi:10.1016/S0370-2693(00)00512-8
[arXiv:hep-ph/0004043 [hep-ph]].

\bibitem{Giudice:2004tc}
G.~F.~Giudice and A.~Romanino,
``Split supersymmetry,''
Nucl. Phys. B \textbf{699}, 65-89 (2004)
[erratum: Nucl. Phys. B \textbf{706}, 487-487 (2005)]
%doi:10.1016/j.nuclphysb.2004.08.001
[arXiv:hep-ph/0406088].

\bibitem{Profumo:2004at}
S.~Profumo and C.~E.~Yaguna,
``A Statistical analysis of supersymmetric dark matter in the MSSM after WMAP,''
Phys. Rev. D \textbf{70}, 095004 (2004)
%doi:10.1103/PhysRevD.70.095004
[arXiv:hep-ph/0407036].

\bibitem{Hisano:2004ds}
J.~Hisano, S.~Matsumoto, M.~M.~Nojiri and O.~Saito,
``Non-perturbative effect on dark matter annihilation and gamma ray signature from galactic center,''
Phys. Rev. D \textbf{71}, 063528 (2005)
%doi:10.1103/PhysRevD.71.063528
[arXiv:hep-ph/0412403].

\bibitem{Baer:2011ec}
H.~Baer, V.~Barger and P.~Huang,
``Hidden SUSY at the LHC: the light higgsino-world scenario and the role of a lepton collider,''
JHEP \textbf{11}, 031 (2011)
%doi:10.1007/JHEP11(2011)031
[arXiv:1107.5581 [hep-ph]].

\bibitem{Baer:2012cf}
H.~Baer, V.~Barger, P.~Huang, D.~Mickelson, A.~Mustafayev and X.~Tata,
``Radiative natural supersymmetry: Reconciling electroweak 
fine-tuning and the Higgs boson mass,''
Phys. Rev. D \textbf{87}, no.11, 115028 (2013)
%doi:10.1103/PhysRevD.87.115028
[arXiv:1212.2655 [hep-ph]].

\bibitem{Baer:2013yha}
H.~Baer, V.~Barger, P.~Huang, D.~Mickelson, A.~Mustafayev, W.~Sreethawong and X.~Tata,
``Same sign diboson signature from supersymmetry models with light higgsinos at the LHC,''
Phys. Rev. Lett. \textbf{110}, no.15, 151801 (2013)
%doi:10.1103/PhysRevLett.110.151801
[arXiv:1302.5816 [hep-ph]].

\bibitem{Schwaller:2013baa}
P.~Schwaller and J.~Zurita,
``Compressed electroweakino spectra at the LHC,''
JHEP \textbf{03}, 060 (2014)
%doi:10.1007/JHEP03(2014)060
[arXiv:1312.7350 [hep-ph]].

\bibitem{Baer:2014cua}
H.~Baer, A.~Mustafayev and X.~Tata,
``Monojets and mono-photons from light higgsino pair production at LHC14,''
Phys. Rev. D \textbf{89}, no.5, 055007 (2014)
%doi:10.1103/PhysRevD.89.055007
[arXiv:1401.1162 [hep-ph]].

\bibitem{Han:2014kaa}
Z.~Han, G.~D.~Kribs, A.~Martin and A.~Menon,
``Hunting quasidegenerate Higgsinos,''
Phys. Rev. D \textbf{89}, no.7, 075007 (2014)
%doi:10.1103/PhysRevD.89.075007
[arXiv:1401.1235 [hep-ph]].

\bibitem{Low:2014cba}
M.~Low and L.~T.~Wang,
``Neutralino dark matter at 14 TeV and 100 TeV,''
JHEP \textbf{08}, 161 (2014)
doi:10.1007/JHEP08(2014)161
[arXiv:1404.0682 [hep-ph]].

\bibitem{Nagata:2014wma}
N.~Nagata and S.~Shirai,
``Higgsino Dark Matter in High-Scale Supersymmetry,''
JHEP \textbf{01}, 029 (2015)
%doi:10.1007/JHEP01(2015)029
[arXiv:1410.4549 [hep-ph]].

\bibitem{Evans:2014pxa}
J.~L.~Evans, M.~Ibe, K.~A.~Olive and T.~T.~Yanagida,
``Light Higgsinos in Pure Gravity Mediation,''
Phys. Rev. D \textbf{91}, 055008 (2015)
%doi:10.1103/PhysRevD.91.055008
[arXiv:1412.3403 [hep-ph]].

\bibitem{Bae:2015jea}
K.~J.~Bae, H.~Baer, V.~Barger, M.~R.~Savoy and H.~Serce,
``Supersymmetry with radiatively-driven naturalness: implications for WIMP and axion searches,''
Symmetry \textbf{7}, no.2, 788-814 (2015)
%doi:10.3390/sym7020788
[arXiv:1503.04137 [hep-ph]].

\bibitem{Mahbubani:2017gjh}
R.~Mahbubani, P.~Schwaller and J.~Zurita,
``Closing the window for compressed Dark Sectors with disappearing charged tracks,''
JHEP \textbf{06}, 119 (2017)
[erratum: JHEP \textbf{10}, 061 (2017)]
%doi:10.1007/JHEP06(2017)119
[arXiv:1703.05327 [hep-ph]].

\bibitem{Fukuda:2017jmk}
H.~Fukuda, N.~Nagata, H.~Otono and S.~Shirai,
``Higgsino Dark Matter or Not: Role of Disappearing Track Searches at the LHC and Future Colliders,''
Phys. Lett. B \textbf{781}, 306-311 (2018)
%doi:10.1016/j.physletb.2018.03.088
[arXiv:1703.09675 [hep-ph]].

\bibitem{Kowalska:2018toh}
K.~Kowalska and E.~M.~Sessolo,
``The discreet charm of higgsino dark matter - a pocket review,''
Adv. High Energy Phys. \textbf{2018}, 6828560 (2018)
%doi:10.1155/2018/6828560
[arXiv:1802.04097 [hep-ph]].

\bibitem{Baer:2018rhs}
H.~Baer, V.~Barger, D.~Sengupta and X.~Tata,
``Is natural higgsino-only dark matter excluded?,''
Eur. Phys. J. C \textbf{78}, no.10, 838 (2018)
%doi:10.1140/epjc/s10052-018-6306-y
[arXiv:1803.11210 [hep-ph]].

\bibitem{Han:2018wus}
T.~Han, S.~Mukhopadhyay and X.~Wang,
``Electroweak Dark Matter at Future Hadron Colliders,''
Phys. Rev. D \textbf{98}, no.3, 035026 (2018)
%doi:10.1103/PhysRevD.98.035026
[arXiv:1805.00015 [hep-ph]].

\bibitem{Fukuda:2019kbp}
H.~Fukuda, N.~Nagata, H.~Oide, H.~Otono and S.~Shirai,
``Cornering Higgsinos Using Soft Displaced Tracks,''
Phys. Rev. Lett. \textbf{124}, no.10, 101801 (2020)
%doi:10.1103/PhysRevLett.124.101801
[arXiv:1910.08065 [hep-ph]].

\bibitem{Baer:2020sgm}
H.~Baer, V.~Barger, S.~Salam, D.~Sengupta and X.~Tata,
``The LHC higgsino discovery plane for present and future SUSY searches,''
Phys. Lett. B \textbf{810}, 135777 (2020)
%doi:10.1016/j.physletb.2020.135777
[arXiv:2007.09252 [hep-ph]].

\bibitem{Rinchiuso:2020skh}
L.~Rinchiuso, O.~Macias, E.~Moulin, N.~L.~Rodd and T.~R.~Slatyer,
``Prospects for detecting heavy WIMP dark matter with the Cherenkov Telescope Array: The Wino and Higgsino,''
Phys. Rev. D \textbf{103}, no.2, 023011 (2021)
%doi:10.1103/PhysRevD.103.023011
[arXiv:2008.00692 [astro-ph.HE]].

\bibitem{Delgado:2020url}
A.~Delgado and M.~Quir\'os,
``Higgsino Dark Matter in the MSSM,''
Phys. Rev. D \textbf{103}, no.1, 015024 (2021)
%doi:10.1103/PhysRevD.103.015024
[arXiv:2008.00954 [hep-ph]].

\bibitem{Co:2021ion}
R.~T.~Co, B.~Sheff and J.~D.~Wells,
``Race to find split Higgsino dark matter,''
Phys. Rev. D \textbf{105}, no.3, 035012 (2022)
%doi:10.1103/PhysRevD.105.035012
[arXiv:2105.12142 [hep-ph]].

\bibitem{Baer:2021srt}
H.~Baer, V.~Barger, D.~Sengupta and X.~Tata,
``New angular and other cuts to improve the Higgsino signal at the LHC,''
Phys. Rev. D \textbf{105}, no.9, 095017 (2022)
%doi:10.1103/PhysRevD.105.095017
[arXiv:2109.14030 [hep-ph]].

\bibitem{Carpenter:2021jbd}
L.~M.~Carpenter, H.~Gilmer and J.~Kawamura,
``Exploring nearly degenerate higgsinos using mono-Z/W signal,''
Phys. Lett. B \textbf{831}, 137191 (2022)
%doi:10.1016/j.physletb.2022.137191
[arXiv:2110.04185 [hep-ph]].

\bibitem{Evans:2022gom}
J.~L.~Evans and K.~A.~Olive,
``Higgsino dark matter in pure gravity mediated supersymmetry,''
Phys. Rev. D \textbf{106}, no.5, 055026 (2022)
%doi:10.1103/PhysRevD.106.055026
[arXiv:2202.07830 [hep-ph]].

\bibitem{Baer:2022qrw}
H.~Baer, V.~Barger, D.~Sengupta and X.~Tata,
``Angular cuts to reduce the $\tau \bar \tau j$ background to the higgsino signal at the LHC,''
[arXiv:2203.03700 [hep-ph]].

\bibitem{Dessert:2022evk}
C.~Dessert, J.~W.~Foster, Y.~Park, B.~R.~Safdi and W.~L.~Xu,
``Higgsino Dark Matter Confronts 14~Years of Fermi \ensuremath{\gamma}-Ray Data,''
Phys. Rev. Lett. \textbf{130}, no.20, 201001 (2023)
%doi:10.1103/PhysRevLett.130.201001
[arXiv:2207.10090 [hep-ph]].

\bibitem{Carpenter:2023agq}
L.~M.~Carpenter, H.~Gilmer, J.~Kawamura and T.~Murphy,
``Taking aim at the wino-Higgsino plane with the LHC,''
Phys. Rev. D \textbf{109}, no.1, 015012 (2024)
%doi:10.1103/PhysRevD.109.015012
[arXiv:2309.07213 [hep-ph]].

\bibitem{Bisal:2023fgb}
S.~Bisal, A.~Chatterjee, D.~Das, S.A.~Pasha,
``Radiative Corrections to Aid the Direct Detection of the Higgsino-like Neutralino Dark Matter: Spin-Independent Interactions,''
[arXiv:2311.09937 [hep-ph]].

\bibitem{Bisal:2024ezn}
S.~Bisal, A.~Chatterjee, D.~Das and S.~A.~Pasha,
``Radiative corrections to the direct detection of the Higgsino-(and Wino-)like neutralino dark matter: Spin-dependent interactions,''
[arXiv:2410.18205 [hep-ph]].

\bibitem{Ibe:2023dcu}
M.~Ibe, Y.~Nakayama and S.~Shirai,
``Precise estimate of charged Higgsino/Wino decay rate,''
JHEP \textbf{03}, 012 (2024)
%doi:10.1007/JHEP03(2024)012
[arXiv:2312.08087 [hep-ph]].

\bibitem{Rodd:2024qsi}
N.~L.~Rodd, B.~R.~Safdi and W.~L.~Xu,
``CTA and SWGO can discover Higgsino dark matter annihilation,''
Phys. Rev. D \textbf{110}, no.4, 043003 (2024)
%doi:10.1103/PhysRevD.110.043003
[arXiv:2405.13104 [hep-ph]].



\bibitem{IceCube:2016dgk}
M.~G.~Aartsen \textit{et al.} [IceCube],
``Search for annihilating dark matter in the Sun with 3 years of IceCube data,''
Eur. Phys. J. C \textbf{77}, no.3, 146 (2017)
[erratum: Eur. Phys. J. C \textbf{79}, no.3, 214 (2019)]
%doi:10.1140/epjc/s10052-017-4689-9
[arXiv:1612.05949 [astro-ph.HE]].


\bibitem{HESS:2016mib}
H.~Abdallah \textit{et al.} [H.E.S.S.],
``Search for dark matter annihilations towards the inner Galactic halo from 10 years of observations with H.E.S.S.,''
Phys. Rev. Lett. \textbf{117}, no.11, 111301 (2016)
%doi:10.1103/PhysRevLett.117.111301
[arXiv:1607.08142 [astro-ph.HE]].

\bibitem{HESS:2018cbt}
H.~Abdallah \textit{et al.} [HESS],
``Search for $\gamma$-Ray Line Signals from Dark Matter Annihilations in the Inner Galactic Halo from 10 Years of Observations with H.E.S.S.,''
Phys. Rev. Lett. \textbf{120}, no.20, 201101 (2018)
%doi:10.1103/PhysRevLett.120.201101
[arXiv:1805.05741 [astro-ph.HE]].

\bibitem{HESS:2022ygk}
H.~Abdalla \textit{et al.} [H.E.S.S.],
``Search for Dark Matter Annihilation Signals in the H.E.S.S. Inner Galaxy Survey,''
Phys. Rev. Lett. \textbf{129}, no.11, 111101 (2022)
%doi:10.1103/PhysRevLett.129.111101
[arXiv:2207.10471 [astro-ph.HE]].

\bibitem{Fermi-LAT:2015ycq}
A.~Drlica-Wagner \textit{et al.} [Fermi-LAT and DES],
``Search for Gamma-Ray Emission from DES Dwarf Spheroidal Galaxy Candidates with Fermi-LAT Data,''
Astrophys. J. Lett. \textbf{809}, no.1, L4 (2015)
%doi:10.1088/2041-8205/809/1/L4
[arXiv:1503.02632 [astro-ph.HE]].

\bibitem{Fermi-LAT:2015att}
M.~Ackermann \textit{et al.} [Fermi-LAT],
``Searching for Dark Matter Annihilation from Milky Way Dwarf Spheroidal Galaxies with Six Years of Fermi Large Area Telescope Data,''
Phys. Rev. Lett. \textbf{115}, no.23, 231301 (2015)
%doi:10.1103/PhysRevLett.115.231301
[arXiv:1503.02641 [astro-ph.HE]].

\bibitem{MAGIC:2016xys}
M.~L.~Ahnen \textit{et al.} [MAGIC and Fermi-LAT],
``Limits to Dark Matter Annihilation Cross-Section from a Combined Analysis of MAGIC and Fermi-LAT Observations of Dwarf Satellite Galaxies,''
JCAP \textbf{02}, 039 (2016)
%doi:10.1088/1475-7516/2016/02/039
[arXiv:1601.06590 [astro-ph.HE]].

\bibitem{Fermi-LAT:2016uux}
A.~Albert \textit{et al.} [Fermi-LAT and DES],
``Searching for Dark Matter Annihilation in Recently Discovered Milky Way Satellites with Fermi-LAT,''
Astrophys. J. \textbf{834}, no.2, 110 (2017)
%doi:10.3847/1538-4357/834/2/110
[arXiv:1611.03184 [astro-ph.HE]].

\bibitem{CTAConsortium:2017dvg}
B.~S.~Acharya \textit{et al.} [CTA Consortium],
``Science with the Cherenkov Telescope Array,''
%WSP, 2018,
%ISBN 978-981-327-008-4
%doi:10.1142/10986
[arXiv:1709.07997 [astro-ph.IM]].

\bibitem{Albert:2019afb}
A.~Albert, R.~Alfaro, H.~Ashkar, C.~Alvarez, J.~Alvarez, J.~C.~Arteaga-Vel\'azquez, H.~A.~Ayala Solares, R.~Arceo, J.~A.~Bellido and S.~BenZvi, \textit{et al.}
``Science Case for a Wide Field-of-View Very-High-Energy Gamma-Ray Observatory in the Southern Hemisphere,''
[arXiv:1902.08429 [astro-ph.HE]].


\bibitem{PandaX-4T:2021bab}
Y.~Meng \textit{et al.} [PandaX-4T],
``Dark Matter Search Results from the PandaX-4T Commissioning Run,''
Phys. Rev. Lett. \textbf{127}, no.26, 261802 (2021)
%doi:10.1103/PhysRevLett.127.261802
[arXiv:2107.13438 [hep-ex]].

\bibitem{XENON:2023cxc}
E.~Aprile \textit{et al.} [XENON],
``First Dark Matter Search with Nuclear Recoils from the XENONnT Experiment,''
Phys. Rev. Lett. \textbf{131}, no.4, 041003 (2023)
%doi:10.1103/PhysRevLett.131.041003
[arXiv:2303.14729 [hep-ex]].

\bibitem{LZ:2022lsv}
J.~Aalbers \textit{et al.} [LZ],
``First Dark Matter Search Results from the LUX-ZEPLIN (LZ) Experiment,''
Phys. Rev. Lett. \textbf{131}, no.4, 041002 (2023)
%doi:10.1103/PhysRevLett.131.041002
[arXiv:2207.03764 [hep-ex]].

\bibitem{PandaX:2024qfu}
Z.~Bo \textit{et al.} [PandaX],
``Dark Matter Search Results from 1.54 Tonne$\cdot$Year Exposure of PandaX-4T,''
[arXiv:2408.00664 [hep-ex]].

\bibitem{LZCollaboration:2024lux}
J.~Aalbers \textit{et al.} [LZ Collaboration],
``Dark Matter Search Results from 4.2 Tonne-Years of Exposure of the LUX-ZEPLIN (LZ) Experiment,''
[arXiv:2410.17036 [hep-ex]].

\bibitem{Martin:2024pxx}
S.~P.~Martin,
``Implications of purity constraints on light Higgsinos,''
Phys. Rev. D \textbf{109}, no.9, 095045 (2024)
%doi:10.1103/PhysRevD.109.095045
[arXiv:2403.19598 [hep-ph]].

\bibitem{Billard:2013qya}
J.~Billard, L.~Strigari and E.~Figueroa-Feliciano,
``Implication of neutrino backgrounds on the reach of next generation dark matter direct detection experiments,''
Phys. Rev. D \textbf{89}, no.2, 023524 (2014)
%doi:10.1103/PhysRevD.89.023524
[arXiv:1307.5458 [hep-ph]].

\bibitem{OHare:2021utq}
C.~A.~J.~O'Hare,
``New Definition of the Neutrino Floor for Direct Dark Matter Searches,''
Phys. Rev. Lett. \textbf{127}, no.25, 251802 (2021)
%doi:10.1103/PhysRevLett.127.251802
[arXiv:2109.03116 [hep-ph]].

\bibitem{Martin:1997ns}
S.~P.~Martin,
``A Supersymmetry primer,''
%Adv. Ser. Direct. High Energy Phys. \textbf{18}, 1-98 (1998)
%doi:10.1142/9789812839657\_0001
[arXiv:hep-ph/9709356 [hep-ph]].

\bibitem{Roussy:2022cmp}
T.~S.~Roussy, L.~Caldwell, T.~Wright, W.~B.~Cairncross, Y.~Shagam, K.~B.~Ng, N.~Schlossberger, S.~Y.~Park, A.~Wang and J.~Ye, \textit{et al.}
``An improved bound on the electron\textquoteright{}s electric dipole moment,''
Science \textbf{381}, no.6653, adg4084 (2023)
%doi:10.1126/science.adg4084
[arXiv:2212.11841 [physics.atom-ph]].

\bibitem{Cesarotti:2018huy}
C.~Cesarotti, Q.~Lu, Y.~Nakai, A.~Parikh and M.~Reece,
``Interpreting the Electron EDM Constraint,''
JHEP \textbf{05}, 059 (2019)
%doi:10.1007/JHEP05(2019)059
[arXiv:1810.07736 [hep-ph]].

\bibitem{Cheung:2012qy}
C.~Cheung, L.~J.~Hall, D.~Pinner and J.~T.~Ruderman,
``Prospects and Blind Spots for Neutralino Dark Matter,''
JHEP \textbf{05}, 100 (2013)
%doi:10.1007/JHEP05(2013)100
[arXiv:1211.4873 [hep-ph]].

\bibitem{Moroi:1992zk}
T.~Moroi and Y.~Okada,
``Upper bound of the lightest neutral Higgs mass in extended supersymmetric Standard Models,''
Phys. Lett. B \textbf{295}, 73-78 (1992)
doi:10.1016/0370-2693(92)90091-H

\bibitem{Moroi:1991mg}
T.~Moroi and Y.~Okada,
``Radiative corrections to Higgs masses in the supersymmetric model with an extra family and antifamily,''
Mod. Phys. Lett. A \textbf{7}, 187-200 (1992)
doi:10.1142/S0217732392000124

\bibitem{Babu:2004xg}
K.S.~Babu, I.~Gogoladze and C.~Kolda,
``Perturbative unification and Higgs boson mass bounds,''
[arXiv:hep-ph/0410085].

\bibitem{Babu:2008ge}
K.S.~Babu, I.~Gogoladze, M.U.~Rehman and Q.~Shafi,
``Higgs Boson Mass, Sparticle Spectrum and Little Hierarchy Problem in Extended MSSM,''
Phys. Rev. D \textbf{78}, 055017 (2008)
%doi:10.1103/PhysRevD.78.055017
[arXiv:0807.3055 [hep-ph]].

\bibitem{Martin:2009bg}
S.~P.~Martin,
``Extra vector-like matter and the lightest Higgs scalar boson mass in low-energy supersymmetry,''
Phys. Rev. D \textbf{81}, 035004 (2010)
%doi:10.1103/PhysRevD.81.035004
[arXiv:0910.2732 [hep-ph]].

\bibitem{Graham:2009gy}
P.~W.~Graham, A.~Ismail, S.~Rajendran and P.~Saraswat,
``A Little Solution to the Little Hierarchy Problem: A Vector-like Generation,''
Phys. Rev. D \textbf{81}, 055016 (2010)
%doi:10.1103/PhysRevD.81.055016
[arXiv:0910.3020 [hep-ph]].


\bibitem{Ellis:2000ds}
J.~R.~Ellis, A.~Ferstl and K.~A.~Olive,
``Reevaluation of the elastic scattering of supersymmetric dark matter,''
Phys. Lett. B \textbf{481}, 304-314 (2000)
%doi:10.1016/S0370-2693(00)00459-7
[arXiv:hep-ph/0001005 [hep-ph]].

\bibitem{Baer:2003jb}
H.~Baer, C.~Balazs, A.~Belyaev and J.~O'Farrill,
``Direct detection of dark matter in supersymmetric models,''
JCAP \textbf{09}, 007 (2003)
%doi:10.1088/1475-7516/2003/09/007
[arXiv:hep-ph/0305191 [hep-ph]].

\bibitem{Baer:2006te}
H.~Baer, A.~Mustafayev, E.~K.~Park and X.~Tata,
``Target dark matter detection rates in models with a well-tempered neutralino,''
JCAP \textbf{01}, 017 (2007)
%doi:10.1088/1475-7516/2007/01/017
[arXiv:hep-ph/0611387 [hep-ph]].

\bibitem{Huang:2014xua}
P.~Huang and C.~E.~M.~Wagner,
``Blind Spots for neutralino Dark Matter in the MSSM with an intermediate $m_A$,''
Phys. Rev. D \textbf{90}, no.1, 015018 (2014)
%doi:10.1103/PhysRevD.90.015018
[arXiv:1404.0392 [hep-ph]].

\bibitem{Baum:2023inl}
S.~Baum, M.~Carena, T.~Ou, D.~Rocha, N.~R.~Shah and C.~E.~M.~Wagner,
``Lighting up the LHC with Dark Matter,''
JHEP \textbf{11}, 037 (2023)
%doi:10.1007/JHEP11(2023)037
[arXiv:2303.01523 [hep-ph]].

\bibitem{Arganda:2024tqo}
E.~Arganda, M.~Carena, M.~de los Rios, A.~D.~Perez, D.~Rocha, R.~M.~Sand\'a Seoane and C.~E.~M.~Wagner,
``Machine-Learning Analysis of Radiative Decays to Dark Matter at the LHC,''
[arXiv:2410.13799 [hep-ph]].

\bibitem{Allanach:2001kg}
B.~C.~Allanach,
``SOFTSUSY: a program for calculating supersymmetric spectra,''
Comput. Phys. Commun. \textbf{143}, 305-331 (2002)
%doi:10.1016/S0010-4655(01)00460-X
[arXiv:hep-ph/0104145].

\bibitem{Belanger:2001fz}
G.~Belanger, F.~Boudjema, A.~Pukhov and A.~Semenov,
``MicrOMEGAs: A Program for calculating the relic density in the MSSM,''
Comput. Phys. Commun. \textbf{149}, 103-120 (2002)
%doi:10.1016/S0010-4655(02)00596-9
[arXiv:hep-ph/0112278].

\bibitem{Belanger:2004yn}
G.~Belanger, F.~Boudjema, A.~Pukhov and A.~Semenov,
``micrOMEGAs: Version 1.3,''
Comput. Phys. Commun. \textbf{174}, 577-604 (2006)
%doi:10.1016/j.cpc.2005.12.005
[arXiv:hep-ph/0405253].

\bibitem{Belanger:2020gnr}
G.~Belanger, A.~Mjallal and A.~Pukhov,
``Recasting direct detection limits within micrOMEGAs and implication for non-standard Dark Matter scenarios,''
Eur. Phys. J. C \textbf{81}, no.3, 239 (2021)
%doi:10.1140/epjc/s10052-021-09012-z
[arXiv:2003.08621 [hep-ph]].

\bibitem{Alguero:2023zol}
G.~Alguero, G.~Belanger, F.~Boudjema, S.~Chakraborti, A.~Goudelis, S.~Kraml, A.~Mjallal and A.~Pukhov,
``micrOMEGAs 6.0: N-component dark matter,''
Comput. Phys. Commun. \textbf{299}, 109133 (2024)
%doi:10.1016/j.cpc.2024.109133
[arXiv:2312.14894 [hep-ph]].

\bibitem{Gondolo:2004sc}
P.~Gondolo, J.~Edsjo, P.~Ullio, L.~Bergstrom, M.~Schelke and E.~A.~Baltz,
``DarkSUSY: Computing supersymmetric dark matter properties numerically,''
JCAP \textbf{07}, 008 (2004)
%doi:10.1088/1475-7516/2004/07/008
[arXiv:astro-ph/0406204 [astro-ph]].

\bibitem{Harz:2023llw}
J.~Harz, B.~Herrmann, M.~Klasen, K.~Kova\v{r}\'\i{}k and L.~P.~Wiggering,
``Precision predictions for dark matter with DM@NLO in the MSSM,''
Eur. Phys. J. C \textbf{84}, no.4, 342 (2024)
%doi:10.1140/epjc/s10052-024-12660-6
[arXiv:2312.17206 [hep-ph]].

\bibitem{Randall:1998uk}
L.~Randall and R.~Sundrum,
``Out of this world supersymmetry breaking,''
Nucl. Phys. B \textbf{557}, 79-118 (1999)
%doi:10.1016/S0550-3213(99)00359-4
[arXiv:hep-th/9810155].

\bibitem{Giudice:1998xp}
G.~F.~Giudice, M.~A.~Luty, H.~Murayama and R.~Rattazzi,
``Gaugino mass without singlets,''
JHEP \textbf{12}, 027 (1998)
%doi:10.1088/1126-6708/1998/12/027
[arXiv:hep-ph/9810442].

\bibitem{XLZD:2024nsu}
J.~Aalbers \textit{et al.} [XLZD],
``The XLZD Design Book: Towards the Next-Generation Liquid Xenon Observatory for Dark Matter and Neutrino Physics,''
[arXiv:2410.17137 [hep-ex]].

\bibitem{ATLAS:2024umc}
G.~Aad \textit{et al.} [ATLAS],
``Search for Nearly Mass-Degenerate Higgsinos Using Low-Momentum Mildly Displaced Tracks in pp Collisions at s=13\,\,TeV with the ATLAS Detector,''
Phys. Rev. Lett. \textbf{132}, no.22, 221801 (2024)
%doi:10.1103/PhysRevLett.132.221801
[arXiv:2401.14046 [hep-ex]].

\bibitem{ATLAS:2022rme}
G.~Aad \textit{et al.} [ATLAS],
``Search for long-lived charginos based on a disappearing-track signature using 136 fb$^{-1}$ of pp collisions at $\sqrt{s}$~=~13~TeV with the ATLAS detector,''
Eur. Phys. J. C \textbf{82}, no.7, 606 (2022)
%doi:10.1140/epjc/s10052-022-10489-5
[arXiv:2201.02472 [hep-ex]].
%Also known as  CERN-EP-2021-209, ATLAS-CONF-2021-015

\bibitem{CMS:2023mny}
A.~Hayrapetyan \textit{et al.} [CMS],
``Search for supersymmetry in final states with disappearing tracks in proton-proton collisions at s=13\,\,TeV,''
Phys. Rev. D \textbf{109}, no.7, 072007 (2024)
%doi:10.1103/PhysRevD.109.072007
[arXiv:2309.16823 [hep-ex]].
%Also known as CERN-EP-2023-209, CMS-SUS-21-006-003, CMS-PAS-SUS-21-006.

\bibitem{ATLAS:2021moa}
G.~Aad \textit{et al.} [ATLAS],
``Search for chargino\textendash{}neutralino pair production in final states with three leptons and missing transverse momentum in $\sqrt{s} = 13$~TeV pp collisions with the ATLAS detector,''
Eur. Phys. J. C \textbf{81}, no.12, 1118 (2021)
%doi:10.1140/epjc/s10052-021-09749-7
[arXiv:2106.01676 [hep-ex]].
%Also known as CERN-EP-2021-059, ATLAS-CONF-2020-015.

\bibitem{CMS:2023qhl}
 [CMS],
``Combined search for electroweak production of winos, binos, higgsinos, and sleptons in proton-proton collisions at $sqrt{s}=$ 13 TeV,''
CMS-PAS-SUS-21-008.

\bibitem{CMS:2021edw}
A.~Tumasyan \textit{et al.} [CMS],
``Search for supersymmetry in final states with two or three soft leptons and missing transverse momentum in proton-proton collisions at $ \sqrt{s} $ = 13 TeV,''
JHEP \textbf{04}, 091 (2022)
%doi:10.1007/JHEP04(2022)091
[arXiv:2111.06296 [hep-ex]].
%Also known as CMS-SUS-18-004, CERN-EP-2021-168.

\bibitem{Agin:2023yoq}
D.~Agin, B.~Fuks, M.~D.~Goodsell and T.~Murphy,
``Monojets reveal overlapping excesses for light compressed higgsinos,''
[arXiv:2311.17149 [hep-ph]].

\bibitem{Chakraborti:2024pdn}
M.~Chakraborti, S.~Heinemeyer and I.~Saha,
``Consistent Excesses in the Search for $\tilde \chi_2^{\rm 0} \tilde \chi_1^{\rm \pm}$ : Wino/bino vs. Higgsino Dark Matter,''
[arXiv:2403.14759 [hep-ph]].

\bibitem{Agin:2024yfs}
D.~Agin, B.~Fuks, M.~D.~Goodsell and T.~Murphy,
``Seeking a coherent explanation of LHC excesses for compressed spectra,''
[arXiv:2404.12423 [hep-ph]].

\bibitem{ATLAS:2018jjf}
[ATLAS],
``ATLAS sensitivity to winos and higgsinos with a highly compressed mass spectrum at the HL-LHC,'' ATL-PHYS-PUB-2018-031, \color{blue}
\href{https://cds.cern.ch/record/2647294}{https://cds.cern.ch/record/2647294}\color{black}

\bibitem{Saito:2019rtg}
M.~Saito, R.~Sawada, K.~Terashi and S.~Asai,
``Discovery reach for wino and higgsino dark matter with a disappearing track signature at a 100 TeV $pp$ collider,''
Eur. Phys. J. C \textbf{79}, no.6, 469 (2019)
%doi:10.1140/epjc/s10052-019-6974-2
[arXiv:1901.02987 [hep-ph]].

\bibitem{Capdevilla:2021fmj}
R.~Capdevilla, F.~Meloni, R.~Simoniello and J.~Zurita,
``Hunting wino and higgsino dark matter at the muon collider with disappearing tracks,''
JHEP \textbf{06}, 133 (2021)
doi:10.1007/JHEP06(2021)133
[arXiv:2102.11292 [hep-ph]].

\bibitem{Han:2020uak}
T.~Han, Z.~Liu, L.~T.~Wang and X.~Wang,
``WIMPs at High Energy Muon Colliders,''
Phys. Rev. D \textbf{103}, no.7, 075004 (2021)
%doi:10.1103/PhysRevD.103.075004
[arXiv:2009.11287 [hep-ph]].

\bibitem{Han:2022ubw}
T.~Han, Z.~Liu, L.~T.~Wang and X.~Wang,
``WIMP Dark Matter at High Energy Muon Colliders $-$A White Paper for Snowmass 2021,''
[arXiv:2203.07351 [hep-ph]].

\bibitem{Capdevilla:2024bwt}
R.~Capdevilla, F.~Meloni and J.~Zurita,
``Discovering Electroweak Interacting Dark Matter at Muon Colliders using Soft Tracks,''
[arXiv:2405.08858 [hep-ph]].

\end{thebibliography}
\end{document}